\documentclass[12pt]{article}
\usepackage{amssymb}
\usepackage{graphicx}
\usepackage{xcolor}

\definecolor{primary}{RGB}{0, 139, 139}      % Deep Ocean Blue
\definecolor{secondary}{RGB}{180, 40, 40}   % Crimson Red
\definecolor{accent}{RGB}{40, 120, 120}      % Elegant Teal
\usepackage[colorlinks=true]{hyperref}
\hypersetup{linkcolor=primary,citecolor=accent,urlcolor=secondary}

\begin{document}
\newcommand{\beq}{\begin{equation}}
\newcommand{\eeq}{\end{equation}}
\newcommand{\beqa}{\begin{eqnarray}}
\newcommand{\eeqa}{\end{eqnarray}}
\newcommand{\beqar}{\begin{eqnarray*}}
\newcommand{\eeqar}{\end{eqnarray*}}
\newcommand{\al}{\alpha}
\newcommand{\be}{\beta}
\newcommand{\del}{\delta}
\newcommand{\D}{\Delta}
\newcommand{\eps}{\epsilon}
\newcommand{\ga}{\gamma}
\newcommand{\Ga}{\Gamma}
\newcommand{\ka}{\kappa}
\newcommand{\nn}{\nonumber}
\newcommand{\inn}{\!\cdot\!}
\newcommand{\h}{\eta}
\newcommand{\ii}{\iota}
\newcommand{\kk}{\varphi}
\newcommand\F{{}_3F_2}
\newcommand{\la}{\lambda}
\newcommand{\La}{\Lambda}
\newcommand{\na}{\prt}
\newcommand{\Om}{\Omega}
\newcommand{\om}{\omega}
\newcommand{\p}{\Phi}
\newcommand{\sig}{\sigma}
\renewcommand{\t}{\theta}
\newcommand{\z}{\zeta}
\newcommand{\ssc}{\scriptscriptstyle}
\newcommand{\eg}{{\it e.g.,}\ }
\newcommand{\ie}{{\it i.e.,}\ }
\newcommand{\labell}[1]{\label{#1}} %{\label{#1}} %
\newcommand{\reef}[1]{(\ref{#1})}
\newcommand\prt{\partial}
\newcommand\veps{\varepsilon}
\newcommand{\pol}{\varepsilon}
\newcommand\vp{\varphi}
\newcommand\ls{\ell_s}
\newcommand\cF{{\cal F}}
\newcommand\cA{{\cal A}}
\newcommand\cS{{\cal S}}
\newcommand\cT{{\cal T}}
\newcommand\cV{{\cal V}}
\newcommand\cL{{\cal L}}
\newcommand\cM{{\cal M}}
\newcommand\cN{{\cal N}}
\newcommand\cG{{\cal G}}
\newcommand\cK{{\cal K}}
\newcommand\cH{{\cal H}}
\newcommand\cI{{\cal I}}
\newcommand\cJ{{\cal J}}
\newcommand\cl{{\iota}}
\newcommand\cP{{\cal P}}
\newcommand\cQ{{\cal Q}}
\newcommand\cg{{\tilde {{\cal G}}}}
\newcommand\cR{{\cal R}}
\newcommand\cB{{\cal B}}
\newcommand\cO{{\cal O}}
\newcommand\tcO{{\tilde {{\cal O}}}}
\newcommand\bz{\bar{z}}
\newcommand\bb{\bar{b}}
\newcommand\ba{\bar{a}}
\newcommand\bg{\bar{g}}
\newcommand\bc{\bar{c}}
\newcommand\bom{\bar{\omega}}
\newcommand\bw{\bar{w}}
\newcommand\bX{\bar{X}}
\newcommand\bK{\bar{K}}
\newcommand\bA{\bar{A}}
\newcommand\bH{\bar{H}}
\newcommand\bF{\bar{F}}
\newcommand\bxi{\bar{\xi}}
\newcommand\bphi{\bar{\phi}}
\newcommand\bpsi{\bar{\psi}}
\newcommand\bprt{\bar{\prt}}
\newcommand\bet{\bar{\eta}}
\newcommand\btau{\bar{\tau}}
\newcommand\hF{\hat{F}}
\newcommand\hA{\hat{A}}
\newcommand\hT{\hat{T}}
\newcommand\htau{\hat{\tau}}
\newcommand\hD{\hat{D}}
\newcommand\hf{\hat{f}}
\newcommand\hK{\hat{K}}
\newcommand\hg{\hat{g}}
\newcommand\hp{\hat{\Phi}}
\newcommand\hi{\hat{i}}
\newcommand\ha{\hat{a}}
\newcommand\hb{\hat{b}}
\newcommand\hQ{\hat{Q}}
\newcommand\hP{\hat{\Phi}}
\newcommand\hS{\hat{S}}
\newcommand\hX{\hat{X}}
\newcommand\tL{\tilde{\cal L}}
\newcommand\hL{\hat{\cal L}}
\newcommand\tG{{\tilde G}}
\newcommand\ti{{\tilde i}}
\newcommand\tj{{\tilde j}}
\newcommand\tg{{\tilde g}}
\newcommand\tphi{{\widetilde \Phi}}
\newcommand\tPhi{{\widetilde \Phi}}
\newcommand\te{{\tilde e}}
\newcommand\tk{{\tilde k}}
\newcommand\tf{{\tilde f}}
\newcommand\tH{{\tilde H}}
\newcommand\ta{{\tilde a}}
\newcommand\tb{{\tilde b}}
\newcommand\tc{{\tilde c}}
\newcommand\td{{\tilde d}}
\newcommand\tm{{\tilde m}}
\newcommand\tmu{{\tilde \mu}}
\newcommand\tnu{{\tilde \nu}}
\newcommand\talpha{{\tilde \alpha}}
\newcommand\tbeta{{\tilde \beta}}
\newcommand\trho{{\tilde \rho}}
 \newcommand\tR{{\tilde R}}
\newcommand\teta{{\tilde \eta}}
\newcommand\tF{{\widetilde F}}
\newcommand\tK{{\tilde K}}
\newcommand\tE{{\widetilde E}}
\newcommand\tpsi{{\tilde \psi}}
\newcommand\tX{{\widetilde X}}
\newcommand\tD{{\widetilde D}}
\newcommand\tO{{\widetilde O}}
\newcommand\tS{{\tilde S}}
\newcommand\tB{{\tilde B}}
\newcommand\tA{{\widetilde A}}
\newcommand\tT{{\widetilde T}}
\newcommand\tC{{\widetilde C}}
\newcommand\tV{{\widetilde V}}
\newcommand\thF{{\widetilde {\hat {F}}}}
\newcommand\Tr{{\rm Tr}}
\newcommand\tr{{\rm tr}}
\newcommand\STr{{\rm STr}}
\newcommand\hR{\hat{R}}
\newcommand\M[2]{M^{#1}{}_{#2}}
\newcommand\MZ{\mathbb{Z}}
\newcommand\MR{\mathbb{R}}
\newcommand\bS{\textbf{ S}}
\newcommand\bI{\textbf{ I}}
\newcommand\bJ{\textbf{ J}}

%\begin{document}
\begin{titlepage}
\begin{center}

\vskip 0.5 cm
{\LARGE \bf  
 Heterotic string couplings at  order $\alpha'^3$ \\[6pt] in NS-NS sector }\\
\vskip 1.25 cm
Alireza  Pahlavan\footnote{alirezapahlavan08@gmail.com}, Mehdi Ameri\footnote{amerimehdi9755@gmail.com}, Mohammad R. Garousi\footnote{garousi@um.ac.ir}

\vskip 1 cm
{{\it Department of Physics, Faculty of Science, Ferdowsi University of Mashhad\\}{\it P.O. Box 1436, Mashhad, Iran}\\}
\vskip .1 cm
 \end{center}

\begin{abstract}

We utilize the standard T-duality procedure to derive the classical effective action of heterotic string theory at the eight-derivative order within the NS–NS sector, which comprises the metric, the \(B\)-field, and the dilaton. Starting from the minimal basis at this order—consisting of 872 even-parity and 477 odd-parity couplings—we perform a dimensional reduction on a circle and impose invariance under T-duality transformations, specifically the Buscher rules supplemented by higher-derivative corrections. This invariance uniquely fixes a subset of the couplings to match those of type II theory up to an overall factor, while the remaining couplings are determined in terms of the coefficients at order \(\alpha'\). For these latter couplings, we adopt both the Metsaev–Tseytlin and Meissner schemes. Subsequently, through field redefinitions, we recast the action into a canonical form in which the dilaton appears solely through the overall factor \(e^{-2\Phi}\). In the Meissner scheme, the pure gravity sector precisely reproduces the known S-matrix results. In both schemes, the pure gravity terms can be expressed as the double trace \((\Tr(R^2))^2\), and when combined with the corresponding Yang–Mills terms \((\Tr(F^2))^2\), they take the unified form \((\Tr(R^2 - F^2))^2\), as anticipated in the literature.

\end{abstract}

\vspace{2cm}
\end{titlepage}
\tableofcontents

\newpage
\section{Introduction}

String theory remains the most comprehensive and promising framework for quantum gravity. In this theory, the particles observed in our universe are believed to correspond to the massless states arising from the fluctuations of a fundamental free string, which lives in the critical dimension \(D\). The interactions among these massless fields are described by string theory S-matrix elements, defined as correlation functions of the corresponding vertex operators on the worldsheet—a two-dimensional Riemann surface of arbitrary genus \cite{Becker:2007zj}. At weak coupling, the sphere-level S-matrix elements determine the dynamics of these massless fields, while massive fields contribute through various physical channels in these amplitudes. At low energies, the contributions from massive poles are expanded, yielding S-matrix elements that contain only massless poles along with a series of higher-momentum contact terms. One may then propose a higher-derivative field theory that reproduces both the massless pole structure and the first few higher-momentum contact terms \cite{Gross:1986iv,Gross:1986mw}. In principle, this S-matrix approach can be employed to systematically derive the weak-coupling, low-energy effective action of string theory.

The spectrum of free string theory in spacetime exhibits various symmetries, such as supersymmetry in the case of superstrings, and T-duality when \(n\) spatial directions are compactified on a torus \(T^{(n)}\), applicable to bosonic, super, and heterotic strings \cite{Becker:2007zj}. In the latter case, the \((D-n)\)-dimensional spectrum is invariant under T-duality, which manifests as an \(O(n,n,\MZ)\) symmetry. It is widely believed in string theory that such symmetries should persist in the interacting theory as well. Consequently, one may employ these symmetries to derive the effective action, as an alternative to the more direct S-matrix approach described earlier (see, e.g., \cite{Ozkan:2024euj} for the use of supersymmetry in this context). 
To utilize T-duality for determining the effective action, one must first compactify the theory on at least one circle. The symmetry observed in the \((D-1)\)-dimensional spectrum is then expected to be a symmetry of the corresponding \((D-1)\)-dimensional effective action. At first glance, it might seem that this symmetry cannot be used to constrain the original \(D\)-dimensional effective action. However, at the sphere level (classical level), the effective action is background-independent \cite{Garousi:2022ovo}. This is because the corresponding string theory S-matrix elements involve no loops and therefore no internal momentum integrals. Consequently, the \(D\)-dimensional and \((D-1)\)-dimensional S-matrix elements coincide, implying that the associated effective actions are identical.
In contrast, at the torus level (one-loop order), the effective action becomes background-dependent \cite{Garousi:2025xdz}. In this case, the S-matrix elements involve a loop integral over internal momenta in \(D\) dimensions. When one spacetime dimension is compactified on a circle, this integral is replaced by a sum over Kaluza–Klein momenta, along with an additional sum over winding momenta—contributions that are absent in the \(D\)-dimensional S-matrix elements. As a result, the \(D\)-dimensional and \((D-1)\)-dimensional effective actions are no longer identical \cite{Garousi:2025xdz,Garousi:2025gsl}. Since T-duality symmetry exists only in the compactified \((D-1)\)-dimensional theory, it cannot be used to directly determine the loop-level effective action in the original \(D\) dimensions \cite{Garousi:2025gsl}.

 At the classical level, however, T-duality can be used to determine the effective action in \(D\) dimensions. There are two primary approaches to employing T-duality for this purpose. One method is Double Field Theory (DFT)  \cite{Siegel:1993xq,Siegel:1993th,Siegel:1993bj,Hull:2009mi,Aldazabal:2013sca}, which exploits background independence at the classical level by doubling the spacetime dimensions—one set corresponding to the usual Kaluza–Klein momenta and the other to winding momenta—while requiring that the fields do not depend on the winding coordinates. Using the fact that the T-duality symmetry of the full quantum theory compactified on \(T^{(n)}\), which is \(O(n,n,\mathbb{Z})\), is enlarged to an \(O(n,n,\mathbb{R})\) symmetry at the classical level \cite{Sen:1991zi,Hohm:2014sxa}, one then demands that the classical effective action in this \(2D\)-dimensional framework exhibit a manifest \(O(D,D,\mathbb{R})\) T-duality symmetry, though this requires deforming the other spacetime symmetries \cite{Hohm:2014xsa,Marques:2015vua}. This approach has been successfully applied to derive the effective action of string theory at orders \(\alpha'\) and \(\alpha'^2\) \cite{Baron:2018lve,Baron:2020xel,Hassler:2024yis,Gitsis:2024gfb,Gitsis:2025clo,Gitsis:2026ufz}. However, there are indications that this symmetry may not persist at order \(\alpha'^3\) \cite{Hronek:2020xxi,Hsia:2024kpi}.

 The other method for using T-duality to derive the classical effective action is a more direct approach: one reduces the \(D\)-dimensional effective action on a circle and then imposes T-duality on the resulting \((D-1)\)-dimensional effective action—which is equivalent to the original \(D\)-dimensional action after decomposing the massless fields into their \((D-1)\)-dimensional components \cite{Garousi:2019wgz}. In this approach, however, the T-duality symmetry is not manifest, as the Buscher rules governing the transformations of the \((D-1)\)-dimensional massless fields must be deformed by higher-derivative corrections. Importantly, the other spacetime symmetries—namely diffeomorphism and local Lorentz invariance—remain manifest. This method has proven successful in determining the effective action at orders \(\alpha'\), \(\alpha'^2\), and \(\alpha'^3\) in bosonic string theory \cite{Garousi:2019wgz,Garousi:2019mca,Wulff:2024ips,Ameri:2025bei}, at orders \(\alpha'\) and \(\alpha'^2\) in heterotic string theory  \cite{Garousi:2019wgz,Garousi:2023kxw}, and at order \(\alpha'^3\) in the NS–NS sector of type II superstring theory \cite{Garousi:2020gio,Garousi:2020lof,Garousi:2022ghs}. In this paper, we aim to apply this method to derive the effective action of heterotic string theory in the NS–NS sector at order \(\alpha'^3\).
 
 The structure of the manuscript is as follows. In the next section, we review the T-duality method for constructing the effective action. In Section 3, we apply this method to derive the effective action of heterotic string theory at orders \(\alpha'^2\) and \(\alpha'^3\) in the Metsaev–Tseytlin scheme, where the \(\alpha'\)-order couplings are the known ones from the Metsaev–Tseytlin action. In Section 4, we apply the T-duality method to find the \(\alpha'^3\) couplings in the Meissner scheme, where the \(\alpha'\)-order couplings are those of the Meissner action, and the corresponding \(\alpha'^2\) couplings have already been determined in \cite{Garousi:2023kxw,Gholian:2023kjj} via T-duality. In Section 5, we briefly discuss our results. All calculations were performed using the "xAct" package \cite{Nutma:2013zea} for tensor algebra and perturbation theory.

\section{T-duality constraint}

In this section, we summarize the constraint that spacetime T-duality imposes on the classical effective action \({\bf S}(\Psi)\), where \(\Psi\) collectively denotes the massless fields in \(D\) dimensions. When one of the spatial directions is a circle, the T-duality constraint on the circularly reduced effective action \(S(\psi)\) takes the form \cite{ Kaloper:1997ux, Garousi:2019wgz,Garousi:2023kxw}
\beqa
S(\psi) \sim S(\psi')\,, \labell{SS}
\eeqa
where \(\psi\) represents the components of the \(D\)-dimensional field \(\Psi\) reduced to \((D-1)\) dimensions, and \(\psi'\) denotes its image under the T-duality transformation, given by the Buscher rules \cite{Buscher:1987sk,Rocek:1991ps} modified by higher-derivative corrections. The symbol \(\sim\) indicates equality up to total derivative terms and the use of Bianchi identities.

To solve the constraint \reef{SS}, one expands the effective action in derivatives, equivalently in powers of \(\alpha'\), as
\beqa
\bS=\sum_{n=0}^{\infty}\alpha'^n\bS^{(n)}&;& \bS^{(n)}=-\frac{2}{\kappa^2}\int d^Dx\sqrt{-G}\,e^{-2\Phi}\cL_n\,,\labell{Sn}
\eeqa 
where at each order \(\cL_n\) consists of a minimal basis of massless-field terms with unknown coupling constants. The reduction of this \(D\)-dimensional action similarly admits an \(\alpha'\)-expansion \(S=\sum_{n=0}^{\infty}\alpha'^nS^{(n)}\). The unknown coupling constants of the \(D\)-dimensional action \({\bS}^{(n)}\) also appear in the reduced \((D-1)\)-dimensional action \(S^{(n)}\).

At leading order, the T-duality constraint reduces to
\beqa
S^{(0)}(\psi)&\sim & S^{(0)}(\psi_0)\labell{S0}
\eeqa
where \(\psi_0\) denotes the Buscher rules  \cite{Buscher:1987sk,Rocek:1991ps}. These Buscher transformations for the NS-NS fields simplify considerably when using the following circular reduction ansatz \cite{Maharana:1992my}:  
\beqa  
G_{\mu\nu} = \left(\matrix{\bg_{ab} + e^{\varphi} g_{a} g_{b} &  e^{\varphi} g_{a} \cr e^{\varphi} g_{b} & e^{\varphi} &}\!\!\!\!\!\right), \quad  
B_{\mu\nu} = \left(\matrix{\bb_{ab} + b_{[a} g_{b]} & b_{a} \cr -b_{b} & 0 &}\!\!\!\!\!\right), \quad  
\Phi = \bar{\phi} + \varphi/4,  
\labell{reduction}  
\eeqa  
where indices \(a,b\) denote directions orthogonal to the Killing coordinate \(y\). In the above reduction, \(\bg_{ab}\) is the \((D-1)\)-dimensional base-space metric, \(\bb_{ab}\) is an antisymmetric tensor, \(\bar{\phi}\) is the dilaton, \(\varphi\) is a scalar, and \(g_a, b_a\) are two vectors. Their T-duality transformation \(\psi'=\psi_0\), i.e., the Buscher rules, are then given by:
\beqa  
&&\bg_{ab}' = \bg_{ab}\,, \quad 
\bar{H}_{abc}' = \bar{H}_{abc}\,, \quad 
\bar{\phi}' = \bar{\phi}\,, \nn\\
&&\varphi' = -\varphi\,, \quad 
g_a' = b_a\,, \quad 
b_a' = g_a\,.  
\labell{flac}  
\eeqa  
where the field strength \(\bH\) is expressed in terms of the field strengths \(W = 2db\) and \(V = 2dg\) as \cite{Kaloper:1997ux}:
\beqa
\bar{H} &= 3d\bb- \frac{3}{2}g\wedge W - \frac{3}{2}b\wedge V\,,
\eeqa
These field strengths satisfy the Bianchi identities:
\begin{equation}
dW=0\,,\,\,\,\,\, dV=0\,,\,\,\,\,\, d\bar{H} = -\frac{3}{2}V\wedge W\,. \labell{bian0}
\end{equation}
Using the reductions \reef{reduction}, one finds that the leading-order Lagrangian
\begin{equation}
{\bf S}^{(0)}=-\frac{2}{\kappa^2}\int \mathrm{d}^{10}x e^{-2\Phi}\sqrt{-G}\big(R+4\nabla_a\Phi\nabla^a\Phi-\frac{1}{12}H^2\big)\labell{leading}
\end{equation}
satisfies the T-duality constraint \reef{S0} (see, e.g., \cite{Garousi:2019wgz}).

Beyond leading order, the inclusion of higher-derivative terms in the effective action necessitates an \(\alpha'\)-extension of the Buscher rules \cite{Garousi:2019wgz}. This extension takes the general form
\beqa
\psi' = \psi_0 + \sum_{n=1}^{\infty} \alpha'^n \psi^{(n)} \,,\labell{psi}
\eeqa
where \(\psi_0\) represents the leading-order Buscher rules given in \reef{flac}, and \(\psi_n\) denotes the higher-derivative correction at order \(\alpha'^n\). These corrections must satisfy the Bianchi identity \reef{bian0}, which relates \(\bH^{(n)}\) to \(g^{(n)}=e^{\vp/2}\Delta g^{(n)}\) and \(b^{(n)}=e^{-\vp/2}\Delta b^{(n)}\) as follows \cite{Garousi:2023kxw, Ameri:2025bei}:
\beqa
\bar{H}^{(n)} &=& 3d\bar{B}^{(n)}- 3 W \wedge b^{(n)} - 3 g^{(n)} \wedge V \labell{dHbar1} \\
&& - 3\sum_{0<k<n} \Big[ g^{(k)} \wedge db^{(n-k)} + dg^{(k)} \wedge b^{(n-k)} \Big]\,.\nn
\eeqa 
In this equation, the  \(\bar{B}^{(n)}\)  is a 2-form at order $\alpha'^{n}$. Each \(\psi^{(n)}=\{\varphi^{(n)},\bphi^{(n)},\bg^{(n)},\Delta b^{(n)},\Delta g^{(n)},\bar{B}^{(n)}\}\) includes all possible gauge invariant contractions of the \((D-1)\)-dimensional fields at order \(\alpha'^n\), with unknown parameters associated to each term.  Each \(V\) appears as \(e^{\varphi/2}V\) and each \(W\) as \(e^{-\varphi/2}W\) in the contractions.
Then, using the Taylor expansion of \(S^{(n)}(\psi')\) as
\beqa
S^{(n)}(\psi') = S^{(n)}(\psi_0) + \sum_{m=1}^{\infty} \alpha'^m S^{(n,m)}(\psi_0) \,,\labell{Tay}
\eeqa
one can rewrite the T-duality constraint \reef{SS} at order \(\alpha'^n\) for \(n>0\) as \cite{Garousi:2023kxw, Ameri:2025bei}:
\beqa
\fbox{$\displaystyle S^{(0,n)}_{(n)}(\psi_0) \sim S^{(n)}(\psi) - S^{(n)}(\psi_0) - S^{(0,n)}_{(n')}(\psi_0) - \sum_{0<k<n} S^{(k,n-k)}(\psi_0)$}\,.\labell{Tn}
\eeqa
Here, the Taylor expansion of the leading-order action is split into two parts:
\beqa
S^{(0,n)}(\psi_0) = S^{(0,n)}_{(n)}(\psi_0) + S^{(0,n)}_{(n')}(\psi_0)\labell{S0n}\,,
\eeqa
where the first term on the right-hand side is linear in the correction \(\psi^{(n)}\), while the second term depends nonlinearly on the lower-order corrections \(\psi^{(1)}, \psi^{(2)}, \dots, \psi^{(n-1)}\). 

The constraint \reef{Tn} can be used iteratively to determine both the effective actions and the corresponding higher-derivative corrections to the Buscher rules. Specifically, at order \(n=1\), the second term in \reef{S0n} vanishes. The parameters of \(\psi^{(1)}\) then appear on the left-hand side of \reef{Tn}, while the coupling constants in \(S^{(1)}\) appear on the right-hand side. Solving this equation determines the coupling constants in \(S^{(1)}\) and fixes the parameters in \(\psi^{(1)}\) up to one overall coupling constant \cite{Garousi:2019wgz}. Some parameters in \(\psi^{(1)}\) may appear only in the homogeneous solution of \reef{Tn}, where the right-hand side is zero. Since we are not interested in such homogeneous solutions, we discard these parameters, thereby fixing \(\psi^{(1)}\) in terms of the remaining unfixed coupling constant in \(S^{(1)}\) \cite{Garousi:2019wgz}. At order \(n=2\), the second term in \reef{S0n} is nonzero but contains no new parameters. Thus, the unknown parameters of \(\psi^{(2)}\) appear solely on the left-hand side of \reef{Tn}, while the coupling constants of \(S^{(2)}\) and the unfixed coupling constant from the previous order appear on the right-hand side. The inhomogeneous solution of \reef{Tn} fixes the couplings in \(S^{(2)}\) in terms of the unfixed coupling constant from the previous order, and determines the parameters in \(\psi^{(2)}\) in terms of this unfixed coupling \cite{Garousi:2019mca,Garousi:2023kxw}. This same procedure applies at higher orders \(n\), yielding the effective action and its associated T-duality transformations up to a small number of unfixed coupling constants, which must ultimately be determined by other methods, such as \(S\)-matrix computations. Importantly, the coupling-constant relations derived from the T-duality constraint \reef{Tn} are independent of the base-space geometry \cite{Garousi:2019mca}. This allows us to considerably simplify the calculation by assuming a flat base space, which we adopt in the present work.

The T-duality constraint \reef{Tn} serves two main purposes:  
(i) determining the T-duality corrections for known effective actions at lower orders in \(\alpha'\) (where all parameters are already fixed) and subsequently using these corrections to derive the effective action at higher orders; and  
(ii) simultaneously solving for the minimal-basis parameters of the effective action along with their corresponding T-duality corrections. Our interest lies in employing the T-duality constraint to obtain the effective action at order \(\alpha'^3\) in the NS-NS sector of heterotic string theory, namely
\[
\mathbf{S} = \mathbf{S}^{(0)} + \alpha' \, \mathbf{S}^{(1)} + \alpha'^2 \, \mathbf{S}^{(2)} + \alpha'^3 \, \mathbf{S}^{(3)}.
\]
The leading-order action is given in \reef{leading}. The actions at order \(\alpha'\) depend on the scheme chosen for the couplings; however, they are identical across all schemes up to field redefinitions \cite{Gross:1986iv,Metsaev:1987zx}. We consider the \(\alpha'\) couplings in two specific schemes: the Metsaev–Tseytlin scheme \cite{Metsaev:1987zx}, which minimizes the number of couplings at order \(\alpha'\), and the Meissner scheme \cite{Meissner:1996sa}, where the propagators derived from the leading-order action remain unaffected by \(\alpha'\) corrections. It is well known that the effective action at order \(\alpha'^2\) and beyond inherits scheme dependence from the \(\alpha'\) order \cite{Bento:1990nv}.  
In the following section, we derive the \(\alpha'^3\) couplings that are specifically compatible with the Metsaev–Tseytlin scheme.

\section{Effective action in Metsaev-Tseytlin  scheme}

Unlike the effective action of bosonic string theory and the NS-NS sector of superstring theory, where the $B$-field gauge transformation takes the standard form $\delta B_{\mu\nu}=\partial_\mu\lambda_{\nu}-\partial_\nu\lambda_{\mu}$, rendering the field strength $H=dB$ gauge invariant, the corresponding gauge transformation in the heterotic theory is anomalous due to the Green–Schwarz anomaly cancellation mechanism \cite{Green:1984sg}. Consequently, $H$ is not gauge invariant in the heterotic theory. In the absence of Yang–Mills fields, a gauge-invariant field strength is obtained by replacing $H$ with the following modified expression \cite{Green:1984sg}:
\beqa
 H_{\mu\nu\alpha}&\to &H_{\mu\nu\alpha}-\frac{3}{2}\alpha'\Omega_{\mu\nu\alpha}(\omega) \labell{Green-Schwarz}
\eeqa
Here, $\Omega_{\mu\nu\alpha}$ denotes the gravitational Chern–Simons three-form, defined as
\beqa
\Omega_{\mu\nu\alpha}(\omega)&=&\omega_{[\mu i}{}^j\partial_\nu\omega_{\alpha] j}{}^i+\frac{2}{3}\omega_{[\mu i}{}^j\omega_{\nu j}{}^k\omega_{\alpha]k}{}^i\,;~~~~\omega_{\mu i}{}^j=\partial_\mu e_\nu{}^j e^\nu{}_i-\Gamma_{\mu\nu}{}^\rho e_\alpha{}^j e^\nu{}_i,
\eeqa
where $\omega_{\mu i}{}^j$ is the spin connection, $e_{\mu}{}^i$ is the vielbein satisfying $G_{\mu\nu}=e_{\mu}{}^i e_{\nu}{}^j \eta_{ij}$, and $\Gamma_{\mu\nu}{}^\rho$ is the Christoffel connection. The spin connection with purely curved indices is given by $\omega_{\mu\nu\alpha}=\omega_{\mu i}{}^j e_{\nu j} e_\alpha{}^i$, which is antisymmetric in its last two indices. We adopt this tensor throughout our calculations. In terms of this tensor, the Chern–Simons form and the Riemann curvature are expressed as
\beqa
\Omega_{\alpha\beta\rho}(\omega)&=&\omega_{[\alpha}{}^{\mu\nu}\nabla_\beta\omega_{\rho]\nu\mu}-\frac{4}{3}\omega_{[\alpha}{}^{\mu\nu}\omega_{\beta\nu}{}^\gamma\omega_{\rho]\gamma\mu},\nn\\
R_{\alpha\beta\gamma\mu}(\omega)&=&\nabla_{\gamma}\omega _{\mu\alpha\beta}-\nabla_{\mu}\omega _{\gamma\alpha\beta}+\omega_{\gamma\alpha}{}^\nu\omega_{\mu\beta\nu}-\omega_{\gamma\beta}{}^\nu\omega_{\mu\alpha\nu},
\eeqa
where in the first line the indices $\alpha,\beta,\rho$ are to be fully antisymmetrized.

In the heterotic theory, the fundamental field is the vielbein rather than the metric. The reduction of the vielbein consistent with the metric reduction in \reef{reduction} is given by
\beqa
&&e_\mu{}^i=\left(\matrix{\bar{e}_a{}^\ti & 0 &\cr e^{\vp/2}g_{a }&e^{\vp/2}&}\right),
\eeqa
where $\bar{e}_a{}^\ti \bar{e}_b{}^\tj\eta_{\ti\tj}=\bg_{ab}$.
Using this reduction, one obtains the reduced components of $\omega$ and $\nabla\omega$ as follows:
\beqa
\omega_{abc}&\!\!\!=\!\!\!&\bom_{abc}\,,\,\,\omega_{aby}=\frac{1}{2}e^\vp V_{ab}\,,\,\,\omega_{ybc}=-\frac{1}{2}e^\vp V_{bc}\,,\,\,\omega_{yby}=-\frac{1}{2}e^\vp\nabla_b\vp,\nn\\
\nabla_d\omega_{abc}&=&\nabla_d\bom_{abc}+\frac{1}{4}e^\vp V_{ad}V_{bc}+\frac{1}{4}e^\vp V_{ac}V_{bd}-\frac{1}{4}e^\vp V_{ab}V_{cd},\nn\\
\nabla_d\omega_{aby}&=&\frac{1}{2}e^\vp V^c{}_d\bom_{abc}+\frac{1}{4}e^\vp V_{ad}\nabla_b\vp+\frac{1}{4}e^\vp V_{ab}\nabla_d\vp+\frac{1}{2}e^\vp\nabla_dV_{ab},\nn\\
\nabla_d\omega_{yba}&=&\frac{1}{2}e^\vp V^c{}_d\bom_{cba}-\frac{1}{4}e^\vp V_{bd}\nabla_a\vp+\frac{1}{4}e^\vp V_{ad}\nabla_b\vp+\frac{1}{4}e^\vp V_{ab}\nabla_d\vp+\frac{1}{2}\nabla_d V_{ab},\nn\\
\nabla_d\omega_{yby}&=&\frac{1}{2}e^{2\vp}V_b{}^aV_{da}-\frac{1}{2}e^\vp\nabla_d\nabla_b\vp\,,\,\,\nabla_y\omega_{yby}=-\frac{1}{4}e^{2\vp}V_{ba}\nabla^a\vp,\label{Redo}\\
\nabla_y\omega_{aby}&=&\frac{1}{4}e^\vp\nabla_a\vp\nabla_b\vp-\frac{1}{2}e^\vp\bom_{acb}\nabla^c\vp\,,\,\,\nabla_y\omega_{ybc}=\frac{1}{2}e^\vp\bom_{abc}\nabla^a\vp,\nn\\
\nabla_y\omega_{abc}&=&\frac{1}{2}e^\vp V^d{}_c\bom_{abd}-\frac{1}{2}e^\vp V^d{}_b\bom_{acd}-\frac{1}{2}e^\vp V^d{}_a\bom_{dcb}+\frac{1}{4}e^\vp V_{bc}\nabla_a\vp+\frac{1}{4}e^\vp V_{ac}\nabla_b\vp-\frac{1}{4}e^\vp V_{ab}\nabla_c\vp.\nn
\eeqa
Here, the covariant derivatives on the right-hand side are taken with respect to the $(D-1)$-dimensional base space, and $\bom_{abc}$ denotes the spin connection on the base space. In the above reductions, we have omitted terms involving the vector $g_a$ without derivatives, which, as argued in \cite{Garousi:2019mca}, cancel when reducing invariant quantities such as $R_{\mu\nu\alpha\beta}R^{\mu\nu\alpha\beta}$. The reductions of $H$, $\nabla H$, $\nabla\Phi$, and $\nabla\nabla\Phi$ can be found in \cite{Garousi:2019mca}.

\subsection{Couplings at order $\alpha'$}

The four-derivative effective action of heterotic string theory in the Metsaev–Tseytlin (MT) scheme is given by \cite{Metsaev:1987zx}
\beqa
{\bf S}_{\mathrm{MT}}^{(1)}&=&-\frac{2}{8\kappa^2}\int d^{10}x \sqrt{-G}e^{-2\Phi}\Big[R_{\alpha \beta \gamma \delta } R^{\alpha \beta \gamma \delta }+ \frac{1}{24} H_{\alpha }{}^{\delta \epsilon } H^{\alpha \beta \gamma } H_{\beta \delta }{}^{\varepsilon } H_{\gamma \epsilon \varepsilon } \nn\\&&\qquad\qquad\qquad\qquad\qquad  -  \frac{1}{8} H_{\alpha \beta }{}^{\delta } H^{\alpha \beta \gamma } H_{\gamma }{}^{\epsilon \varepsilon } H_{\delta \epsilon \varepsilon } -  \frac{1}{2} H_{\alpha }{}^{\delta \epsilon } H^{\alpha \beta \gamma } R_{\beta \gamma \delta \epsilon }\Big]\,.\labell{fourmin}
\eeqa
This action is even under parity. There is also an odd-parity coupling arising from substituting the Green–Schwarz modification \reef{Green-Schwarz} into the leading-order action \reef{leading}:
\beqa
 {\bf S}^{(1)}_{\mathrm{GS}}&=&-\frac{2}{\kappa^2}\int\mathrm{d}^{10}x\sqrt{-G}e^{-2\Phi}\big[\frac{1}{4}H_{\alpha\beta\gamma}\Omega^{\alpha\beta\gamma}\big].\labell{GS1}
\eeqa
The reduction of the full effective action at this order,
\beqa
 {\bf S}^{(1)}&=&{\bf S}_{\mathrm{MT}}^{(1)}+{\bf S}^{(1)}_{\mathrm{GS}}\,,\labell{S1}
 \eeqa
must satisfy the T-duality constraint \reef{Tn} at order $\alpha'$, which takes the form
\beqa  
S^{(0,1)}_{(1)}(\psi_0) \sim S^{(1)}(\psi) - S^{(1)}(\psi_0). \labell{T1}  
\eeqa 
Using the reductions in \reef{Redo} together with the corresponding reductions for $H$, one can compute $S^{(1)}(\psi)$ when the base space is flat, and subsequently determine the T-duality-transformed action $S^{(1)}(\psi_0)$. To evaluate the right-hand side of the above equation, one needs the reduction of the leading-order action \reef{leading} for a non-flat base space, which reads
\beqa
S^{(0)}(\psi)&=&-\frac{2}{\kappa^2}\int\mathrm{d}^{9}x\sqrt{-\bg}e^{-2\Phi}\big[-2\nabla_b\bom^a{}_a{}^b-\bom^{abc}\bom_{abc}-\bom^a{}_a{}^b\bom^c{}_{bc}-\nabla_a\nabla^a\vp+4\nabla_a\bphi\nabla^a\bphi\nn\\&&
+2\nabla_a\vp\nabla^a\bphi-\frac{1}{4}\nabla_a\vp\nabla^a\vp-\frac{1}{4}e^\vp V_{ab}V^{ab}-\frac{1}{4}e^{-\vp}W_{ab}W^{ab}-\frac{1}{12}\bH_{abc}\bH^{abc}\big].\labell{S00}
\eeqa
This must then be perturbed and the perturbation transformed under the Buscher rules \reef{flac}. Since the fundamental field in the base space is $\bar{e}_a{}^\ti$, the base-space spin connection expressed in terms of this field is
\beqa
\bom_{abc}&=&\frac{1}{2}\bar{e}_c{}^\ti\prt_a\bar{e}_{b\ti}-\frac{1}{2}\bar{e}_b{}^\ti\prt_a\bar{e}_{c\ti}-\frac{1}{2}\bar{e}_c{}^\ti\prt_b\bar{e}_{a\ti}-\frac{1}{2}\bar{e}_a{}^\ti\prt_b\bar{e}_{c\ti}+\frac{1}{2}\bar{e}_b{}^\ti\prt_c\bar{e}_{a\ti}+\frac{1}{2}\bar{e}_a{}^\ti\prt_c\bar{e}_{b\ti}.\label{s0}
\eeqa
In a flat base space, the vielbein is constant, $\bar{e}_a{}^\ti =\delta_a^\ti$, so $\prt_a \bar{e}_b{}^\ti=0$, though its perturbation is not constant. In flat base space, one has $\bg_{ab}^{(n)}=2\bar{e}_a{}^\ti \bar{e}_{b\ti}^{(n)}$. Consequently, the first-order perturbation of the base-space spin connection takes the form
\beqa
\bom^{(1)}_{abc}&=&\frac{1}{2}\prt_c \bg^{(1)}_{ab}-\frac{1}{2}\prt_b \bg^{(1)}_{ac}.
\eeqa
Using this perturbation for the base-space spin connection, along with the perturbations of the other fields in \reef{S00}, one can compute $S_{(1)}^{(0,1)}(\psi)$, and then apply the Buscher rules to obtain the left-hand side of the constraint equation \reef{T1}. This yields the result that the constraint \reef{T1} is satisfied by the inhomogeneous solution of  this equation, provided the Buscher rules are supplemented with the following corrections, first derived in \cite{Garousi:2019wgz}:
\beqa
   \bar{g}_{ab}^{(1)}&=&\frac{1}{4}\Big(e^\vp V_a {}^c V_{bc}-e^{-\vp}W_a {}^c W_{bc}\Big), \nn\\
 \bphi^{(1)}&=&\frac{1}{16}\Big(e^\vp V^2- e^{-\vp}W^2\Big), \nn\\
 \vp^{(1)}&=&\frac{1}{4}\Big(\prt_a\vp\prt^a\vp+e^\vp V^2+e^{-\vp}W^2-V_{ab}W^{ab}\Big), \nn\\
  \Delta g_{a}^{(1)}&=&\frac{1}{8}\Big(2e^{-\vp/2}\prt^b W_{ab}+e^{\vp/2}\bH_{abc} V^{bc}-4e^{-\vp/2}\prt^b\bphi W_{ab}+e^{\vp/2}\prt^b\vp V_{ab}-\frac{1}{2}e^{-\vp/2}\bH_{abc}W^{bc}\Big), \nn\\
   \Delta b_{a}^{(1)}&=&-\frac{1}{8}\Big(2e^{\vp/2}\prt^b V_{ab}+e^{-\vp/2}\bH_{abc} W^{bc}-4e^{\vp/2}\prt^b\bphi V_{ab}-e^{-\vp/2}\prt^b\vp W_{ab}-\frac{1}{2}e^{\vp/2}\bH_{abc}V^{bc}\Big), \nn\\
   \bar{B}_{ab}^{(1)}&=&\frac{1}{4}\Big(V_a{}^cW_{bc}-V_b{}^cW_{ac}\Big).\labell{dbH1}
\eeqa
Note that homogeneous solutions of \reef{T1} may be added, allowing these corrections to be expressed in alternative forms; however, such additions do not affect the coupling constants of the effective action. One may first solve the homogeneous part of \reef{T1} to find independent corrections to the Buscher rules, and then the inhomogeneous solution of \reef{T1} can fix all parameters in the corrections to the Buscher rules uniquely. In the Appendix \reef{AppAA}, we find such independent corrections to the Buscher rules.

 \subsection{Couplings at order $\alpha'^2$}

It is known that the effective action of the heterotic theory contains no pure gravity coupling of the form \(R^3\) at order \(\alpha'^2\) \cite{Metsaev:1986yb}. Instead, it features a pure gravity coupling \(\Omega^2\), which arises from substituting the Green–Schwarz modification \reef{Green-Schwarz} into the leading-order action \reef{leading}. As observed in \cite{Garousi:2023kxw}, this latter coupling is inconsistent with T-duality. Therefore, additional couplings involving the \(H\)-field must appear at order \(\alpha'^2\) to restore T-duality invariance. Such couplings have already been found in \cite{Garousi:2023kxw} within the Meissner scheme. In this subsection, we derive the corresponding couplings in the MT scheme.

The effective action at order \(\alpha'^2\) in the MT scheme takes the form
\beqa
 {\bf S}^{(2)}&=&{\bf S}_{\mathrm{MT}}^{(2)}+{\bf S}^{(2)}_{\mathrm{GS}}\,,
 \eeqa
where \({\bf S}^{(2)}_{\mathrm{GS}}\) denotes the couplings resulting from replacing the modified field strength \reef{Green-Schwarz} in the leading-order action \reef{leading} and in the next-to-leading-order action \reef{fourmin}. These order-\(\alpha'^2\) couplings are explicitly given by
\beqa
	{\bf S}_{\mathrm{GS}}^{(2)}&=&-\frac{2}{\kappa^2}\int \mathrm{d}^{10} x \sqrt{-G} \,e^{-2\Phi}\Big[- \frac{1}{32} H_{\alpha }{}^{\delta \epsilon } H^{\alpha \beta \gamma } H_{\beta \delta }{}^{\varepsilon } \Omega_{\gamma \epsilon \varepsilon }+\frac{3}{32} H_{\alpha \beta }{}^{\delta } H^{\alpha \beta \gamma } H_{\gamma }{}^{\epsilon \varepsilon } \Omega_{\delta \epsilon \varepsilon }\nn\\&&
\qquad\qquad\qquad\qquad\qquad	+\frac{3}{16} H^{\alpha \beta \gamma } \Omega_{\alpha }{}^{\delta \epsilon } R_{\beta \gamma \delta \epsilon }- \frac{3}{16} \Omega_{\alpha \beta \gamma }\Omega^{\alpha \beta \gamma }\Big].
\eeqa
It should be noted that the action includes both odd-parity and even-parity couplings.

Since the couplings in \({\bf S}_{\mathrm{MT}}^{(2)}\) are not known in the MT scheme, we adopt the minimal basis at this order, which consists of 73 couplings with undetermined coefficients \cite{Garousi:2019cdn,Garousi:2023kxw}: 60 even-parity and 13 odd-parity couplings. We then impose the T-duality constraint \reef{Tn} on \({\bf S}^{(2)}\), which for \(n=2\) is given by :
\beqa
S^{(0,2)}_{(2)}(\psi_0) \sim S^{(2)}(\psi) - S^{(2)}(\psi_0) - S^{(0,2)}_{(2')}(\psi_0) - S^{(1,1)}(\psi_0)\,.\label{T2}
\eeqa
The terms on the right-hand side of \reef{T2} are defined as follows. The last term corresponds to the Buscher transformation of the first-order perturbation of the reduction of \reef{S1}, which involves \(\psi^{(1)}\) as given in \reef{dbH1}. The third term is the Buscher transformation of the second-order perturbation of the reduction of the leading-order action, given in \reef{S00}, which involves only terms quadratic in \(\psi^{(1)}\) from \reef{dbH1}. The first term is the reduction of \({\bf S}^{(2)}\), while the second term is its transformation under the Buscher rules. All terms on the right-hand side depend only on the coupling constants of \({\bf S}^{(2)}\). The left-hand side of \reef{T2}, by contrast, is the Buscher transformation of the second-order perturbation of the reduction of the leading-order action \reef{S00}, which involves only \(\psi^{(2)}\). These perturbations include all contractions of the base-space fields at order \(\alpha'^2\) with unknown coefficients, which therefore appear exclusively on the left-hand side of the equation.

To compute the left-hand side, one requires the second-order perturbation of the base-space spin connection. For a flat base space, this is given by
\beqa
 \bom^{(2)}_{abc}&=&\frac{1}{4}\bg^{(1)}_c{}^d\prt_a\bg^{(1)}_{bd}-\frac{1}{4}\bg^{(1)}_b{}^d\prt_a\bg^{(1)}_{cd}+\frac{1}{2}\prt_c \bg^{(2)}_{ab}-\frac{1}{2}\prt_b \bg^{(2)}_{ac}.
 \eeqa
The first two terms in this expression contribute to \(S^{(0,2)}_{(2')}\), while the last two appear on the left-hand side of the constraint equation \reef{T2}. By using these second-order perturbations together with the other second-order perturbations arising from the expansion of \reef{S00}, we find that the inhomogeneous part of the constraint equation \reef{T2} fixes all 73 coupling constants as well as the corresponding parameters in the second-order perturbations.

Our analysis yields 5 odd-parity and 25 even-parity couplings at order \(\alpha'^2\). We then apply field redefinitions at order \(\alpha'^2\), while keeping the order-\(\alpha'\) couplings in \reef{S1} fixed, to reduce these 30 couplings. We find that there exists a field redefinition such that the order-\(\alpha'^2\) couplings can be expressed in terms of the following 18 terms:
\beqa
\bS^{(2)}_{\mathrm{MT}}&\!\!\!\!\!=\!\!\!\!\!&-\frac{2}{\kappa^2}\int d^{10} x\sqrt{-G} e^{-2\Phi}\Big[- \frac{1}{768} H_{\alpha }{}^{\delta \epsilon } H^{\alpha \beta \gamma } H_{\beta \delta }{}^{\varepsilon } H_{\gamma }{}^{\lambda \mu } H_{\epsilon \lambda }{}^{\nu } H_{\varepsilon \mu \nu }\nn\\
&&+\frac{1}{256} H_{\alpha \beta }{}^{\delta } H^{\alpha \beta \gamma } H_{\gamma }{}^{\epsilon \varepsilon } H_{\delta }{}^{\lambda \mu } H_{\epsilon \lambda }{}^{\nu } H_{\varepsilon \mu \nu }- \frac{1}{256} H_{\alpha \beta }{}^{\delta } H^{\alpha \beta \gamma } H_{\gamma }{}^{\epsilon \varepsilon } H_{\delta \epsilon }{}^{\lambda } H_{\varepsilon }{}^{\mu \nu } H_{\lambda \mu \nu }\nn\\
&&- \frac{1}{8} H^{\alpha \beta \gamma } H^{\delta \epsilon \varepsilon } R_{\alpha \delta \beta }{}^{\lambda } R_{\gamma \epsilon \varepsilon \lambda }- \frac{1}{32} H_{\alpha }{}^{\delta \epsilon } H^{\alpha \beta \gamma } R_{\beta \delta }{}^{\varepsilon \lambda } R_{\gamma \varepsilon \epsilon \lambda }+\frac{1}{32} H_{\alpha \beta }{}^{\delta } H^{\alpha \beta \gamma } R_{\gamma }{}^{\epsilon \varepsilon \lambda } R_{\delta \epsilon \varepsilon \lambda }\nn\\
&&- \frac{1}{32} H_{\alpha }{}^{\delta \epsilon } H^{\alpha \beta \gamma } R_{\beta }{}^{\varepsilon }{}_{\gamma }{}^{\lambda } R_{\delta \varepsilon \epsilon \lambda }\!+\!\frac{1}{16} H_{\alpha \beta }{}^{\delta } H^{\alpha \beta \gamma } H_{\gamma }{}^{\epsilon \varepsilon } H_{\epsilon }{}^{\lambda \mu } R_{\delta \lambda \varepsilon \mu }\!- \!\frac{1}{64} H_{\alpha \beta }{}^{\delta } H^{\alpha \beta \gamma } H_{\gamma }{}^{\epsilon \varepsilon } H_{\delta }{}^{\lambda \mu } R_{\epsilon \lambda \varepsilon \mu }\nn\\
&&+\frac{1}{64} H^{\alpha \beta \gamma } H^{\delta \epsilon \varepsilon } \nabla_{\gamma }H_{\epsilon \varepsilon \lambda } \nabla_{\delta }H_{\alpha \beta }{}^{\lambda }- \frac{1}{64} H_{\alpha }{}^{\delta \epsilon } H^{\alpha \beta \gamma } \nabla_{\delta }H_{\beta }{}^{\varepsilon \lambda } \nabla_{\epsilon }H_{\gamma \varepsilon \lambda }\nn\\
&&- \frac{9}{256} H_{\alpha }{}^{\delta \epsilon } H^{\alpha \beta \gamma } \nabla_{\gamma }H_{\beta }{}^{\varepsilon \lambda } \nabla_{\epsilon }H_{\delta \varepsilon \lambda }+\frac{7}{128} H^{\alpha \beta \gamma } H^{\delta \epsilon \varepsilon } \nabla_{\beta }H_{\alpha \delta }{}^{\lambda } \nabla_{\varepsilon }H_{\gamma \epsilon \lambda }\nn\\
&&+\frac{1}{64} H_{\alpha \beta }{}^{\delta } H^{\alpha \beta \gamma } \nabla_{\delta }H_{\epsilon \varepsilon \lambda } \nabla^{\lambda }H_{\gamma }{}^{\epsilon \varepsilon }\!-\! \frac{3}{64} H_{\alpha }{}^{\delta \epsilon } H^{\alpha \beta \gamma } R_{\delta \epsilon \varepsilon \lambda } \nabla_{\gamma }H_{\beta }{}^{\varepsilon \lambda }\!-\! \frac{1}{32} H^{\alpha \beta \gamma } H^{\delta \epsilon \varepsilon } R_{\gamma \epsilon \varepsilon \lambda } \nabla_{\delta }H_{\alpha \beta }{}^{\lambda }\nn\\
	&&- \frac{1}{128} H_{\alpha }{}^{\delta \epsilon } H^{\alpha \beta \gamma } H_{\beta \delta }{}^{\varepsilon } H_{\gamma }{}^{\lambda \mu } \nabla_{\varepsilon }H_{\epsilon \lambda \mu }- \frac{1}{32} H^{\alpha \beta \gamma } H^{\delta \epsilon \varepsilon } R_{\gamma \epsilon \varepsilon \lambda } \nabla^{\lambda }H_{\alpha \beta \delta }\Big].\labell{S2}
\eeqa
Among these, the last four terms are odd under parity, while the remaining ones are even.

We take the above 18 couplings as our standard set in the MT scheme. Using these as fixed starting couplings in the T-duality constraint \reef{T2}, we determine the corresponding order-\(\alpha'^2\) T-duality transformations required for the study of the effective action at order \(\alpha'^3\) in the next subsection. These corrections are given by
\beqa
   \bar{g}_{ab}^{(2)}&=&-\frac{1}{4}e^\vp\prt_cV_{bd}\prt^cV_a{}^{d}+\cdots, \nn\\
 \bphi^{(2)}&=&-\frac{5}{256}\prt_a\bH^{abc}\prt^d\bH_{bcd}+\cdots, \nn\\
 \vp^{(2)}&=&-\frac{1}{16}\prt_a\bH_{bcd}\prt^a\bH^{bcd}+\cdots, \nn\\
  \Delta g_{a}^{(2)}&=&-\frac{1}{2}e^{\vp/2}\prt_c\bH_{abd}\prt^b V^{cd}+\cdots, \nn\\
   \Delta b_{a}^{(2)}&=&\frac{29}{24}e^{\vp/2}\prt_c\bH_{abd}\prt^b V^{cd}+\cdots, \nn\\
   \bar{B}_{ab}^{(2)}&=&-\frac{5}{16}(\prt_c V^{cd}\prt_a W_{bd}-\prt_c V^{cd}\prt_b W_{ad})+\cdots.\label{dbH2}
\eeqa
Here, the dots denote many additional terms that we have omitted from the main text due to their length. These corrections, together with those in \reef{dbH1}, are strictly necessary for the computation of the effective action at order \(\alpha'^3\).

\subsection{Couplings at order $\alpha'^3$}

Having found the couplings at order \(\alpha'^2\) in the MT scheme \reef{S2} and the corresponding T-duality transformation corrections at the same order, we now seek the effective action at order \(\alpha'^3\) in the MT scheme. The effective action at this order takes the form
\beqa
 {\bf S}^{(3)}&=&{\bf S}_{\mathrm{MT}}^{(3)}+{\bf S}^{(3)}_{\mathrm{GS}}\,,
 \eeqa
where \({\bf S}^{(3)}_{\mathrm{GS}}\) denotes the known couplings obtained by replacing the modified field strength \reef{Green-Schwarz} in the action \reef{fourmin} at order \(\alpha'\), and in the action \reef{S2} at order \(\alpha'^2\), to produce couplings at order \(\alpha'^3\). These order-\(\alpha'^3\) couplings are explicitly given by
\beqa
\bS^{(3)}_{\mathrm{GS}}&=&-\frac{2}{\kappa^2}\int \mathrm{d}^{10} x \sqrt{-G} \,e^{-2\Phi}\Big[- \frac{3}{256} H_{\alpha \beta }{}^{\delta } H^{\alpha \beta \gamma } H_{\gamma }{}^{\epsilon \varepsilon } H_{\epsilon }{}^{\lambda \sigma } H_{\lambda \sigma }{}^{\psi } \Omega_{\delta \varepsilon \psi }\nn\\
&&+\frac{3}{256} H_{\alpha }{}^{\delta \epsilon } H^{\alpha \beta \gamma } H_{\beta \delta }{}^{\varepsilon } H_{\gamma }{}^{\lambda \sigma } H_{\epsilon \lambda }{}^{\psi } \Omega_{\varepsilon \sigma \psi }- \frac{3}{256} H_{\alpha \beta }{}^{\delta } H^{\alpha \beta \gamma } H_{\gamma }{}^{\epsilon \varepsilon } H_{\delta }{}^{\lambda \sigma } H_{\epsilon \lambda }{}^{\psi } \Omega_{\varepsilon \sigma \psi }\nn\\
&&- \frac{3}{256} H_{\alpha }{}^{\delta \epsilon } H^{\alpha \beta \gamma } H_{\beta \delta }{}^{\varepsilon } H_{\gamma \epsilon }{}^{\lambda } H_{\varepsilon }{}^{\sigma \psi } \Omega_{\lambda \sigma \psi }+\frac{3}{128} H_{\alpha \beta }{}^{\delta } H^{\alpha \beta \gamma } H_{\gamma }{}^{\epsilon \varepsilon } H_{\delta \epsilon }{}^{\lambda } H_{\varepsilon }{}^{\sigma \psi } \Omega_{\lambda \sigma \psi }\nn\\
&&+\frac{3}{32} H_{\alpha }{}^{\delta \epsilon } H^{\alpha \beta \gamma } H^{\varepsilon \lambda \sigma } \Omega_{\beta \varepsilon \lambda } R_{\gamma \delta \epsilon \sigma }+\frac{3}{8} H^{\alpha \beta \gamma } \Omega^{\delta \epsilon \varepsilon } R_{\alpha \delta \beta }{}^{\lambda } R_{\gamma \epsilon \varepsilon \lambda }\nn\\
&&+\frac{3}{32} H^{\alpha \beta \gamma } \Omega_{\alpha }{}^{\delta \epsilon } R_{\beta \delta }{}^{\varepsilon \lambda } R_{\gamma \varepsilon \epsilon \lambda }- \frac{3}{32} H_{\alpha }{}^{\delta \epsilon } H^{\alpha \beta \gamma } H_{\beta }{}^{\varepsilon \lambda } \Omega_{\delta \epsilon }{}^{\sigma } R_{\gamma \varepsilon \lambda \sigma }\nn\\
&&- \frac{3}{32} H^{\alpha \beta \gamma } \Omega_{\alpha \beta }{}^{\delta } R_{\gamma }{}^{\epsilon \varepsilon \lambda } R_{\delta \epsilon \varepsilon \lambda }+\frac{3}{32} H^{\alpha \beta \gamma } \Omega_{\alpha }{}^{\delta \epsilon } R_{\beta }{}^{\varepsilon }{}_{\gamma }{}^{\lambda } R_{\delta \varepsilon \epsilon \lambda }\nn\\
&&+\frac{3}{32} H_{\alpha \beta }{}^{\delta } H^{\alpha \beta \gamma } H^{\epsilon \varepsilon \lambda } \Omega_{\gamma \epsilon }{}^{\sigma } R_{\delta \varepsilon \lambda \sigma }+\frac{3}{64} H_{\alpha }{}^{\delta \epsilon } H^{\alpha \beta \gamma } H^{\varepsilon \lambda \sigma } \Omega_{\beta \gamma \varepsilon } R_{\delta \lambda \epsilon \sigma }\nn\\
&&- \frac{3}{32} H_{\alpha \beta }{}^{\delta } H^{\alpha \beta \gamma } H_{\gamma }{}^{\epsilon \varepsilon } \Omega_{\epsilon }{}^{\lambda \sigma } R_{\delta \lambda \varepsilon \sigma }+\frac{3}{64} H_{\alpha \beta }{}^{\delta } H^{\alpha \beta \gamma } H_{\gamma }{}^{\epsilon \varepsilon } \Omega_{\delta }{}^{\lambda \sigma } R_{\epsilon \lambda \varepsilon \sigma }\nn\\
&&- \frac{3}{64} H^{\alpha \beta \gamma } \Omega^{\delta \epsilon \varepsilon } R_{\beta \varepsilon \gamma \lambda } \nabla_{\alpha }H_{\delta \epsilon }{}^{\lambda }+\frac{9}{128} H^{\alpha \beta \gamma } \Omega_{\alpha }{}^{\delta \epsilon } R_{\delta \epsilon \varepsilon \lambda } \nabla_{\gamma }H_{\beta }{}^{\varepsilon \lambda }\nn\\
&&+\frac{3}{128} H^{\alpha \beta \gamma } \Omega_{\alpha }{}^{\delta \epsilon } \nabla_{\beta }H_{\delta }{}^{\varepsilon \lambda } \nabla_{\gamma }H_{\epsilon \varepsilon \lambda }+\frac{9}{128} H_{\alpha }{}^{\delta \epsilon } H^{\alpha \beta \gamma } R_{\delta \epsilon \varepsilon \lambda } \nabla_{\gamma }\Omega_{\beta }{}^{\varepsilon \lambda }\nn\\
&&+\frac{3}{64} H^{\alpha \beta \gamma } \Omega^{\delta \epsilon \varepsilon } R_{\gamma \epsilon \varepsilon \lambda } \nabla_{\delta }H_{\alpha \beta }{}^{\lambda }- \frac{3}{64} H^{\alpha \beta \gamma } \Omega^{\delta \epsilon \varepsilon } \nabla_{\gamma }H_{\epsilon \varepsilon \lambda } \nabla_{\delta }H_{\alpha \beta }{}^{\lambda }\nn\\
&&- \frac{3}{64} H^{\alpha \beta \gamma } H^{\delta \epsilon \varepsilon } \nabla_{\gamma }\Omega_{\epsilon \varepsilon \lambda } \nabla_{\delta }H_{\alpha \beta }{}^{\lambda }+\frac{3}{64} H^{\alpha \beta \gamma } H^{\delta \epsilon \varepsilon } R_{\gamma \epsilon \varepsilon \lambda } \nabla_{\delta }\Omega_{\alpha \beta }{}^{\lambda }\nn\\
&&+\frac{3}{128} H^{\alpha \beta \gamma } \Omega_{\alpha }{}^{\delta \epsilon } \nabla_{\delta }H_{\beta }{}^{\varepsilon \lambda } \nabla_{\epsilon }H_{\gamma \varepsilon \lambda }+\frac{3}{256} H_{\alpha }{}^{\delta \epsilon } H^{\alpha \beta \gamma } H^{\varepsilon \lambda \sigma } \Omega_{\beta \delta \varepsilon } \nabla_{\epsilon }H_{\gamma \lambda \sigma }\nn\\
&&+\frac{27}{256} H^{\alpha \beta \gamma } \Omega_{\alpha }{}^{\delta \epsilon } \nabla_{\gamma }H_{\beta }{}^{\varepsilon \lambda } \nabla_{\epsilon }H_{\delta \varepsilon \lambda }+\frac{9}{128} H^{\alpha \beta \gamma } \Omega_{\alpha }{}^{\delta \epsilon } R_{\beta \gamma \varepsilon \lambda } \nabla_{\epsilon }H_{\delta }{}^{\varepsilon \lambda }\nn\\
&&+\frac{3}{256} H_{\alpha }{}^{\delta \epsilon } H^{\alpha \beta \gamma } H_{\beta }{}^{\varepsilon \lambda } \Omega_{\gamma \delta }{}^{\sigma } \nabla_{\epsilon }H_{\varepsilon \lambda \sigma }+\frac{3}{64} H_{\alpha }{}^{\delta \epsilon } H^{\alpha \beta \gamma } \nabla_{\delta }H_{\beta }{}^{\varepsilon \lambda } \nabla_{\epsilon }\Omega_{\gamma \varepsilon \lambda }\nn\\
&&+\frac{27}{256} H_{\alpha }{}^{\delta \epsilon } H^{\alpha \beta \gamma } \nabla_{\gamma }H_{\beta }{}^{\varepsilon \lambda } \nabla_{\epsilon }\Omega_{\delta \varepsilon \lambda }- \frac{21}{128} H^{\alpha \beta \gamma } \Omega^{\delta \epsilon \varepsilon } \nabla_{\beta }H_{\alpha \delta }{}^{\lambda } \nabla_{\varepsilon }H_{\gamma \epsilon \lambda }\nn\\
&&+\frac{3}{256} H_{\alpha }{}^{\delta \epsilon } H^{\alpha \beta \gamma } H_{\beta \delta }{}^{\varepsilon } \Omega_{\gamma }{}^{\lambda \sigma } \nabla_{\varepsilon }H_{\epsilon \lambda \sigma }- \frac{21}{128} H^{\alpha \beta \gamma } H^{\delta \epsilon \varepsilon } \nabla_{\beta }H_{\alpha \delta }{}^{\lambda } \nabla_{\varepsilon }\Omega_{\gamma \epsilon \lambda }\nn\\
&&+\frac{3}{256} H_{\alpha }{}^{\delta \epsilon } H^{\alpha \beta \gamma } H_{\beta \delta }{}^{\varepsilon } H_{\gamma }{}^{\lambda \sigma } \nabla_{\varepsilon }\Omega_{\epsilon \lambda \sigma }- \frac{3}{128} H_{\alpha \beta }{}^{\delta } H^{\alpha \beta \gamma } \nabla_{\gamma }H^{\epsilon \varepsilon \lambda } \nabla_{\lambda }\Omega_{\delta \epsilon \varepsilon }\nn\\
&&+\frac{3}{64} H^{\alpha \beta \gamma } \Omega^{\delta \epsilon \varepsilon } R_{\gamma \epsilon \varepsilon \lambda } \nabla^{\lambda }H_{\alpha \beta \delta }- \frac{3}{64} H^{\alpha \beta \gamma } \Omega^{\delta \epsilon \varepsilon } R_{\beta \varepsilon \gamma \lambda } \nabla^{\lambda }H_{\alpha \delta \epsilon }\nn\\
&&- \frac{3}{128} H^{\alpha \beta \gamma } \Omega_{\alpha \beta }{}^{\delta } \nabla_{\delta }H_{\epsilon \varepsilon \lambda } \nabla^{\lambda }H_{\gamma }{}^{\epsilon \varepsilon }- \frac{3}{128} H_{\alpha \beta }{}^{\delta } H^{\alpha \beta \gamma } \nabla_{\delta }\Omega_{\epsilon \varepsilon \lambda } \nabla^{\lambda }H_{\gamma }{}^{\epsilon \varepsilon }\nn\\
&&- \frac{3}{128} H^{\alpha \beta \gamma } \Omega_{\alpha \beta }{}^{\delta } \nabla_{\gamma }H_{\epsilon \varepsilon \lambda } \nabla^{\lambda }H_{\delta }{}^{\epsilon \varepsilon }+\frac{3}{64} H^{\alpha \beta \gamma } H^{\delta \epsilon \varepsilon } R_{\gamma \epsilon \varepsilon \lambda } \nabla^{\lambda }\Omega_{\alpha \beta \delta }\nn\\
&&- \frac{3}{256} H_{\alpha }{}^{\delta \epsilon } H^{\alpha \beta \gamma } H_{\beta }{}^{\varepsilon \lambda } \Omega_{\gamma \delta }{}^{\sigma } \nabla_{\sigma }H_{\epsilon \varepsilon \lambda }- \frac{3}{256} H_{\alpha \beta }{}^{\delta } H^{\alpha \beta \gamma } H_{\gamma }{}^{\epsilon \varepsilon } H_{\epsilon }{}^{\lambda \sigma } H_{\varepsilon \lambda }{}^{\psi } \Omega_{\delta \sigma \psi }
\nn\\ &&- \frac{9}{128} H^{\alpha \beta \gamma } H^{\delta \epsilon \varepsilon } \Omega_{\alpha \beta \delta } \Omega_{\gamma \epsilon \varepsilon }+\frac{9}{128} H_{\alpha }{}^{\delta \epsilon } H^{\alpha \beta \gamma } \Omega_{\beta \delta }{}^{\varepsilon } \Omega_{\gamma \epsilon \varepsilon }\nn\\
&&- \frac{9}{128} H_{\alpha }{}^{\delta \epsilon } H^{\alpha \beta \gamma } \Omega_{\beta \gamma }{}^{\varepsilon } \Omega_{\delta \epsilon \varepsilon }- \frac{9}{128} H_{\alpha \beta }{}^{\delta } H^{\alpha \beta \gamma } \Omega_{\gamma }{}^{\epsilon \varepsilon } \Omega_{\delta \epsilon \varepsilon }- \frac{9}{64} \Omega_{\alpha }{}^{\delta \epsilon } \Omega^{\alpha \beta \gamma } R_{\beta \gamma \delta \epsilon }\Big]\,.
\eeqa
It should be noted that the action includes both odd-parity and even-parity couplings.

The couplings in \({\bf S}_{\mathrm{MT}}^{(3)}\) are not known in the MT scheme, so we adopt the minimal basis at this order, which consists of 872 even-parity couplings \cite{Garousi:2020mqn} and 477 odd-parity couplings \cite{Garousi:2024rzh}. We then impose the T-duality constraint \reef{Tn} on \({\bf S}^{(3)}\), which for \(n=3\) is given by
\beqa
S^{(0,3)}_{(3)}(\psi_0) \sim S^{(3)}(\psi) - S^{(3)}(\psi_0) - S^{(0,3)}_{(3')}(\psi_0) - S^{(1,2)}(\psi_0)- S^{(2,1)}(\psi_0)\,.\labell{T3}
\eeqa
The terms on the right-hand side of \reef{T3} are defined as follows. The last term corresponds to the Buscher transformation of the first-order perturbation of the reduction of \reef{S2}, which involves \(\psi^{(1)}\) as given in \reef{dbH1}. The fourth term is the Buscher transformation of the second-order perturbation of the reduction of \reef{S1}, which involves \((\psi^{(1)})^2\) as given in \reef{dbH1} and \(\psi^{(2)}\) as given in \reef{dbH2}. The third term is the Buscher transformation of the third-order perturbation of the reduction of the leading-order action, given in \reef{S00}, which involves the known \(\psi^{(1)}\) and \(\psi^{(2)}\). The first term is the reduction of \({\bf S}^{(3)}\), while the second term is its transformation under the Buscher rules. All terms on the right-hand side depend only on the coupling constants of \({\bf S}^{(3)}\). The left-hand side of \reef{T3}, by contrast, is the Buscher transformation of the third-order perturbation of the reduction of the leading-order action \reef{S00}, which involves only \(\psi^{(3)}\). These perturbations include all contractions of the base-space fields at order \(\alpha'^3\) with unknown coefficients, which therefore appear exclusively on the left-hand side of the above equation.

To compute the left-hand side, one requires the third-order perturbation of the base-space spin connection. For a flat base space, this is given by
\beqa
 \bom^{(3)}_{abc}&=&\frac{3}{8}\bg^{(1)}_c{}^d\prt_a\bg^{(2)}_{bd}-\frac{3}{8}\bg^{(1)}_b{}^d\prt_a\bg^{(2)}_{cd}+\frac{3}{8}\bg^{(2)}_c{}^d\prt_a\bg^{(1)}_{bd}-\frac{3}{8}\bg^{(2)}_b{}^d\prt_a\bg^{(1)}_{cd}\nn\\
 &&-\frac{3}{8}\bg^{(1)}_c{}^d\bg^{(1)}_d{}^e\prt_a\bg^{(1)}_{be}+\frac{3}{8}\bg^{(1)}_b{}^d\bg^{(1)}_d{}^e\prt_a\bg^{(1)}_{ce}+\frac{1}{2}\prt_c\bg^{(3)}_{ab}-\frac{1}{2}\prt_b\bg^{(3)}_{ac}\,.
 \eeqa
The last two terms in this expression contribute to the left-hand side of the constraint equation \reef{T3}, while all other terms appear in \(S^{(0,3)}_{(3')}\). By using these third-order perturbations together with the other third-order perturbations arising from the expansion of \reef{S00}, we find that the inhomogeneous part of the constraint equation \reef{T3}  fixes all 1349 coupling constants, as well as the corresponding parameters in the third-order perturbations, up to one unknown coupling constant. Our calculation yields
\beqa
\bS^{(3)}_{\mathrm{MT}}&=&-\frac{2}{\kappa^2}\int d^{10} x\sqrt{-G} e^{-2\Phi}\Big[\frac{1}{64} R_{\alpha \beta }{}^{\epsilon \varepsilon } R^{\alpha \beta \gamma \delta } R_{\gamma }{}^{\zeta }{}_{\epsilon }{}^{\eta } R_{\delta \zeta \varepsilon \eta }+2a R_{\alpha }{}^{\epsilon }{}_{\gamma }{}^{\varepsilon } R^{\alpha \beta \gamma \delta } R_{\beta }{}^{\zeta }{}_{\epsilon }{}^{\eta } R_{\delta \eta \varepsilon \zeta }\nn\\
&&-\left(\frac{1}{64}- a\right) R_{\alpha \beta }{}^{\epsilon \varepsilon } R^{\alpha \beta \gamma \delta } R_{\gamma }{}^{\zeta }{}_{\epsilon }{}^{\eta } R_{\delta \eta \varepsilon \zeta }- \frac{1}{64} R_{\alpha \beta }{}^{\epsilon \varepsilon } R^{\alpha \beta \gamma \delta } R_{\gamma }{}^{\zeta }{}_{\delta }{}^{\eta } R_{\epsilon \zeta \varepsilon \eta }\nn\\
&&- \frac{1}{16} R_{\alpha \gamma \beta }{}^{\epsilon } R^{\alpha \beta \gamma \delta } R_{\delta }{}^{\varepsilon \zeta \eta } R_{\epsilon \zeta \varepsilon \eta }+\dots\Big]\,,\labell{S3}
\eeqa
The dots represent 603 couplings, including contractions of \(R\), \(H\), \(\nabla\Phi\) and their covariant derivatives. Their coefficients are either fixed numbers or the unfixed parameter \(a\). The odd-parity terms do not contain the parameter \(a\). These 603 terms can be changed under field redefinitions.

The fact that T-duality cannot fix the parameter \(a\) is consistent with the observation that both the NS–NS sector of the heterotic effective action and that of type II superstring theory are invariant under T-duality, yet they are not identical effective actions. If the heterotic theory could fix all parameters, its effective action would coincide with that of type II theory, which is not the case. Therefore, the parameter \(a\) must remain unfixed by T-duality.  

We find that the couplings proportional to the coefficient \(a\) coincide exactly with those appearing in superstring theory \cite{Garousi:2020gio}. These same couplings also arise in bosonic theory; however, as in the heterotic case, the bosonic theory contains additional couplings whose coefficients depend on those at order \(\alpha'\) \cite{Ameri:2025bei}. Nevertheless, these additional couplings differ from their heterotic counterparts.

In superstring theory, the S-matrix calculation fixes the overall coefficient of the T-duality-invariant couplings at order \(\alpha'^3\) to be proportional to \(\zeta(3)\). A comparison with the explicit calculation in bosonic theory \cite{Ameri:2025bei} yields the following expression for the parameter \(a\) in the above equation:
\beqa
a &=& \frac{\zeta(3)}{16} + c \labell{ac}
\eeqa
where \(c\) is an ordinary number. A corresponding calculation in bosonic theory reveals that \(c\) is nonzero.

To determine the number \(c\) for the heterotic theory, one may calculate the S-matrix element of four gravitons in the  effective action \reef{S3} and then compare it with the corresponding result from the string theory S-matrix. This calculation is straightforward in the Meissner scheme, in which the graviton propagator is not modified by couplings at order \(\alpha'\). In the MT scheme, instead of using the S-matrix method, we use the fact that in the heterotic theory, the Yang–Mills couplings and the gravity couplings appearing in the double trace should be combined into \((\Tr(F^2-\alpha' R^2/2))^2\) \cite{Romans:1985xd,Gross:1986mw}.

The pure YM couplings at four-derivative order in the MT scheme have been found in \cite{Garousi:2024avb} to be
\beqa
{\bf S}^{(1)}_{\rm MT}&=&-\frac{2 }{8\kappa^2}\int d^{10}x \sqrt{-G}e^{-2\Phi}\Big[\frac{1}{4} F_{\alpha }{}^{\gamma kl} F^{\alpha \beta ij} F_{\beta }{}^{\delta }{}_{kl} F_{\gamma \delta ij} -  \frac{1}{2} F_{\alpha }{}^{\gamma }{}_{ij} F^{\alpha \beta ij} F_{\beta }{}^{\delta kl} F_{\gamma \delta kl}\nn\\&& \qquad\qquad\qquad\qquad\qquad-  \frac{1}{8} F_{\alpha \beta }{}^{kl} F^{\alpha \beta ij} F_{\gamma \delta kl} F^{\gamma \delta }{}_{ij} +\cdots \Big]\,.\labell{fourmin1}
\eeqa
where dots represent the couplings that involve NS-NS fields. If one replaces  in the above couplings \(F_{\mu\nu\alpha\beta}\rightarrow \sqrt{\alpha'/2}R_{\mu\nu\alpha\beta}\)  and uses the Bianchi identity for the Riemann tensors, one finds exactly the pure gravity couplings in \reef{S3} that have no parameter \(a\). Hence, the number \(c\) in \reef{ac} is zero. Moreover, T-duality constrains the pure YM couplings at four-derivative order to be expressible solely in terms of the double trace \((\Tr(F^2))^2\) \cite{Garousi:2024avb}. Using the Bianchi identity, one finds that the pure gravity terms in \reef{S3} with coefficient \(a\) cannot be written in double-trace form. Hence, these are pure gravity couplings in the heterotic theory that have no counterpart in the pure YM couplings. In other words, there are pure gravity couplings in the heterotic theory that cannot be combined with YM couplings to be written as \(\Tr(F^2-\alpha' R^2/2)\). This is consistent with the fact that there are no YM couplings at four-derivative order with coefficient \(\zeta(3)\).

Therefore, the NS-NS couplings \reef{S3} in the heterotic theory, where the parameter \(a\) is also fixed, are
\beqa
\bS^{(3)}_{\mathrm{MT}}&=&-\frac{2}{\kappa^2}\int d^{10} x\sqrt{-G} e^{-2\Phi}\Big[\frac{1}{128}\Tr( R_{\alpha }{}^{\gamma }  R_{\beta }{}^{\delta } )\Tr(R^{\alpha \beta }R_{\gamma \delta }) -  \frac{1}{64} \Tr(R_{\alpha }{}^{\gamma } R^{\alpha \beta })\Tr( R_{\beta }{}^{\delta } R_{\gamma \delta })\nn\\&& \qquad\qquad\qquad\qquad\qquad-  \frac{1}{256} \Tr(R_{\alpha \beta } R_{\gamma \delta })\Tr(R^{\alpha \beta } R^{\gamma \delta }) +\cdots\nn\\
&&+\frac{\zeta(3)}{16}\Big ( R_{\alpha \beta }{}^{\epsilon \varepsilon } R^{\alpha \beta \gamma \delta } R_{\gamma }{}^{\zeta }{}_{\epsilon }{}^{\eta } R_{\delta \eta \varepsilon \zeta }+2 R_{\alpha }{}^{\epsilon }{}_{\gamma }{}^{\varepsilon } R^{\alpha \beta \gamma \delta } R_{\beta }{}^{\zeta }{}_{\epsilon }{}^{\eta } R_{\delta \eta \varepsilon \zeta }+\dots\Big)\Big]\,,\labell{S4}
\eeqa
The couplings in the last line are exactly the couplings that appear in superstring theory \cite{Garousi:2020gio}. Using field redefinitions, these couplings have been written in canonical form in which the dilaton factor appears only as the overall factor \(e^{-2\Phi}\) \cite{Garousi:2020lof}. Using such field redefinitions, we write the other couplings in the canonical form as
\beqa
{\bf S}_{\mathrm{MT}}^{(3)}&\!\!\!\!=\!\!\!\!&-\frac{2}{\kappa^2}\int d^{10}x \sqrt{-G}e^{-2\Phi}\Big[ \frac{\zeta(3)}{16}\cL_{\rm II}+[R^4]_{3}+[R^2\nabla H^2]_{17}+[\nabla H^4]_{8}+[H^8]_{2}\nn\\&&+[H^6R]_{4}+[H^4R^2]_{9}+[H^2 R^3 ]_{12}+[H^2\nabla H^2 R]_{85}+[H^4\nabla H^2]_{99}\labell{S3MT}\\&&+[H^6 \nabla H]_{9}+[H^4 R\nabla H]_{41}+[H^2 R^2\nabla H]_{24}
	+[R^3\nabla H]_{3}+[R\nabla H^3]_{7}+[H^2\nabla H^3]_{31}\Big]\,,\nn
\eeqa
where
\beqa
[R^4]_{3}&=&\frac{1}{128}\Tr( R_{\alpha }{}^{\gamma }  R_{\beta }{}^{\delta } )\Tr(R^{\alpha \beta }R_{\gamma \delta }) -  \frac{1}{64} \Tr(R_{\alpha }{}^{\gamma } R^{\alpha \beta })\Tr( R_{\beta }{}^{\delta } R_{\gamma \delta })\nn\\&& -  \frac{1}{256} \Tr(R_{\alpha \beta } R_{\gamma \delta })\Tr(R^{\alpha \beta } R^{\gamma \delta })\,,\nn\\
{[}H^8{]}_{2}&=&\frac{1}{2048} H_{\alpha }{}^{\delta \epsilon } H^{\alpha \beta \gamma } H_{\beta \delta }{}^{\varepsilon } H_{\gamma }{}^{\theta \lambda } H_{\epsilon \theta }{}^{\mu } H_{\varepsilon }{}^{\nu \rho } H_{\lambda \nu }{}^{\sigma } H_{\mu \rho \sigma }\nn\\
&&- \frac{1}{4096} H_{\alpha }{}^{\delta \epsilon } H^{\alpha \beta \gamma } H_{\beta \delta }{}^{\varepsilon } H_{\gamma \epsilon }{}^{\theta } H_{\varepsilon }{}^{\lambda \mu } H_{\theta }{}^{\nu \rho } H_{\lambda \nu }{}^{\sigma } H_{\mu \rho \sigma }\,,\nn\\
{[}R^2\nabla H^2{]}_{17}&=&\frac{1}{384} R_{\epsilon \theta \varepsilon \lambda } R^{\epsilon \varepsilon \theta \lambda } \nabla_{\delta }H_{\alpha \beta \gamma } \nabla^{\delta }H^{\alpha \beta \gamma }- \frac{5}{64} R_{\gamma }{}^{\varepsilon \theta \lambda } R_{\epsilon \theta \varepsilon \lambda } \nabla_{\delta }H_{\alpha \beta }{}^{\epsilon } \nabla^{\delta }H^{\alpha \beta \gamma }\nn\\
&&- \frac{1}{64} R_{\beta }{}^{\theta }{}_{\epsilon }{}^{\lambda } R_{\gamma \theta \varepsilon \lambda } \nabla_{\delta }H_{\alpha }{}^{\epsilon \varepsilon } \nabla^{\delta }H^{\alpha \beta \gamma }+\frac{1}{64} R_{\beta }{}^{\theta }{}_{\epsilon }{}^{\lambda } R_{\gamma \lambda \varepsilon \theta } \nabla_{\delta }H_{\alpha }{}^{\epsilon \varepsilon } \nabla^{\delta }H^{\alpha \beta \gamma }\nn\\
&&- \frac{1}{256} R_{\beta \gamma }{}^{\theta \lambda } R_{\epsilon \varepsilon \theta \lambda } \nabla_{\delta }H_{\alpha }{}^{\epsilon \varepsilon } \nabla^{\delta }H^{\alpha \beta \gamma }- \frac{1}{64} R_{\gamma }{}^{\varepsilon \theta \lambda } R_{\epsilon \theta \varepsilon \lambda } \nabla^{\delta }H^{\alpha \beta \gamma } \nabla^{\epsilon }H_{\alpha \beta \delta }\nn\\
&&- \frac{1}{16} R_{\beta }{}^{\theta }{}_{\epsilon }{}^{\lambda } R_{\gamma \theta \varepsilon \lambda } \nabla^{\delta }H^{\alpha \beta \gamma } \nabla^{\varepsilon }H_{\alpha \delta }{}^{\epsilon }+\frac{1}{16} R_{\beta }{}^{\theta }{}_{\epsilon }{}^{\lambda } R_{\gamma \lambda \varepsilon \theta } \nabla^{\delta }H^{\alpha \beta \gamma } \nabla^{\varepsilon }H_{\alpha \delta }{}^{\epsilon }\nn\\
&&+\frac{1}{32} R_{\beta }{}^{\theta }{}_{\gamma }{}^{\lambda } R_{\epsilon \theta \varepsilon \lambda } \nabla^{\delta }H^{\alpha \beta \gamma } \nabla^{\varepsilon }H_{\alpha \delta }{}^{\epsilon }- \frac{1}{64} R_{\beta \gamma \epsilon }{}^{\lambda } R_{\delta \lambda \varepsilon \theta } \nabla^{\delta }H^{\alpha \beta \gamma } \nabla^{\theta }H_{\alpha }{}^{\epsilon \varepsilon }\nn\\
&&- \frac{1}{32} R_{\beta \delta \gamma }{}^{\lambda } R_{\epsilon \theta \varepsilon \lambda } \nabla^{\delta }H^{\alpha \beta \gamma } \nabla^{\theta }H_{\alpha }{}^{\epsilon \varepsilon }- \frac{1}{16} R_{\alpha \epsilon \beta }{}^{\lambda } R_{\gamma \theta \varepsilon \lambda } \nabla^{\delta }H^{\alpha \beta \gamma } \nabla^{\theta }H_{\delta }{}^{\epsilon \varepsilon }\nn\\
&&+\frac{1}{32} R_{\alpha \theta \beta }{}^{\lambda } R_{\gamma \lambda \epsilon \varepsilon } \nabla^{\delta }H^{\alpha \beta \gamma } \nabla^{\theta }H_{\delta }{}^{\epsilon \varepsilon }+\frac{3}{32} R_{\alpha \epsilon \beta }{}^{\lambda } R_{\gamma \lambda \varepsilon \theta } \nabla^{\delta }H^{\alpha \beta \gamma } \nabla^{\theta }H_{\delta }{}^{\epsilon \varepsilon }\nn\\
&&+\frac{3}{64} R_{\alpha \epsilon \beta \varepsilon } R_{\gamma \theta \delta \lambda } \nabla^{\delta }H^{\alpha \beta \gamma } \nabla^{\lambda }H^{\epsilon \varepsilon \theta }- \frac{1}{64} R_{\alpha \epsilon \beta \varepsilon } R_{\gamma \lambda \delta \theta } \nabla^{\delta }H^{\alpha \beta \gamma } \nabla^{\lambda }H^{\epsilon \varepsilon \theta }\nn\\
&&+\frac{1}{128} R_{\alpha \beta \delta \epsilon } R_{\gamma \lambda \varepsilon \theta } \nabla^{\delta }H^{\alpha \beta \gamma } \nabla^{\lambda }H^{\epsilon \varepsilon \theta }\,,\nn\\
{[}\nabla H^4{]}_{8}&=&- \frac{1}{512} \nabla^{\delta }H^{\alpha \beta \gamma } \nabla^{\varepsilon }H_{\alpha \beta }{}^{\epsilon } \nabla_{\lambda }H_{\epsilon \varepsilon \theta } \nabla^{\lambda }H_{\gamma \delta }{}^{\theta }+\frac{1}{256} \nabla^{\delta }H^{\alpha \beta \gamma } \nabla_{\epsilon }H_{\delta \theta \lambda } \nabla^{\varepsilon }H_{\alpha \beta }{}^{\epsilon } \nabla^{\lambda }H_{\gamma \varepsilon }{}^{\theta }\nn\\
&&- \frac{1}{256} \nabla^{\delta }H^{\alpha \beta \gamma } \nabla^{\varepsilon }H_{\alpha \beta }{}^{\epsilon } \nabla_{\theta }H_{\delta \epsilon \lambda } \nabla^{\lambda }H_{\gamma \varepsilon }{}^{\theta }- \frac{1}{512} \nabla^{\delta }H^{\alpha \beta \gamma } \nabla^{\varepsilon }H_{\alpha \beta }{}^{\epsilon } \nabla_{\lambda }H_{\delta \epsilon \theta } \nabla^{\lambda }H_{\gamma \varepsilon }{}^{\theta }\nn\\
&&- \frac{17}{1024} \nabla_{\delta }H_{\alpha \beta }{}^{\epsilon } \nabla^{\delta }H^{\alpha \beta \gamma } \nabla_{\lambda }H_{\epsilon \varepsilon \theta } \nabla^{\lambda }H_{\gamma }{}^{\varepsilon \theta }- \frac{1}{256} \nabla^{\delta }H^{\alpha \beta \gamma } \nabla^{\epsilon }H_{\alpha \beta \gamma } \nabla_{\lambda }H_{\epsilon \varepsilon \theta } \nabla^{\lambda }H_{\delta }{}^{\varepsilon \theta }\nn\\
&&+\frac{1}{1024} \nabla_{\gamma }H_{\varepsilon \theta \lambda } \nabla^{\delta }H^{\alpha \beta \gamma } \nabla^{\epsilon }H_{\alpha \beta \delta } \nabla^{\lambda }H_{\epsilon }{}^{\varepsilon \theta }+\frac{1}{512} \nabla_{\delta }H_{\alpha \beta \gamma } \nabla^{\delta }H^{\alpha \beta \gamma } \nabla_{\theta }H_{\epsilon \varepsilon \lambda } \nabla^{\lambda }H^{\epsilon \varepsilon \theta },\nn\\
{[}H^6R{]}_{4}&=&\frac{55}{1536} H_{\alpha \beta }{}^{\delta } H^{\alpha \beta \gamma } H_{\gamma }{}^{\epsilon \varepsilon } H_{\delta \epsilon }{}^{\theta } H_{\varepsilon }{}^{\lambda \mu } H_{\lambda }{}^{\nu \rho } R_{\theta \nu \mu \rho }\nn\\
&&+\frac{25}{3072} H_{\alpha \beta }{}^{\delta } H^{\alpha \beta \gamma } H_{\gamma }{}^{\epsilon \varepsilon } H_{\delta }{}^{\theta \lambda } H_{\epsilon \varepsilon }{}^{\mu } H_{\theta }{}^{\nu \rho } R_{\lambda \nu \mu \rho }\nn\\
&&+\frac{3}{512} H_{\alpha }{}^{\delta \epsilon } H^{\alpha \beta \gamma } H_{\beta \delta }{}^{\varepsilon } H_{\gamma \epsilon }{}^{\theta } H_{\varepsilon }{}^{\lambda \mu } H_{\theta }{}^{\nu \rho } R_{\lambda \nu \mu \rho }\nn\\
&&- \frac{9}{1024} H_{\alpha \beta }{}^{\delta } H^{\alpha \beta \gamma } H_{\gamma }{}^{\epsilon \varepsilon } H_{\delta \epsilon }{}^{\theta } H_{\varepsilon }{}^{\lambda \mu } H_{\theta }{}^{\nu \rho } R_{\lambda \nu \mu \rho }\,,\nn\\
{[}H^4R^2{]}_{9}&=&- \frac{25}{512} H_{\alpha \beta }{}^{\delta } H^{\alpha \beta \gamma } H^{\epsilon \varepsilon \theta } H^{\lambda \mu \nu } R_{\gamma \epsilon \delta \lambda } R_{\varepsilon \mu \theta \nu }- \frac{1}{256} H_{\alpha }{}^{\delta \epsilon } H^{\alpha \beta \gamma } H_{\beta }{}^{\varepsilon \theta } H^{\lambda \mu \nu } R_{\gamma \lambda \delta \epsilon } R_{\varepsilon \mu \theta \nu }\nn\\
&&- \frac{1}{64} H_{\alpha }{}^{\delta \epsilon } H^{\alpha \beta \gamma } H_{\beta \delta }{}^{\varepsilon } H_{\gamma }{}^{\theta \lambda } R_{\epsilon }{}^{\mu }{}_{\theta }{}^{\nu } R_{\varepsilon \nu \lambda \mu }- \frac{1}{128} H_{\alpha }{}^{\delta \epsilon } H^{\alpha \beta \gamma } H_{\beta \delta }{}^{\varepsilon } H_{\gamma }{}^{\theta \lambda } R_{\epsilon }{}^{\mu }{}_{\varepsilon }{}^{\nu } R_{\theta \mu \lambda \nu }\nn\\
&&- \frac{1}{128} H_{\alpha \beta }{}^{\delta } H^{\alpha \beta \gamma } H_{\gamma }{}^{\epsilon \varepsilon } H_{\delta }{}^{\theta \lambda } R_{\epsilon }{}^{\mu }{}_{\varepsilon }{}^{\nu } R_{\theta \mu \lambda \nu }- \frac{1}{128} H_{\alpha }{}^{\delta \epsilon } H^{\alpha \beta \gamma } H_{\beta \delta }{}^{\varepsilon } H_{\gamma \epsilon }{}^{\theta } R_{\varepsilon }{}^{\lambda \mu \nu } R_{\theta \mu \lambda \nu }\nn\\
&&- \frac{1}{32} H_{\alpha \beta }{}^{\delta } H^{\alpha \beta \gamma } H_{\gamma }{}^{\epsilon \varepsilon } H_{\delta \epsilon }{}^{\theta } R_{\varepsilon }{}^{\lambda \mu \nu } R_{\theta \mu \lambda \nu }+\frac{21}{512} H_{\alpha \beta }{}^{\delta } H^{\alpha \beta \gamma } H_{\epsilon }{}^{\lambda \mu } H^{\epsilon \varepsilon \theta } R_{\gamma \varepsilon \delta }{}^{\nu } R_{\theta \nu \lambda \mu }\nn\\
&&+\frac{1}{512} H_{\alpha \beta }{}^{\delta } H^{\alpha \beta \gamma } H_{\gamma }{}^{\epsilon \varepsilon } H_{\delta \epsilon \varepsilon } R_{\theta \mu \lambda \nu } R^{\theta \lambda \mu \nu }\,,\nn\\
{[}H^2 R^3 {]}_{12}&=&- \frac{1}{32} H_{\alpha }{}^{\delta \epsilon } H^{\alpha \beta \gamma } R_{\beta \delta }{}^{\varepsilon \theta } R_{\gamma \varepsilon }{}^{\lambda \mu } R_{\epsilon \lambda \theta \mu }+\frac{1}{32} H_{\alpha }{}^{\delta \epsilon } H^{\alpha \beta \gamma } R_{\beta }{}^{\varepsilon }{}_{\gamma }{}^{\theta } R_{\delta }{}^{\lambda }{}_{\varepsilon }{}^{\mu } R_{\epsilon \lambda \theta \mu }\nn\\
&&- \frac{1}{32} H_{\alpha }{}^{\delta \epsilon } H^{\alpha \beta \gamma } R_{\beta }{}^{\varepsilon }{}_{\gamma }{}^{\theta } R_{\delta }{}^{\lambda }{}_{\varepsilon }{}^{\mu } R_{\epsilon \mu \theta \lambda }+\frac{5}{32} H^{\alpha \beta \gamma } H^{\delta \epsilon \varepsilon } R_{\alpha \delta \beta \epsilon } R_{\gamma }{}^{\theta \lambda \mu } R_{\varepsilon \lambda \theta \mu }\nn\\
&&+\frac{1}{32} H_{\alpha }{}^{\delta \epsilon } H^{\alpha \beta \gamma } R_{\beta }{}^{\varepsilon }{}_{\delta }{}^{\theta } R_{\gamma }{}^{\lambda }{}_{\epsilon }{}^{\mu } R_{\varepsilon \lambda \theta \mu }- \frac{1}{16} H_{\alpha }{}^{\delta \epsilon } H^{\alpha \beta \gamma } R_{\beta }{}^{\varepsilon }{}_{\gamma }{}^{\theta } R_{\delta }{}^{\lambda }{}_{\epsilon }{}^{\mu } R_{\varepsilon \lambda \theta \mu }\nn\\
&&+\frac{7}{16} H_{\alpha }{}^{\delta \epsilon } H^{\alpha \beta \gamma } R_{\beta \delta \gamma }{}^{\varepsilon } R_{\epsilon }{}^{\theta \lambda \mu } R_{\varepsilon \lambda \theta \mu }- \frac{1}{32} H_{\alpha \beta }{}^{\delta } H^{\alpha \beta \gamma } R_{\gamma }{}^{\epsilon }{}_{\delta }{}^{\varepsilon } R_{\epsilon }{}^{\theta \lambda \mu } R_{\varepsilon \lambda \theta \mu }\nn\\
&&+\frac{1}{64} H^{\alpha \beta \gamma } H^{\delta \epsilon \varepsilon } R_{\alpha \beta }{}^{\theta \lambda } R_{\gamma \delta \epsilon }{}^{\mu } R_{\varepsilon \mu \theta \lambda }+\frac{1}{16} H^{\alpha \beta \gamma } H^{\delta \epsilon \varepsilon } R_{\alpha \delta \beta }{}^{\theta } R_{\gamma }{}^{\lambda }{}_{\epsilon }{}^{\mu } R_{\varepsilon \mu \theta \lambda }\nn\\
&&- \frac{1}{32} H_{\alpha }{}^{\delta \epsilon } H^{\alpha \beta \gamma } R_{\beta }{}^{\varepsilon }{}_{\delta }{}^{\theta } R_{\gamma }{}^{\lambda }{}_{\epsilon }{}^{\mu } R_{\varepsilon \mu \theta \lambda }- \frac{1}{64} H_{\alpha }{}^{\delta \epsilon } H^{\alpha \beta \gamma } R_{\beta \delta \gamma \epsilon } R_{\varepsilon \lambda \theta \mu } R^{\varepsilon \theta \lambda \mu}\nn\\
{[}R^3\nabla H{]}_{3}&=&- \frac{1}{8} R_{\alpha }{}^{\epsilon }{}_{\beta }{}^{\varepsilon } R_{\gamma }{}^{\lambda }{}_{\epsilon }{}^{\mu } R_{\delta \lambda \varepsilon \mu } \nabla^{\delta }H^{\alpha \beta \gamma }+\frac{1}{8} R_{\alpha }{}^{\epsilon }{}_{\beta }{}^{\varepsilon } R_{\gamma }{}^{\lambda }{}_{\epsilon }{}^{\mu } R_{\delta \mu \varepsilon \lambda } \nabla^{\delta }H^{\alpha \beta \gamma }\nn\\
&&- \frac{1}{8} R_{\alpha \delta \beta }{}^{\epsilon } R_{\gamma }{}^{\varepsilon \lambda \mu } R_{\epsilon \lambda \varepsilon \mu } \nabla^{\delta }H^{\alpha \beta \gamma }\,,\nn\\
{[}R\nabla H^3{]}_{7}&=&\frac{1}{32} R_{\gamma \epsilon \lambda \mu } \nabla^{\delta }H^{\alpha \beta \gamma } \nabla_{\varepsilon }H_{\delta }{}^{\lambda \mu } \nabla^{\varepsilon }H_{\alpha \beta }{}^{\epsilon }+\frac{1}{32} R_{\gamma \epsilon \lambda \mu } \nabla_{\beta }H_{\varepsilon }{}^{\lambda \mu } \nabla^{\delta }H^{\alpha \beta \gamma } \nabla^{\varepsilon }H_{\alpha \delta }{}^{\epsilon }\nn\\
&&+\frac{1}{64} R_{\beta \lambda \gamma \mu } \nabla^{\delta }H^{\alpha \beta \gamma } \nabla_{\varepsilon }H_{\epsilon }{}^{\lambda \mu } \nabla^{\varepsilon }H_{\alpha \delta }{}^{\epsilon }- \frac{1}{32} R_{\gamma \delta \varepsilon \mu } \nabla^{\delta }H^{\alpha \beta \gamma } \nabla_{\epsilon }H_{\beta \lambda }{}^{\mu } \nabla^{\lambda }H_{\alpha }{}^{\epsilon \varepsilon }\nn\\
&&+\frac{1}{64} R_{\gamma \varepsilon \lambda \mu } \nabla_{\delta }H_{\alpha }{}^{\epsilon \varepsilon } \nabla^{\delta }H^{\alpha \beta \gamma } \nabla^{\mu }H_{\beta \epsilon }{}^{\lambda }+\frac{5}{64} R_{\epsilon \mu \varepsilon \lambda } \nabla_{\beta }H_{\alpha \delta }{}^{\epsilon } \nabla^{\delta }H^{\alpha \beta \gamma } \nabla^{\mu }H_{\gamma }{}^{\varepsilon \lambda }\nn\\
&&+\frac{1}{16} R_{\gamma \mu \varepsilon \lambda } \nabla^{\delta }H^{\alpha \beta \gamma } \nabla^{\epsilon }H_{\alpha \beta \delta } \nabla^{\mu }H_{\epsilon }{}^{\varepsilon \lambda }\,,\nn\\
{[}H^6 \nabla H{]}_{9}&=&\frac{1}{512} H_{\alpha }{}^{\delta \epsilon } H^{\alpha \beta \gamma } H_{\beta \delta }{}^{\varepsilon } H_{\gamma }{}^{\lambda \mu } H_{\epsilon \lambda }{}^{\nu } H_{\mu }{}^{\rho \sigma } \nabla_{\varepsilon }H_{\nu \rho \sigma }\nn\\
&&+\frac{1}{1024} H_{\alpha \beta }{}^{\delta } H^{\alpha \beta \gamma } H_{\gamma }{}^{\epsilon \varepsilon } H_{\delta }{}^{\lambda \mu } H_{\epsilon }{}^{\nu \rho } H_{\varepsilon \nu }{}^{\sigma } \nabla_{\mu }H_{\lambda \rho \sigma }\nn\\
&&+\frac{1}{512} H_{\alpha }{}^{\delta \epsilon } H^{\alpha \beta \gamma } H_{\beta \delta }{}^{\varepsilon } H_{\gamma \epsilon }{}^{\lambda } H_{\varepsilon }{}^{\mu \nu } H_{\mu }{}^{\rho \sigma } \nabla_{\nu }H_{\lambda \rho \sigma }\nn\\
&&+\frac{3}{128} H_{\alpha \beta }{}^{\delta } H^{\alpha \beta \gamma } H_{\gamma }{}^{\epsilon \varepsilon } H_{\delta \epsilon }{}^{\lambda } H_{\varepsilon }{}^{\mu \nu } H_{\mu }{}^{\rho \sigma } \nabla_{\nu }H_{\lambda \rho \sigma }\nn\\
&&+\frac{1}{256} H_{\alpha }{}^{\delta \epsilon } H^{\alpha \beta \gamma } H_{\beta \delta }{}^{\varepsilon } H_{\gamma }{}^{\lambda \mu } H_{\epsilon \lambda }{}^{\nu } H_{\varepsilon }{}^{\rho \sigma } \nabla_{\nu }H_{\mu \rho \sigma }\nn\\
&&+\frac{1}{512} H_{\alpha }{}^{\delta \epsilon } H^{\alpha \beta \gamma } H_{\beta \delta }{}^{\varepsilon } H_{\gamma }{}^{\lambda \mu } H_{\epsilon }{}^{\nu \rho } H_{\lambda \nu }{}^{\sigma } \nabla_{\rho }H_{\varepsilon \mu \sigma }\nn\\
&&+\frac{1}{512} H_{\alpha \beta }{}^{\delta } H^{\alpha \beta \gamma } H_{\gamma }{}^{\epsilon \varepsilon } H_{\epsilon }{}^{\lambda \mu } H_{\lambda }{}^{\nu \rho } H_{\mu \nu }{}^{\sigma } \nabla_{\sigma }H_{\delta \varepsilon \rho }\nn\\
&&- \frac{1}{512} H_{\alpha \beta }{}^{\delta } H^{\alpha \beta \gamma } H_{\gamma }{}^{\epsilon \varepsilon } H_{\epsilon }{}^{\lambda \mu } H_{\varepsilon \lambda }{}^{\nu } H_{\mu }{}^{\rho \sigma } \nabla_{\sigma }H_{\delta \nu \rho }\nn\\
&&- \frac{1}{512} H_{\alpha }{}^{\delta \epsilon } H^{\alpha \beta \gamma } H_{\beta \delta }{}^{\varepsilon } H_{\gamma }{}^{\lambda \mu } H_{\epsilon \lambda }{}^{\nu } H_{\mu }{}^{\rho \sigma } \nabla_{\sigma }H_{\varepsilon \nu \rho }\,,\nn\\
{[}H^2 R^2\nabla H{]}_{24}&=&- \frac{1}{16} H^{\alpha \beta \gamma } H^{\delta \epsilon \varepsilon } R_{\gamma }{}^{\mu }{}_{\epsilon }{}^{\nu } R_{\varepsilon \nu \lambda \mu } \nabla_{\beta }H_{\alpha \delta }{}^{\lambda }+\frac{1}{64} H^{\alpha \beta \gamma } H^{\delta \epsilon \varepsilon } R_{\gamma \lambda \delta }{}^{\nu } R_{\epsilon \varepsilon \mu \nu } \nabla_{\beta }H_{\alpha }{}^{\lambda \mu }\nn\\
&&- \frac{1}{64} H_{\alpha }{}^{\delta \epsilon } H^{\alpha \beta \gamma } R_{\gamma }{}^{\nu }{}_{\varepsilon \lambda } R_{\delta \mu \epsilon \nu } \nabla_{\beta }H^{\varepsilon \lambda \mu }- \frac{3}{32} H_{\alpha }{}^{\delta \epsilon } H^{\alpha \beta \gamma } R_{\gamma \delta \varepsilon }{}^{\nu } R_{\epsilon \lambda \mu \nu } \nabla_{\beta }H^{\varepsilon \lambda \mu }\nn\\
&&- \frac{1}{256} H^{\alpha \beta \gamma } H^{\delta \epsilon \varepsilon } R_{\gamma }{}^{\mu }{}_{\lambda }{}^{\nu } R_{\epsilon \varepsilon \mu \nu } \nabla_{\delta }H_{\alpha \beta }{}^{\lambda }- \frac{7}{128} H^{\alpha \beta \gamma } H^{\delta \epsilon \varepsilon } R_{\gamma \epsilon }{}^{\mu \nu } R_{\varepsilon \mu \lambda \nu } \nabla_{\delta }H_{\alpha \beta }{}^{\lambda }\nn\\
&&+\frac{1}{32} H_{\alpha }{}^{\delta \epsilon } H^{\alpha \beta \gamma } R_{\epsilon }{}^{\lambda \mu \nu } R_{\varepsilon \mu \lambda \nu } \nabla_{\delta }H_{\beta \gamma }{}^{\varepsilon }- \frac{1}{192} H^{\alpha \beta \gamma } H^{\delta \epsilon \varepsilon } R_{\delta \lambda }{}^{\mu \nu } R_{\epsilon \mu \varepsilon \nu } \nabla^{\lambda }H_{\alpha \beta \gamma }\nn\\
&&+\frac{13}{256} H^{\alpha \beta \gamma } H^{\delta \epsilon \varepsilon } R_{\gamma \lambda }{}^{\mu \nu } R_{\epsilon \mu \varepsilon \nu } \nabla^{\lambda }H_{\alpha \beta \delta }+\frac{11}{128} H^{\alpha \beta \gamma } H^{\delta \epsilon \varepsilon } R_{\gamma }{}^{\mu }{}_{\epsilon }{}^{\nu } R_{\varepsilon \lambda \mu \nu } \nabla^{\lambda }H_{\alpha \beta \delta }\nn\\
&&- \frac{1}{32} H^{\alpha \beta \gamma } H^{\delta \epsilon \varepsilon } R_{\gamma }{}^{\mu }{}_{\epsilon }{}^{\nu } R_{\varepsilon \nu \lambda \mu } \nabla^{\lambda }H_{\alpha \beta \delta }- \frac{9}{128} H_{\alpha }{}^{\delta \epsilon } H^{\alpha \beta \gamma } R_{\delta \varepsilon }{}^{\mu \nu } R_{\epsilon \mu \lambda \nu } \nabla^{\lambda }H_{\beta \gamma }{}^{\varepsilon }\nn\\
&&+\frac{17}{128} H_{\alpha }{}^{\delta \epsilon } H^{\alpha \beta \gamma } R_{\delta }{}^{\mu }{}_{\epsilon }{}^{\nu } R_{\varepsilon \mu \lambda \nu } \nabla^{\lambda }H_{\beta \gamma }{}^{\varepsilon }- \frac{1}{32} H_{\alpha }{}^{\delta \epsilon } H^{\alpha \beta \gamma } R_{\gamma \lambda }{}^{\mu \nu } R_{\epsilon \mu \varepsilon \nu } \nabla^{\lambda }H_{\beta \delta }{}^{\varepsilon }\nn\\
&&- \frac{1}{32} H^{\alpha \beta \gamma } H^{\delta \epsilon \varepsilon } R_{\gamma }{}^{\nu }{}_{\delta \lambda } R_{\epsilon \mu \varepsilon \nu } \nabla^{\mu }H_{\alpha \beta }{}^{\lambda }- \frac{1}{64} H^{\alpha \beta \gamma } H^{\delta \epsilon \varepsilon } R_{\gamma }{}^{\nu }{}_{\delta \epsilon } R_{\varepsilon \lambda \mu \nu } \nabla^{\mu }H_{\alpha \beta }{}^{\lambda }\nn\\
&&- \frac{5}{128} H^{\alpha \beta \gamma } H^{\delta \epsilon \varepsilon } R_{\gamma }{}^{\nu }{}_{\delta \epsilon } R_{\varepsilon \nu \lambda \mu } \nabla^{\mu }H_{\alpha \beta }{}^{\lambda }- \frac{1}{32} H_{\alpha }{}^{\delta \epsilon } H^{\alpha \beta \gamma } R_{\gamma \varepsilon \lambda }{}^{\nu } R_{\delta \mu \epsilon \nu } \nabla^{\mu }H_{\beta }{}^{\varepsilon \lambda }\nn\\
&&- \frac{1}{32} H_{\alpha }{}^{\delta \epsilon } H^{\alpha \beta \gamma } R_{\gamma }{}^{\nu }{}_{\delta \mu } R_{\epsilon \varepsilon \lambda \nu } \nabla^{\mu }H_{\beta }{}^{\varepsilon \lambda }- \frac{1}{64} H_{\alpha }{}^{\delta \epsilon } H^{\alpha \beta \gamma } R_{\gamma \mu \delta }{}^{\nu } R_{\epsilon \nu \varepsilon \lambda } \nabla^{\mu }H_{\beta }{}^{\varepsilon \lambda }\nn\\
&&- \frac{1}{8} H_{\alpha }{}^{\delta \epsilon } H^{\alpha \beta \gamma } R_{\gamma }{}^{\nu }{}_{\delta \epsilon } R_{\varepsilon \mu \lambda \nu } \nabla^{\mu }H_{\beta }{}^{\varepsilon \lambda }- \frac{1}{32} H_{\alpha \beta }{}^{\delta } H^{\alpha \beta \gamma } R_{\gamma \epsilon \delta }{}^{\nu } R_{\varepsilon \mu \lambda \nu } \nabla^{\mu }H^{\epsilon \varepsilon \lambda }\nn\\
&&- \frac{3}{32} H^{\alpha \beta \gamma } H^{\delta \epsilon \varepsilon } R_{\beta \delta \gamma \epsilon } R_{\varepsilon \nu \lambda \mu } \nabla^{\nu }H_{\alpha }{}^{\lambda \mu }+\frac{1}{8} H_{\alpha }{}^{\delta \epsilon } H^{\alpha \beta \gamma } R_{\beta \gamma \delta \varepsilon } R_{\epsilon \lambda \mu \nu } \nabla^{\nu }H^{\varepsilon \lambda \mu }\,.
\eeqa
The couplings with structures \([H^2\nabla H^2 R]_{85}\), \([H^4\nabla H^2]_{99}\), \([H^4 R\nabla H]_{41}\), and \([H^2\nabla H^3]_{31}\) appear in Appendix \reef{AppA}.     We note that, although the above couplings are presented in canonical form, they may not be irreducible; a more judicious choice of field redefinitions could potentially reduce them to a smaller set of independent couplings. Thus, while the minimal basis at order $\alpha'^3$  comprises 1349 couplings \cite{Garousi:2020mqn,Garousi:2024rzh}, the true minimum number of T-duality-invariant couplings is currently unknown. It would be worthwhile to investigate this question.

One may be interested in checking the above couplings, found by T-duality, against the string theory S-matrix elements of five- or higher-point functions. The couplings at order \(\alpha'\) in this scheme, which appear in \reef{fourmin}, modify the standard graviton propagators. To obtain complete T-duality-invariant couplings whose corresponding graviton polarizations remain standard, one should find the couplings in the Meissner scheme. In the next section we perform this calculation.

\section{Effective action in Meissner  scheme}

Having obtained the complete set of couplings up to order \(\alpha'^3\) in the MT scheme, one may either apply field redefinitions to recast them into alternative schemes or impose the T-duality constraint directly to derive the couplings in a given scheme. In this section, we focus on the Meissner scheme, where the effective action takes the form
\[
\mathbf{S} = \mathbf{S}_{\mathrm{M}} + \mathbf{S}_{\mathrm{GS}},
\]
with both terms containing even- and odd-parity couplings at all orders in \(\alpha'\). The contributions \(\mathbf{S}_{\mathrm{GS}}\) at each order are obtained by inserting the Green–Schwarz replacement \reef{Green-Schwarz} into the lower-order terms of \(\mathbf{S}_{\mathrm{M}}\).

At order \(\alpha'\), the action is
\[
\mathbf{S}^{(1)} = \mathbf{S}_{\mathrm{M}}^{(1)} + \mathbf{S}_{\mathrm{GS}}^{(1)}.
\]
The even-parity couplings in the Meissner scheme at this order are given by \cite{Meissner:1996sa}:
\beqa
{\bf S}_{\mathrm{M}}^{(1)}&=&-\frac{2}{8\kappa^2}\int d^{10}x \sqrt{-G}e^{-2\Phi}\Big[ \frac{1}{24} H_{\alpha 
}{}^{\delta \epsilon } H^{\alpha \beta \gamma } H_{\beta 
\delta }{}^{\varepsilon } H_{\gamma \epsilon \varepsilon } -  
\frac{1}{8} H_{\alpha \beta }{}^{\delta } H^{\alpha \beta 
\gamma } H_{\gamma }{}^{\epsilon \varepsilon } H_{\delta 
\epsilon \varepsilon } \nn\\&&+ \frac{1}{144} H_{\alpha \beta \gamma 
} H^{\alpha \beta \gamma } H_{\delta \epsilon \varepsilon } H^{
\delta \epsilon \varepsilon } + H_{\alpha }{}^{\gamma \delta } 
H_{\beta \gamma \delta } R^{\alpha \beta } - 4 
R_{\alpha \beta } R^{\alpha \beta } -  
\frac{1}{6} H_{\alpha \beta \gamma } H^{\alpha \beta \gamma } 
R + R^2 \nn\\&&+ R_{\alpha \beta \gamma 
\delta } R^{\alpha \beta \gamma \delta } -  
\frac{1}{2} H_{\alpha }{}^{\delta \epsilon } H^{\alpha \beta 
\gamma } R_{\beta \gamma \delta \epsilon } -  
\frac{2}{3} H_{\beta \gamma \delta } H^{\beta \gamma \delta } 
\nabla_{\alpha }\nabla^{\alpha }\Phi + \frac{2}{3} H_{\beta 
\gamma \delta } H^{\beta \gamma \delta } \nabla_{\alpha }\Phi 
\nabla^{\alpha }\Phi \nn\\&&+ 8 R \nabla_{\alpha }\Phi 
\nabla^{\alpha }\Phi - 16 R_{\alpha \beta } 
\nabla^{\alpha }\Phi \nabla^{\beta }\Phi + 16 \nabla_{\alpha 
}\Phi \nabla^{\alpha }\Phi \nabla_{\beta }\Phi \nabla^{\beta 
}\Phi - 32 \nabla^{\alpha }\Phi \nabla_{\beta }\nabla_{\alpha 
}\Phi \nabla^{\beta }\Phi \nn\\&&+ 2 H_{\alpha }{}^{\gamma \delta } 
H_{\beta \gamma \delta } \nabla^{\beta }\nabla^{\alpha }\Phi \Big]\,.\labell{fourmax}
\eeqa
The odd-parity couplings at this order are identical to those in the MT scheme, as given in \reef{GS1}. The above Meissner-scheme couplings are equivalent to the MT-scheme expressions in \reef{fourmin} up to a specific field redefinition at order \(\alpha'\), which is essential for handling field redefinitions at higher orders. We verify that \(\mathbf{S}^{(1)}\) satisfies the T-duality constraint \reef{T1} and determine the corresponding \(\alpha'\)-corrections to the Buscher rules. These corrections are necessary for studying T-duality at higher orders and have already been computed in \cite{Garousi:2019wgz}.

At order \(\alpha'^2\), the action reads
\[
\mathbf{S}^{(2)} = \mathbf{S}_{\mathrm{M}}^{(2)} + \mathbf{S}_{\mathrm{GS}}^{(2)}.
\]
Here, \(\mathbf{S}_{\mathrm{GS}}^{(2)}\), which includes both even- and odd-parity terms, is obtained by applying the Green–Schwarz replacement \reef{Green-Schwarz} to the leading-order action \reef{leading} and to \(\mathbf{S}_{\mathrm{M}}^{(1)}\) in \reef{fourmax}. The Meissner-scheme contribution \(\mathbf{S}_{\mathrm{M}}^{(2)}\) was derived in \cite{Garousi:2023kxw,Gholian:2023kjj} and is given by
\beqa
\bS^{(2)}_{\mathrm{M}}&=&-\frac{2}{\kappa^2}\int d^{10} x \sqrt{-G} \,e^{-2\Phi}\Big[- \frac{1}{768} H_{\alpha }{}^{\delta \epsilon } H^{\alpha \beta \gamma } H_{\beta \delta }{}^{\varepsilon } H_{\gamma }{}^{\mu \zeta } H_{\epsilon \mu }{}^{\eta } H_{\varepsilon \zeta \eta }\nn\\
&&+\frac{1}{32} H^{\alpha \beta \gamma } H^{\delta \epsilon \varepsilon } R_{\alpha \beta \delta }{}^{\mu } R_{\gamma \mu \epsilon \varepsilon }- \frac{1}{32} H_{\alpha }{}^{\delta \epsilon } H^{\alpha \beta \gamma } R_{\beta }{}^{\varepsilon }{}_{\gamma }{}^{\mu } R_{\delta \varepsilon \epsilon \mu }\nn\\
&&+\frac{1}{64} H_{\alpha }{}^{\delta \epsilon } H^{\alpha \beta \gamma } H_{\beta \delta }{}^{\varepsilon } H_{\gamma }{}^{\mu \zeta } R_{\epsilon \mu \varepsilon \zeta }- \frac{1}{128} H^{\alpha \beta \gamma } H^{\delta \epsilon \varepsilon } \nabla_{\beta }H_{\alpha \delta }{}^{\mu } \nabla_{\varepsilon }H_{\gamma \epsilon \mu }\nn\\
&&+\frac{1}{256} H_{\alpha }{}^{\delta \epsilon } H^{\alpha \beta \gamma } \nabla_{\epsilon }H_{\delta \mu \varepsilon } \nabla^{\mu }H_{\beta \gamma }{}^{\varepsilon }- \frac{1}{32} H_{\alpha }{}^{\delta \epsilon } H^{\alpha \beta \gamma } R_{\beta \delta }{}^{\mu \varepsilon } R_{\gamma \varepsilon \epsilon \mu} \nn\\
&& -\frac{1}{16} H^{\alpha \beta \gamma } H^{\delta \epsilon \varepsilon } R_{\gamma \epsilon \mu \varepsilon } \nabla_{\beta }H_{\alpha \delta }{}^{\mu }- \frac{1}{64} H_{\alpha }{}^{\delta \epsilon } H^{\alpha \beta \gamma } R_{\delta \epsilon \mu \varepsilon } \nabla_{\gamma }H_{\beta }{}^{\mu \varepsilon }\nn\\
&&- \frac{1}{128} H_{\alpha }{}^{\delta \epsilon } H^{\alpha \beta \gamma } H_{\beta \delta }{}^{\varepsilon } H_{\gamma }{}^{\mu \zeta } \nabla_{\varepsilon }H_{\epsilon \mu \zeta }\Big]\,,
\labell{S2M}
\eeqa
where the terms in the last two lines are odd under parity, while all preceding ones are even. One may transform \(\mathbf{S}_{\mathrm{MT}}^{(2)}\) into the above form by applying a field redefinition at order \(\alpha'^2\) with arbitrary parameters, together with the fixed order-\(\alpha'\) redefinition that relates \(\mathbf{S}_{\mathrm{MT}}^{(1)}\) to \(\mathbf{S}_{\mathrm{M}}^{(1)}\). The resulting fixed order-\(\alpha'^2\) redefinition is necessary for subsequent higher-order field redefinitions. We demonstrate that \(\mathbf{S}^{(2)}\) satisfies the T-duality constraint \reef{T2} and extract the corresponding \(\alpha'^2\)-corrections to the Buscher rules, which are required for higher-order T-duality analyses. 

At order \(\alpha'^3\), the action is
\[
\mathbf{S}^{(3)} = \mathbf{S}_{\mathrm{M}}^{(3)} + \mathbf{S}_{\mathrm{GS}}^{(3)}.
\]
The term \(\mathbf{S}_{\mathrm{GS}}^{(3)}\), again containing both parities, is obtained from the Green–Schwarz replacement applied to \(\mathbf{S}_{\mathrm{M}}^{(1)}\) \reef{fourmax} and \(\mathbf{S}_{\mathrm{M}}^{(2)}\) \reef{S2M}; note that substituting into the leading-order action yields no \(\alpha'^3\) contributions. The Meissner-scheme piece \(\mathbf{S}_{\mathrm{M}}^{(3)}\), however, is not yet known. We therefore consider the minimal basis consisting of 1349 independent couplings with undetermined coefficients \cite{Garousi:2020mqn,Garousi:2024rzh}. One could in principle fix these coefficients by employing field redefinitions at order \(\alpha'^3\) with arbitrary parameters, along with the fixed redefinitions at orders \(\alpha'\) and \(\alpha'^2\) already used to map the MT scheme to the Meissner scheme. Rather than performing this explicit calculation, we instead impose the T-duality constraint \reef{T3} to determine the 1349 coupling constants, following the same procedure as in the MT scheme of the previous section. We find that T-duality fixes all coefficients uniquely up to a single free parameter, which corresponds to the overall coefficient of the superstring couplings.

The resulting action, however, is not in canonical form. We therefore apply a field redefinition solely at order \(\alpha'^3\) to rewrite it such that the dilaton appears only through the overall factor \(e^{-2\Phi}\). This yields the canonical-form result presented below:
\beqa
{\bf S}_{\mathrm{M}}^{(3)}&\!\!\!\!=\!\!\!\!&-\frac{2}{\kappa^2}\int d^{10}x \sqrt{-G}e^{-2\Phi}\Big[ \frac{\z(3)}{16}\cL_{\rm II}+[R^4]_{5}+[R^2\nabla H^2]_{14}+[\nabla H^4]_{7}+[H^8]_{3}\nn\\&&+[H^6 R]_{2}+[H^4 R^2]_{10}+[H^2 R^3 ]_{19}+[H^2\nabla H^2 R]_{96}+[H^4\nabla H^2]_{93}\labell{S3M}\\&&
+[H^6 \nabla H]_{15}+[H^4 R\nabla H]_{52}+[H^2 R^2\nabla H]_{31}
	+[R^3\nabla H]_{3}+[R\nabla H^3]_{4}+[H^2\nabla H^3]_{17}\Big]\,,\nn
\eeqa
where
\beqa
[R^4]_{5}&\!\!\!\!\!=\!\!\!\!\!&\frac{1}{64} R_{\alpha \beta }{}^{\epsilon \varepsilon } R^{\alpha \beta \gamma \delta } R_{\gamma }{}^{\zeta }{}_{\epsilon }{}^{\eta } R_{\delta \zeta \varepsilon \eta }- \frac{1}{64} R_{\alpha \beta }{}^{\epsilon \varepsilon } R^{\alpha \beta \gamma \delta } R_{\gamma }{}^{\zeta }{}_{\epsilon }{}^{\eta } R_{\delta \eta \varepsilon \zeta }\nn\\
&&- \frac{1}{64} R_{\alpha \beta }{}^{\epsilon \varepsilon } R^{\alpha \beta \gamma \delta } R_{\gamma }{}^{\zeta }{}_{\delta }{}^{\eta } R_{\epsilon \zeta \varepsilon \eta }+\frac{1}{16} R_{\alpha \gamma \beta }{}^{\epsilon } R^{\alpha \beta \gamma \delta } R_{\delta }{}^{\varepsilon \zeta \eta } R_{\epsilon \zeta \varepsilon \eta }\nn\\
&&- \frac{1}{128} R_{\alpha \gamma \beta \delta } R^{\alpha \beta \gamma \delta } R_{\epsilon \zeta \varepsilon \eta } R^{\epsilon \varepsilon \zeta \eta }\,,\nn\\
{[}H^8{]}_{3}&\!\!\!\!\!=\!\!\!\!\!&\frac{1}{2048} H_{\alpha }{}^{\delta \epsilon } H^{\alpha \beta \gamma } H_{\beta \delta }{}^{\varepsilon } H_{\gamma }{}^{\zeta \eta } H_{\epsilon \zeta }{}^{\theta } H_{\varepsilon }{}^{\iota \kappa } H_{\eta \iota }{}^{\rho } H_{\theta \kappa \rho }\nn\\
&&+\frac{1}{4096} H_{\alpha }{}^{\delta \epsilon } H^{\alpha \beta \gamma } H_{\beta \delta }{}^{\varepsilon } H_{\gamma \epsilon }{}^{\zeta } H_{\varepsilon }{}^{\eta \theta } H_{\zeta }{}^{\iota \kappa } H_{\eta \iota }{}^{\rho } H_{\theta \kappa \rho }\nn\\
&&- \frac{1}{294912} H_{\alpha }{}^{\delta \epsilon } H^{\alpha \beta \gamma } H_{\beta \delta }{}^{\varepsilon } H_{\gamma \epsilon \varepsilon } H_{\zeta }{}^{\iota \kappa } H^{\zeta \eta \theta } H_{\eta \iota }{}^{\rho } H_{\theta \kappa \rho }\,,
\nn\\
{[}R^2\nabla H^2{]}_{14}&\!\!\!\!\!=\!\!\!\!\!&- \frac{1}{32} R_{\gamma }{}^{\varepsilon \zeta \eta } R_{\epsilon \zeta \varepsilon \eta } \nabla^{\delta }H^{\alpha \beta \gamma } \nabla^{\epsilon }H_{\alpha \beta \delta }- \frac{1}{16} R_{\gamma }{}^{\zeta }{}_{\epsilon }{}^{\eta } R_{\delta \zeta \varepsilon \eta } \nabla^{\delta }H^{\alpha \beta \gamma } \nabla^{\varepsilon }H_{\alpha \beta }{}^{\epsilon }\nn\\
&&+\frac{1}{32} R_{\gamma }{}^{\zeta }{}_{\epsilon }{}^{\eta } R_{\delta \eta \varepsilon \zeta } \nabla^{\delta }H^{\alpha \beta \gamma } \nabla^{\varepsilon }H_{\alpha \beta }{}^{\epsilon }+\frac{1}{32} R_{\gamma }{}^{\zeta }{}_{\delta }{}^{\eta } R_{\epsilon \eta \varepsilon \zeta } \nabla^{\delta }H^{\alpha \beta \gamma } \nabla^{\varepsilon }H_{\alpha \beta }{}^{\epsilon }\nn\\
&&- \frac{3}{32} R_{\beta }{}^{\zeta }{}_{\epsilon }{}^{\eta } R_{\gamma \zeta \varepsilon \eta } \nabla^{\delta }H^{\alpha \beta \gamma } \nabla^{\varepsilon }H_{\alpha \delta }{}^{\epsilon }+\frac{3}{32} R_{\beta }{}^{\zeta }{}_{\epsilon }{}^{\eta } R_{\gamma \eta \varepsilon \zeta } \nabla^{\delta }H^{\alpha \beta \gamma } \nabla^{\varepsilon }H_{\alpha \delta }{}^{\epsilon }\nn\\
&&- \frac{1}{32} R_{\beta \zeta \epsilon }{}^{\eta } R_{\gamma \eta \delta \varepsilon } \nabla^{\delta }H^{\alpha \beta \gamma } \nabla^{\zeta }H_{\alpha }{}^{\epsilon \varepsilon }+\frac{1}{16} R_{\beta \epsilon \gamma }{}^{\eta } R_{\delta \zeta \varepsilon \eta } \nabla^{\delta }H^{\alpha \beta \gamma } \nabla^{\zeta }H_{\alpha }{}^{\epsilon \varepsilon }\nn\\
&&- \frac{3}{64} R_{\beta \gamma \epsilon }{}^{\eta } R_{\delta \eta \varepsilon \zeta } \nabla^{\delta }H^{\alpha \beta \gamma } \nabla^{\zeta }H_{\alpha }{}^{\epsilon \varepsilon }+\frac{1}{64} R_{\alpha \zeta \beta }{}^{\eta } R_{\gamma \eta \epsilon \varepsilon } \nabla^{\delta }H^{\alpha \beta \gamma } \nabla^{\zeta }H_{\delta }{}^{\epsilon \varepsilon }\nn\\
&&+\frac{1}{16} R_{\alpha \epsilon \beta }{}^{\eta } R_{\gamma \eta \varepsilon \zeta } \nabla^{\delta }H^{\alpha \beta \gamma } \nabla^{\zeta }H_{\delta }{}^{\epsilon \varepsilon }+\frac{1}{64} R_{\alpha \epsilon \beta \varepsilon } R_{\gamma \zeta \delta \eta } \nabla^{\delta }H^{\alpha \beta \gamma } \nabla^{\eta }H^{\epsilon \varepsilon \zeta }\nn\\
&&- \frac{1}{64} R_{\alpha \epsilon \beta \varepsilon } R_{\gamma \eta \delta \zeta } \nabla^{\delta }H^{\alpha \beta \gamma } \nabla^{\eta }H^{\epsilon \varepsilon \zeta }- \frac{1}{128} R_{\alpha \beta \delta \epsilon } R_{\gamma \eta \varepsilon \zeta } \nabla^{\delta }H^{\alpha \beta \gamma } \nabla^{\eta }H^{\epsilon \varepsilon \zeta }\,,\nn\\
{[}\nabla H^4{]}_{7}&\!\!\!\!\!=\!\!\!\!\!&- \frac{1}{512} \nabla^{\delta }H^{\alpha \beta \gamma } \nabla^{\varepsilon }H_{\alpha \beta }{}^{\epsilon } \nabla_{\eta }H_{\epsilon \varepsilon \zeta } \nabla^{\eta }H_{\gamma \delta }{}^{\zeta }+\frac{1}{256} \nabla^{\delta }H^{\alpha \beta \gamma } \nabla_{\epsilon }H_{\delta \zeta \eta } \nabla^{\varepsilon }H_{\alpha \beta }{}^{\epsilon } \nabla^{\eta }H_{\gamma \varepsilon }{}^{\zeta }\nn\\
&&- \frac{1}{256} \nabla^{\delta }H^{\alpha \beta \gamma } \nabla^{\varepsilon }H_{\alpha \beta }{}^{\epsilon } \nabla_{\zeta }H_{\delta \epsilon \eta } \nabla^{\eta }H_{\gamma \varepsilon }{}^{\zeta }- \frac{1}{512} \nabla^{\delta }H^{\alpha \beta \gamma } \nabla^{\varepsilon }H_{\alpha \beta }{}^{\epsilon } \nabla_{\eta }H_{\delta \epsilon \zeta } \nabla^{\eta }H_{\gamma \varepsilon }{}^{\zeta }\nn\\
&&+\frac{1}{1024} \nabla_{\delta }H_{\alpha \beta }{}^{\epsilon } \nabla^{\delta }H^{\alpha \beta \gamma } \nabla_{\eta }H_{\epsilon \varepsilon \zeta } \nabla^{\eta }H_{\gamma }{}^{\varepsilon \zeta }+\frac{3}{1024} \nabla_{\gamma }H_{\varepsilon \zeta \eta } \nabla^{\delta }H^{\alpha \beta \gamma } \nabla^{\epsilon }H_{\alpha \beta \delta } \nabla^{\eta }H_{\epsilon }{}^{\varepsilon \zeta }\nn\\
&&- \frac{1}{4608} \nabla_{\delta }H_{\alpha \beta \gamma } \nabla^{\delta }H^{\alpha \beta \gamma } \nabla_{\eta }H_{\epsilon \varepsilon \zeta } \nabla^{\eta }H^{\epsilon \varepsilon \zeta }\,,\nn\\
{[}H^6R{]}_{2}&\!\!\!\!\!=\!\!\!\!\!&- \frac{1}{512} H_{\alpha }{}^{\delta \epsilon } H^{\alpha \beta \gamma } H_{\beta \delta }{}^{\varepsilon } H_{\gamma \epsilon }{}^{\zeta } H_{\varepsilon }{}^{\eta \theta } H_{\zeta }{}^{\iota \kappa } R_{\eta \iota \theta \kappa }\nn\\
&&+\frac{1}{3072} H_{\alpha }{}^{\delta \epsilon } H^{\alpha \beta \gamma } H_{\beta \delta }{}^{\varepsilon } H_{\gamma \epsilon \varepsilon } H_{\zeta }{}^{\iota \kappa } H^{\zeta \eta \theta } R_{\eta \iota \theta \kappa }\,,\nn\\
{[}H^4R^2{]}_{10}&\!\!\!\!\!=\!\!\!\!\!&\frac{1}{256} H_{\alpha }{}^{\delta \epsilon } H^{\alpha \beta \gamma } H_{\varepsilon }{}^{\theta \iota } H^{\varepsilon \zeta \eta } R_{\beta \zeta \gamma \eta } R_{\delta \theta \epsilon \iota }+\frac{9}{512} H_{\alpha \beta }{}^{\delta } H^{\alpha \beta \gamma } H^{\epsilon \varepsilon \zeta } H^{\eta \theta \iota } R_{\gamma \epsilon \delta \eta } R_{\varepsilon \theta \zeta \iota }\nn\\
&&+\frac{1}{128} H_{\alpha }{}^{\delta \epsilon } H^{\alpha \beta \gamma } H_{\beta \delta }{}^{\varepsilon } H_{\gamma }{}^{\zeta \eta } R_{\epsilon }{}^{\theta }{}_{\zeta }{}^{\iota } R_{\varepsilon \iota \eta \theta }- \frac{1}{128} H_{\alpha }{}^{\delta \epsilon } H^{\alpha \beta \gamma } H_{\varepsilon }{}^{\theta \iota } H^{\varepsilon \zeta \eta } R_{\beta \delta \gamma \epsilon } R_{\zeta \theta \eta \iota }\nn\\
&&- \frac{1}{256} H_{\alpha \beta }{}^{\delta } H^{\alpha \beta \gamma } H_{\gamma }{}^{\epsilon \varepsilon } H_{\epsilon \varepsilon }{}^{\zeta } R_{\delta }{}^{\eta \theta \iota } R_{\zeta \theta \eta \iota }+\frac{5}{256} H_{\alpha }{}^{\delta \epsilon } H^{\alpha \beta \gamma } H_{\beta \delta }{}^{\varepsilon } H_{\gamma \epsilon }{}^{\zeta } R_{\varepsilon }{}^{\eta \theta \iota } R_{\zeta \theta \eta \iota }\nn\\
&&- \frac{9}{512} H_{\alpha \beta }{}^{\delta } H^{\alpha \beta \gamma } H_{\epsilon }{}^{\eta \theta } H^{\epsilon \varepsilon \zeta } R_{\gamma \varepsilon \delta }{}^{\iota } R_{\zeta \iota \eta \theta }- \frac{3}{512} H_{\alpha }{}^{\delta \epsilon } H^{\alpha \beta \gamma } H_{\beta }{}^{\varepsilon \zeta } H_{\gamma }{}^{\eta \theta } R_{\delta \epsilon \varepsilon }{}^{\iota } R_{\zeta \iota \eta \theta }\nn\\
&&- \frac{1}{3072} H_{\alpha }{}^{\delta \epsilon } H^{\alpha \beta \gamma } H_{\beta \delta }{}^{\varepsilon } H_{\gamma \epsilon \varepsilon } R_{\zeta \theta \eta \iota } R^{\zeta \eta \theta \iota }+\frac{1}{1024} H_{\alpha \beta }{}^{\delta } H^{\alpha \beta \gamma } H_{\gamma }{}^{\epsilon \varepsilon } H_{\delta \epsilon \varepsilon } R_{\zeta \theta \eta \iota } R^{\zeta \eta \theta \iota }\,,
\nn\\
{[}H^2 R^3{]}_{19}&\!\!\!\!\!=\!\!\!\!\!&- \frac{1}{32} H^{\alpha \beta \gamma } H^{\delta \epsilon \varepsilon } R_{\alpha \delta }{}^{\zeta \eta } R_{\beta \zeta \epsilon }{}^{\theta } R_{\gamma \theta \varepsilon \eta }+\frac{3}{32} H^{\alpha \beta \gamma } H^{\delta \epsilon \varepsilon } R_{\alpha }{}^{\zeta }{}_{\beta }{}^{\eta } R_{\gamma \zeta \delta }{}^{\theta } R_{\epsilon \eta \varepsilon \theta }\nn\\
&&- \frac{1}{32} H^{\alpha \beta \gamma } H^{\delta \epsilon \varepsilon } R_{\alpha }{}^{\zeta }{}_{\beta }{}^{\eta } R_{\gamma }{}^{\theta }{}_{\delta \zeta } R_{\epsilon \eta \varepsilon \theta }- \frac{3}{32} H_{\alpha }{}^{\delta \epsilon } H^{\alpha \beta \gamma } R_{\beta \delta }{}^{\varepsilon \zeta } R_{\gamma }{}^{\eta }{}_{\varepsilon }{}^{\theta } R_{\epsilon \eta \zeta \theta }\nn\\
&&+\frac{1}{8} H_{\alpha }{}^{\delta \epsilon } H^{\alpha \beta \gamma } R_{\beta }{}^{\varepsilon }{}_{\delta }{}^{\zeta } R_{\gamma }{}^{\eta }{}_{\varepsilon }{}^{\theta } R_{\epsilon \eta \zeta \theta }- \frac{1}{32} H_{\alpha }{}^{\delta \epsilon } H^{\alpha \beta \gamma } R_{\beta }{}^{\varepsilon }{}_{\gamma }{}^{\zeta } R_{\delta }{}^{\eta }{}_{\varepsilon }{}^{\theta } R_{\epsilon \eta \zeta \theta }\nn\\
&&+\frac{1}{8} H_{\alpha }{}^{\delta \epsilon } H^{\alpha \beta \gamma } R_{\beta \delta }{}^{\varepsilon \zeta } R_{\gamma }{}^{\eta }{}_{\varepsilon }{}^{\theta } R_{\epsilon \theta \zeta \eta }- \frac{1}{16} H_{\alpha }{}^{\delta \epsilon } H^{\alpha \beta \gamma } R_{\beta }{}^{\varepsilon }{}_{\delta }{}^{\zeta } R_{\gamma }{}^{\eta }{}_{\varepsilon }{}^{\theta } R_{\epsilon \theta \zeta \eta }\nn\\
&&+\frac{3}{32} H_{\alpha }{}^{\delta \epsilon } H^{\alpha \beta \gamma } R_{\beta }{}^{\varepsilon }{}_{\gamma }{}^{\zeta } R_{\delta }{}^{\eta }{}_{\varepsilon }{}^{\theta } R_{\epsilon \theta \zeta \eta }- \frac{1}{32} H_{\alpha \beta }{}^{\delta } H^{\alpha \beta \gamma } R_{\gamma }{}^{\epsilon \varepsilon \zeta } R_{\delta }{}^{\eta }{}_{\varepsilon }{}^{\theta } R_{\epsilon \theta \zeta \eta }\nn\\
&&- \frac{1}{32} H^{\alpha \beta \gamma } H^{\delta \epsilon \varepsilon } R_{\alpha \delta \beta \epsilon } R_{\gamma }{}^{\zeta \eta \theta } R_{\varepsilon \eta \zeta \theta }- \frac{3}{16} H^{\alpha \beta \gamma } H^{\delta \epsilon \varepsilon } R_{\alpha \delta \beta }{}^{\zeta } R_{\gamma }{}^{\eta }{}_{\epsilon }{}^{\theta } R_{\varepsilon \eta \zeta \theta }\nn\\
&&- \frac{1}{32} H_{\alpha }{}^{\delta \epsilon } H^{\alpha \beta \gamma } R_{\beta }{}^{\varepsilon }{}_{\delta }{}^{\zeta } R_{\gamma }{}^{\eta }{}_{\epsilon }{}^{\theta } R_{\varepsilon \eta \zeta \theta }+\frac{1}{64} H_{\alpha \beta }{}^{\delta } H^{\alpha \beta \gamma } R_{\gamma }{}^{\epsilon \varepsilon \zeta } R_{\delta \epsilon }{}^{\eta \theta } R_{\varepsilon \eta \zeta \theta }\nn\\
&&- \frac{1}{32} H_{\alpha }{}^{\delta \epsilon } H^{\alpha \beta \gamma } R_{\beta }{}^{\varepsilon }{}_{\gamma }{}^{\zeta } R_{\delta }{}^{\eta }{}_{\epsilon }{}^{\theta } R_{\varepsilon \eta \zeta \theta }- \frac{1}{32} H_{\alpha \beta }{}^{\delta } H^{\alpha \beta \gamma } R_{\gamma }{}^{\epsilon }{}_{\delta }{}^{\varepsilon } R_{\epsilon }{}^{\zeta \eta \theta } R_{\varepsilon \eta \zeta \theta }\nn\\
&&- \frac{1}{32} H^{\alpha \beta \gamma } H^{\delta \epsilon \varepsilon } R_{\alpha \beta }{}^{\zeta \eta } R_{\gamma \delta \epsilon }{}^{\theta } R_{\varepsilon \theta \zeta \eta }+\frac{1}{8} H^{\alpha \beta \gamma } H^{\delta \epsilon \varepsilon } R_{\alpha \delta \beta }{}^{\zeta } R_{\gamma }{}^{\eta }{}_{\epsilon }{}^{\theta } R_{\varepsilon \theta \zeta \eta }\nn\\
&&+\frac{1}{128} H_{\alpha }{}^{\delta \epsilon } H^{\alpha \beta \gamma } R_{\beta \delta \gamma \epsilon } R_{\varepsilon \eta \zeta \theta } R^{\varepsilon \zeta \eta \theta }\,,\nn\\
{[}R^3\nabla H{]}_{3}&\!\!\!\!\!=\!\!\!\!\!&\frac{1}{8} R_{\alpha }{}^{\epsilon }{}_{\beta }{}^{\varepsilon } R_{\gamma }{}^{\zeta }{}_{\epsilon }{}^{\eta } R_{\delta \zeta \varepsilon \eta } \nabla^{\delta }H^{\alpha \beta \gamma }- \frac{1}{8} R_{\alpha }{}^{\epsilon }{}_{\beta }{}^{\varepsilon } R_{\gamma }{}^{\zeta }{}_{\epsilon }{}^{\eta } R_{\delta \eta \varepsilon \zeta } \nabla^{\delta }H^{\alpha \beta \gamma }\nn\\
&&- \frac{1}{8} R_{\alpha \delta \beta }{}^{\epsilon } R_{\gamma }{}^{\varepsilon \zeta \eta } R_{\epsilon \zeta \varepsilon \eta } \nabla^{\delta }H^{\alpha \beta \gamma }\,,\nn\\
{[}R\nabla H^3{]}_{4}&\!\!\!\!\!=\!\!\!\!\!&\frac{1}{64} R_{\delta \zeta \epsilon \eta } \nabla^{\delta }H^{\alpha \beta \gamma } \nabla^{\varepsilon }H_{\alpha \beta }{}^{\epsilon } \nabla^{\eta }H_{\gamma \varepsilon }{}^{\zeta }- \frac{1}{32} R_{\delta \eta \epsilon \zeta } \nabla^{\delta }H^{\alpha \beta \gamma } \nabla^{\varepsilon }H_{\alpha \beta }{}^{\epsilon } \nabla^{\eta }H_{\gamma \varepsilon }{}^{\zeta }\nn\\
&&+\frac{1}{128} R_{\epsilon \eta \varepsilon \zeta } \nabla_{\delta }H_{\alpha \beta }{}^{\epsilon } \nabla^{\delta }H^{\alpha \beta \gamma } \nabla^{\eta }H_{\gamma }{}^{\varepsilon \zeta }- \frac{1}{64} R_{\gamma \zeta \epsilon \eta } \nabla^{\delta }H^{\alpha \beta \gamma } \nabla^{\varepsilon }H_{\alpha \beta }{}^{\epsilon } \nabla^{\eta }H_{\delta \varepsilon }{}^{\zeta }\,,\nn\\
{[}H^6 \nabla H{]}_{15}&\!\!\!\!\!=\!\!\!\!\!&\frac{1}{512} H_{\alpha }{}^{\delta \epsilon } H^{\alpha \beta \gamma } H_{\beta \delta }{}^{\varepsilon } H_{\gamma \epsilon }{}^{\zeta } H_{\varepsilon }{}^{\eta \theta } H_{\eta }{}^{\iota \rho } \nabla_{\theta }H_{\zeta \iota \rho }\nn\\
&&+\frac{1}{512} H_{\alpha \beta }{}^{\delta } H^{\alpha \beta \gamma } H_{\gamma }{}^{\epsilon \varepsilon } H_{\delta \epsilon }{}^{\zeta } H_{\varepsilon }{}^{\eta \theta } H_{\eta }{}^{\iota \rho } \nabla_{\theta }H_{\zeta \iota \rho }\nn\\
&&- \frac{1}{1024} H_{\alpha \beta }{}^{\delta } H^{\alpha \beta \gamma } H_{\gamma }{}^{\epsilon \varepsilon } H_{\delta }{}^{\zeta \eta } H_{\epsilon \zeta }{}^{\theta } H_{\varepsilon }{}^{\iota \rho } \nabla_{\theta }H_{\eta \iota \rho }\nn\\
&&- \frac{9}{4096} H_{\alpha \beta }{}^{\delta } H^{\alpha \beta \gamma } H_{\gamma }{}^{\epsilon \varepsilon } H_{\delta }{}^{\zeta \eta } H_{\epsilon \varepsilon }{}^{\theta } H_{\zeta }{}^{\iota \rho } \nabla_{\theta }H_{\eta \iota \rho }\nn\\
&&- \frac{3}{1024} H_{\alpha \beta }{}^{\delta } H^{\alpha \beta \gamma } H_{\gamma }{}^{\epsilon \varepsilon } H_{\epsilon }{}^{\zeta \eta } H_{\varepsilon }{}^{\theta \iota } H_{\zeta \theta }{}^{\rho } \nabla_{\iota }H_{\delta \eta \rho }\nn\\
&&- \frac{1}{512} H_{\alpha }{}^{\delta \epsilon } H^{\alpha \beta \gamma } H_{\beta \delta }{}^{\varepsilon } H_{\gamma }{}^{\zeta \eta } H_{\epsilon }{}^{\theta \iota } H_{\zeta \theta }{}^{\rho } \nabla_{\iota }H_{\varepsilon \eta \rho }\nn\\
&&- \frac{1}{2048} H_{\alpha \beta }{}^{\delta } H^{\alpha \beta \gamma } H_{\gamma }{}^{\epsilon \varepsilon } H_{\epsilon }{}^{\zeta \eta } H_{\zeta }{}^{\theta \iota } H_{\theta \iota }{}^{\rho } \nabla_{\rho }H_{\delta \varepsilon \eta }\nn\\
&&- \frac{9}{4096} H_{\alpha \beta }{}^{\delta } H^{\alpha \beta \gamma } H_{\gamma }{}^{\epsilon \varepsilon } H_{\epsilon }{}^{\zeta \eta } H_{\varepsilon }{}^{\theta \iota } H_{\zeta \eta }{}^{\rho } \nabla_{\rho }H_{\delta \theta \iota }\nn\\
&&+\frac{3}{1024} H_{\alpha \beta }{}^{\delta } H^{\alpha \beta \gamma } H_{\gamma }{}^{\epsilon \varepsilon } H_{\epsilon }{}^{\zeta \eta } H_{\varepsilon \zeta }{}^{\theta } H_{\eta }{}^{\iota \rho } \nabla_{\rho }H_{\delta \theta \iota }\nn\\
&&- \frac{15}{4096} H_{\alpha \beta }{}^{\delta } H^{\alpha \beta \gamma } H_{\gamma }{}^{\epsilon \varepsilon } H_{\delta }{}^{\zeta \eta } H_{\epsilon }{}^{\theta \iota } H_{\theta \iota }{}^{\rho } \nabla_{\rho }H_{\varepsilon \zeta \eta }\nn\\
&&+\frac{3}{1024} H_{\alpha \beta }{}^{\delta } H^{\alpha \beta \gamma } H_{\gamma }{}^{\epsilon \varepsilon } H_{\delta }{}^{\zeta \eta } H_{\epsilon }{}^{\theta \iota } H_{\zeta \theta }{}^{\rho } \nabla_{\rho }H_{\varepsilon \eta \iota }\nn\\
&&- \frac{1}{512} H_{\alpha }{}^{\delta \epsilon } H^{\alpha \beta \gamma } H_{\beta \delta }{}^{\varepsilon } H_{\gamma }{}^{\zeta \eta } H_{\epsilon \zeta }{}^{\theta } H_{\eta }{}^{\iota \rho } \nabla_{\rho }H_{\varepsilon \theta \iota }\nn\\
&&+\frac{1}{512} H_{\alpha }{}^{\delta \epsilon } H^{\alpha \beta \gamma } H_{\beta \delta }{}^{\varepsilon } H_{\gamma }{}^{\zeta \eta } H_{\epsilon \zeta }{}^{\theta } H_{\varepsilon }{}^{\iota \rho } \nabla_{\rho }H_{\eta \theta \iota }\nn\\
&&+\frac{1}{1024} H_{\alpha \beta }{}^{\delta } H^{\alpha \beta \gamma } H_{\gamma }{}^{\epsilon \varepsilon } H_{\delta }{}^{\zeta \eta } H_{\epsilon \zeta }{}^{\theta } H_{\varepsilon }{}^{\iota \rho } \nabla_{\rho }H_{\eta \theta \iota }\nn\\
&&- \frac{9}{2048} H_{\alpha \beta }{}^{\delta } H^{\alpha \beta \gamma } H_{\gamma }{}^{\epsilon \varepsilon } H_{\delta }{}^{\zeta \eta } H_{\epsilon \varepsilon }{}^{\theta } H_{\zeta }{}^{\iota \rho } \nabla_{\rho }H_{\eta \theta \iota }\,,\nn\\
{[}H^2\nabla H^3{]}_{17}&\!\!\!\!\!=\!\!\!\!\!&\frac{3}{256} H_{\alpha }{}^{\delta \epsilon } H^{\alpha \beta \gamma } \nabla_{\epsilon }H_{\varepsilon \eta \theta } \nabla^{\zeta }H_{\beta \delta }{}^{\varepsilon } \nabla^{\theta }H_{\gamma \zeta }{}^{\eta }- \frac{3}{256} H_{\alpha }{}^{\delta \epsilon } H^{\alpha \beta \gamma } \nabla_{\varepsilon }H_{\epsilon \eta \theta } \nabla^{\zeta }H_{\beta \delta }{}^{\varepsilon } \nabla^{\theta }H_{\gamma \zeta }{}^{\eta }\nn\\
&&- \frac{1}{512} H^{\alpha \beta \gamma } H^{\delta \epsilon \varepsilon } \nabla^{\zeta }H_{\alpha \beta \delta } \nabla_{\theta }H_{\epsilon \varepsilon \eta } \nabla^{\theta }H_{\gamma \zeta }{}^{\eta }- \frac{5}{512} H_{\alpha \beta }{}^{\delta } H^{\alpha \beta \gamma } \nabla^{\zeta }H_{\gamma }{}^{\epsilon \varepsilon } \nabla_{\theta }H_{\varepsilon \zeta \eta } \nabla^{\theta }H_{\delta \epsilon }{}^{\eta }\nn\\
&&+\frac{1}{512} H_{\alpha }{}^{\delta \epsilon } H^{\alpha \beta \gamma } \nabla_{\zeta }H_{\epsilon \eta \theta } \nabla^{\zeta }H_{\beta \gamma }{}^{\varepsilon } \nabla^{\theta }H_{\delta \varepsilon }{}^{\eta }+\frac{1}{256} H_{\alpha }{}^{\delta \epsilon } H^{\alpha \beta \gamma } \nabla^{\zeta }H_{\beta \gamma }{}^{\varepsilon } \nabla_{\theta }H_{\epsilon \zeta \eta } \nabla^{\theta }H_{\delta \varepsilon }{}^{\eta }\nn\\
&&+\frac{3}{512} H_{\alpha }{}^{\delta \epsilon } H^{\alpha \beta \gamma } \nabla_{\epsilon }H_{\varepsilon \eta \theta } \nabla^{\zeta }H_{\beta \gamma }{}^{\varepsilon } \nabla^{\theta }H_{\delta \zeta }{}^{\eta }- \frac{1}{256} H_{\alpha }{}^{\delta \epsilon } H^{\alpha \beta \gamma } \nabla_{\varepsilon }H_{\epsilon \eta \theta } \nabla^{\zeta }H_{\beta \gamma }{}^{\varepsilon } \nabla^{\theta }H_{\delta \zeta }{}^{\eta }\nn\\
&&- \frac{3}{512} H_{\alpha \beta }{}^{\delta } H^{\alpha \beta \gamma } \nabla^{\zeta }H_{\gamma }{}^{\epsilon \varepsilon } \nabla_{\theta }H_{\epsilon \varepsilon \eta } \nabla^{\theta }H_{\delta \zeta }{}^{\eta }+\frac{1}{1024} H^{\alpha \beta \gamma } H^{\delta \epsilon \varepsilon } \nabla_{\gamma }H_{\zeta \eta \theta } \nabla_{\delta }H_{\alpha \beta }{}^{\zeta } \nabla^{\theta }H_{\epsilon \varepsilon }{}^{\eta }\nn\\
&&+\frac{1}{256} H_{\alpha }{}^{\delta \epsilon } H^{\alpha \beta \gamma } \nabla_{\gamma }H_{\zeta \eta \theta } \nabla_{\delta }H_{\beta }{}^{\varepsilon \zeta } \nabla^{\theta }H_{\epsilon \varepsilon }{}^{\eta }+\frac{3}{1024} H^{\alpha \beta \gamma } H^{\delta \epsilon \varepsilon } \nabla_{\gamma }H_{\zeta \eta \theta } \nabla^{\zeta }H_{\alpha \beta \delta } \nabla^{\theta }H_{\epsilon \varepsilon }{}^{\eta }\nn\\
&&- \frac{1}{512} H^{\alpha \beta \gamma } H^{\delta \epsilon \varepsilon } \nabla_{\gamma }H_{\varepsilon \eta \theta } \nabla_{\delta }H_{\alpha \beta }{}^{\zeta } \nabla^{\theta }H_{\epsilon \zeta }{}^{\eta }+\frac{1}{512} H^{\alpha \beta \gamma } H^{\delta \epsilon \varepsilon } \nabla_{\gamma }H_{\varepsilon \eta \theta } \nabla^{\zeta }H_{\alpha \beta \delta } \nabla^{\theta }H_{\epsilon \zeta }{}^{\eta }\nn\\
&&+\frac{9}{512} H_{\alpha \beta }{}^{\delta } H^{\alpha \beta \gamma } \nabla_{\delta }H_{\varepsilon \eta \theta } \nabla^{\zeta }H_{\gamma }{}^{\epsilon \varepsilon } \nabla^{\theta }H_{\epsilon \zeta }{}^{\eta }+\frac{3}{1024} H_{\alpha }{}^{\delta \epsilon } H^{\alpha \beta \gamma } \nabla_{\delta }H_{\beta \gamma }{}^{\varepsilon } \nabla_{\theta }H_{\varepsilon \zeta \eta } \nabla^{\theta }H_{\epsilon }{}^{\zeta \eta }\nn\\
&&+\frac{5}{1024} H_{\alpha }{}^{\delta \epsilon } H^{\alpha \beta \gamma } \nabla_{\delta }H_{\beta \gamma }{}^{\varepsilon } \nabla_{\epsilon }H_{\zeta \eta \theta } \nabla^{\theta }H_{\varepsilon }{}^{\zeta \eta }\,.
\eeqa
 The couplings corresponding to the structures \([H^2\nabla H^2 R]_{96}\), \([H^4\nabla H^2]_{93}\), \([H^2 R^2 \nabla H]_{31}\), and \([H^4 R\nabla H]_{52}\) are listed in Appendix \reef{AppB}.

To compare our results with those in the literature, we employ the \(t_8\) tensor, first introduced in \cite{Schwarz:1982jn}. Its contraction with four arbitrary antisymmetric tensors \(M^1, \dots, M^4\) is defined as
\beqa
t^{\alpha\beta\gamma\delta\mu\nu\rho\sigma}M^1_{\alpha\beta}M^2_{\gamma\delta}M^3_{\mu\nu}M^4_{\rho\sigma}&\!\!\!\!\!=\!\!\!\!\!&8(\tr M^1M^2M^3M^4+\tr M^1M^3M^2M^4+\tr M^1M^3M^4M^2)\labell{t8}\\
&&-2(\tr M^1M^2\tr M^3M^4+\tr M^1M^3\tr M^2M^4+\tr M^1M^4\tr M^2M^3).\nn
\eeqa
% Using this tensor, we express the contraction involving a double trace of the Riemann curvature as
%\beqa
%t^{\mu_1\cdots\mu_8}\Tr(R_{\mu_1\mu_2}R_{\mu_3\mu_4})\Tr(R_{\mu_5\mu_6}R_{\mu_7\mu_8})&=&8 R_{\alpha \beta}{}^{\zeta \eta} R^{\alpha \beta \gamma \delta} R_{\gamma \zeta}{}^{\theta \iota} R_{\delta \eta \theta \iota}   -4 R_{\alpha \beta}{}^{\zeta \eta} R^{\alpha \beta \gamma \delta} R_{\gamma \delta}{}^{\theta \iota} R_{\zeta \eta \theta \iota}\nn\\
%&&  +16 R_{\alpha \beta \gamma}{}^{\zeta} R^{\alpha \beta \gamma \delta} R_{\delta}{}^{\eta \theta \iota} R_{\zeta \eta \theta \iota}-2(R_{\mu\nu\alpha\beta}R^{\mu\nu\alpha\beta})^2 \nn
%\eeqa
By virtue of the Bianchi identities, the pure gravity couplings we have obtained in the Meissner scheme can be compactly written as
\[
[R^4]_5 = \frac{1}{1024} t^{\mu_1\cdots\mu_8} \Tr(R_{\mu_1\mu_2} R_{\mu_3\mu_4}) \Tr(R_{\mu_5\mu_6} R_{\mu_7\mu_8}).
\]
This pure gravity sector has previously been derived via S-matrix methods in \cite{Gross:1986iv,Gross:1986mw} and, under the restriction of vanishing \(U(1)\) gauge fields in the base space, via a restricted T-duality approach in \cite{Razaghian:2018svg}. Meanwhile, the pure Yang–Mills couplings at the four-derivative order in the Meissner scheme admit an analogous \(t_8\)-tensor representation \cite{Gross:1986mw,Garousi:2024avb}:
\[
[F^4]_4 = \frac{1}{256} t^{\mu_1\cdots\mu_8} \Tr(F_{\mu_1\mu_2} F_{\mu_3\mu_4}) \Tr(F_{\mu_5\mu_6} F_{\mu_7\mu_8}).
\]
Consequently, the combined gravity and Yang–Mills terms take the form $
\left( \Tr(F^2 - \alpha' R^2/2) \right)^2
$
in full agreement with the structure found in the MT scheme.  At the six-derivative order, the cross-terms proportional to \(R^2 F^2\) have been demonstrated in \cite{Garousi:2024imy} to follow consistently from the T-duality constraint applied to both the NS–NS and YM sectors within the Meissner scheme.
 \newpage
\section{Discussion}

    In this paper, we have employed the T-duality method, wherein the effective action at each order of $\alpha'$ is taken to be the minimal basis invariant under local Lorentz transformations and diffeomorphisms. The circular reduction of these bases is then required to be invariant under the Buscher rules  \cite{Buscher:1987sk,Rocek:1991ps} supplemented by their higher-derivative corrections. Using this method, we have obtained the eight-derivative effective couplings of heterotic string theory in the NS–NS sector. The T-duality constraint fixes all 872 couplings in the even-parity basis \cite{Garousi:2020mqn} up to a single parameter, and completely fixes the entire 477 couplings in the odd-parity basis \cite{Garousi:2024rzh}. The single unfixed parameter in the even-parity sector is a fortunate feature of the T-duality constraint, as T-duality is a symmetry of both heterotic and superstring theories, whose effective actions in the NS–NS sectors are not identical. Thus, T-duality leaves one parameter unfixed, serving to distinguish between the NS–NS sectors of the heterotic and superstring effective actions.    Our results for the couplings in the Metsaev–Tseytlin scheme are given in \reef{S3MT}, while those in the Meissner scheme appear in \reef{S3M}, where the unfixed parameter is fixed to $\z(3)$ by the S-matrix method. The part of the effective action fully fixed by T-duality is exclusive to the heterotic theory, whereas the remaining part is common to both heterotic and superstring theories.

 Our pure gravity couplings in the Meissner scheme are in complete agreement with the S-matrix results obtained long ago \cite{Gross:1986iv,Gross:1986mw}. The gravity couplings that are specific to the heterotic theory—in both schemes—can be written as double traces of the Riemann curvature. When these are combined with the corresponding pure Yang–Mills couplings, the result is precisely $(\Tr(F^2-\alpha'R^2/2))^2$. This particular structure for the heterotic-exclusive couplings was conjectured in \cite{Romans:1985xd}. Nevertheless, there are also gravity couplings that do not admit a double-trace representation and for which no Yang–Mills analogue exists. As noted in \cite{Garousi:2024imy}, the T-duality constraint on the classical effective action forces the Yang–Mills sector to appear exclusively in the form \((\Tr(F^2))^n\), with \(n\) determined by the \(\alpha'\)-order. By contrast, T-duality imposes no such restriction on the gravity sector. We therefore expect that the heterotic-exclusive part of the gravity couplings should be writable as $(\Tr(F^2-\alpha'R^2/2))^2$.

 The minimal basis of NS–NS couplings at order \(\alpha'^3\) in heterotic theory contains 1349 independent terms. Upon imposing the T-duality constraint, however, a significant number of these vanish. The remaining non-zero couplings do not constitute the minimal set possible in the heterotic theory. Once the T-duality-fixed couplings are obtained, one may employ field redefinitions to recast them into canonical form. In this canonical basis, we find 354 couplings in the Metsaev–Tseytlin scheme and 371 in the Meissner scheme. The canonical form retains several free parameters associated with field redefinitions, which may be exploited to reduce the total number of couplings further. The counts we have obtained are therefore not necessarily minimal. It would be worthwhile to investigate the true minimal set of non-zero couplings at order \(\alpha'^3\) and to derive their explicit expressions.

 As an additional consistency check in our results, one may perform the cosmological reduction of the couplings presented in \reef{S3MT} and \reef{S3M}  and investigate whether they can be expressed in an \(O(d,d)\)-invariant form. This type of analysis has previously been applied to superstring theory at order \(\alpha'^3\) \cite{Garousi:2021ikb} and to bosonic theory at order \(\alpha'^2\) in \cite{Garousi:2021ocs}. We expect the cosmological reduction of our couplings in both schemes to reproduce the canonical \(O(9,9)\)-invariant structure established by Hohm and Zwiebach \cite{Hohm:2015doa,Hohm:2019jgu,Codina:2021cxh}. Furthermore, the effective actions derived in this work offer a promising starting point for deriving higher-derivative corrections to black hole solutions in string theory, in particular the Kerr–Sen black hole, which serves as a string-theoretic generalization of the Kerr geometry \cite{Sen:1992ua,Hu:2025aji}.

We expect that the T-duality method employed in this paper can also be applied to determine the covariant couplings at higher orders in $\alpha'$. At each order, the couplings consist of families parametrized by a single coefficient that remains unfixed by T-duality; these coefficients are expected to be determined by S-matrix elements, yielding the values $\zeta(3), \zeta(5), \ldots$ \cite{Stieberger:2013wea,Schlotterer:2018zce,Ameri:2025bei}. Additionally, T-duality should generate a subset of couplings whose constants are completely fixed in terms of the overall normalization of the order-$\alpha'$ couplings. In this work, we have identified such couplings at orders $\alpha'^2$ and $\alpha'^3$ (see \reef{S3MT} or \reef{S3M}).  

A closed-form expression for all such couplings in a non-covariant form was previously proposed in \cite{Bergshoeff:1988nn,Bergshoeff:1989de}, based on the supersymmetrization of the Lorentz–Chern–Simons form. These couplings are given by \cite{Bergshoeff:1988nn,Bergshoeff:1989de}:
\beqa
\bS_{\rm{BdR}}= -\frac{2}{\kappa^2}\int d^{10}x \, e^{-2\Phi}\sqrt{-G}\, \Big[ R+4\nabla_{\mu}\Phi \nabla^{\mu}\Phi-\frac{1}{12} \hat H^2+\frac{\alpha' }{8}R_{\mu\nu ij}(\hat{\omega})R^{\mu\nu ij}(\hat{\omega})\Big]\,,
\eeqa
where the modified spin connection is defined as $\hat{\omega}_{\mu}{}^{ij}=\omega_{\mu}{}^{ij}-\frac{1}{2}\hat H_{\mu}{}^{ij}$, with the modified field strength given by $\hat{H}_{\mu}{}^{ij}=H_{\mu}{}^{ij}-\frac{3}{2}\alpha'\Omega_{\mu}{}^{ij}(\hat{\omega})$, and the corresponding curvature is
$
R_{\mu\nu}{}^{ij}(\hat{\omega})=\prt_\mu\hat{\omega}_\nu{}^{ij}-\prt_\nu\hat{\omega}_\mu{}^{ij}+\hat{\omega}_{\mu}{}^{ik}\hat{\omega}_{\nu k}{}^j-\hat{\omega}_{\nu}{}^{ik}\hat{\omega}_{\mu k}{}^j\,.
$
This curvature can be expressed in terms of the standard curvature $R_{\mu\nu}{}^{ij}(\omega)$, supplemented by $\hat H^2$-terms and a non-covariant but locally Lorentz-covariant derivative of $\hat H$. As shown in \cite{Chemissany:2007he}, the above Lagrangian at order $\alpha'$ is equivalent to the covariant Metsaev–Tseytlin Lagrangian \reef{fourmin}, up to total derivative terms and non-covariant field redefinitions. It would therefore be interesting to examine whether the covariant couplings at orders $\alpha'^2$ and $\alpha'^3$ obtained in our result \reef{S3MT} coincide with the corresponding terms in the above non-covariant action, again modulo total derivatives and non-covariant field redefinitions.

We have determined the NS-NS couplings at the eight-derivative order by imposing T-duality symmetry on the corresponding NS-NS basis, which contains 1349 independent couplings. A natural extension of this work would be to incorporate the Yang–Mills field into the coupling structure. Such an enlarged set of couplings could, in principle, also be derived using the T-duality method. However, the main challenge in constructing these couplings at the eight-derivative level is that the basis including both NS-NS and Yang–Mills fields becomes exceedingly large, making it difficult to handle with conventional computing resources. Nonetheless, it may be feasible to generate such a basis using high-performance computing and subsequently impose T-duality constraints to fix the coupling constants. Notably, the corresponding couplings at four- and six-derivative orders have already been obtained in \cite{Garousi:2024avb,Garousi:2024imy}.

{\bf Acknowledgments}: 

  This work is supported by Ferdowsi University of Mashhad under grant  371(1403/06/28).

\newpage
\vskip 0.5 cm

\appendix

\section{Independent corrections to the Buscher rules\label{AppAA}}
\vskip 0.5 cm

In this Appendix, we clarify how to find the independent corrections to the Buscher rules at two-derivative order. If one uses the ``xAct'' package \cite{Nutma:2013zea} to generate all two-derivative corrections to the base-space fields \(\psi = \{\varphi, \bar{\phi}, \bar{g}_{ab}, b_a, g_a, \bar{B}_{ab}\}\), it generates the following terms:
\beqa
   \bar{g}_{ab}^{(1)}&\!\!\!\!\!=\!\!\!\!\!& c_{1}^{} \bar{H}_{a}{}^{cd} \bar{H}_{bcd} + c_{2}^{} \bar{H}_{cde} \bar{H}^{cde} \bar{g}_{ab} + c_{3}^{} e^{\varphi} V_{a}{}^{c} V_{bc} + c_{4}^{} e^{\varphi} \bar{g}_{ab} V_{cd} V^{cd} \nn\\[0.2cm]
&&
 + c_{5}^{} (V_{b}{}^{c} W_{ac} + V_{a}{}^{c} W_{bc}) + c_{6}^{} \bar{g}_{ab} V^{cd} W_{cd} + c_{7}^{} e^{-\varphi} W_{a}{}^{c} W_{bc} \nn\\[0.2cm]
&&
+ c_{8}^{} e^{-\varphi} \bar{g}_{ab} W_{cd} W^{cd}+ c_{9}^{} \nabla_{a}\bar{\phi} \nabla_{b}\bar{\phi} + c_{10}^{} \bg_{ab} \nabla_{c}\bar{\phi} \nabla^{c}\bar{\phi} + c_{11}^{} (\nabla_{a}\varphi \nabla_{b}\bar{\phi} + \nabla_{a}\bar{\phi} \nabla_{b}\varphi)  \nn\\[0.2cm]
&&+ c_{12}^{} \bar{g}_{ab} \nabla_{c}\varphi \nabla^{c}\bar{\phi}+ c_{13}^{} \nabla_{a}\varphi \nabla_{b}\varphi  + c_{14}^{} \bar{g}_{ab} \nabla_{c}\varphi \nabla^{c}\varphi+ c_{15}^{} \nabla_{b}\nabla_{a}\bar{\phi} \nn\\[0.2cm]
&&
+ c_{17}^{} \nabla_{b}\nabla_{a}\varphi + c_{16}^{} \bar{g}_{ab} \nabla_{c}\nabla^{c}\bar{\phi} + c_{18}^{} \bar{g}_{ab} \nabla_{c}\nabla^{c}\varphi,  \nn\\[0.4cm]
 \bphi^{(1)}&\!\!\!\!\!=\!\!\!\!\!&b_{1}^{} \bar{H}_{abc} \bar{H}^{abc} + b_{2}^{} ~e^{\varphi} V_{ab} V^{ab} + b_{3}^{} V^{ab} W_{ab} +b_{4}^{} e^{-\varphi}W_{ab} W^{ab}\nn\\[0.2cm]
&&+ b_{5}^{} \nabla_{a}\bar{\phi} \nabla^{a}\bar{\phi} + b_{6}^{} \nabla_{a}\varphi \nabla^{a}\bar{\phi} + b_{7}^{} \nabla_{a}\varphi \nabla^{a}\varphi+ b_{8}^{} \nabla_{a}\nabla^{a}\bar{\phi} + b_{9}^{} \nabla_{a}\nabla^{a}\varphi , \nn\\[0.4cm]
 \vp^{(1)}&\!\!\!\!\!=\!\!\!\!\!& e_{1}^{} \bar{H}_{abc} \bar{H}^{abc} + e_{2}^{} ~e^{\varphi} V_{ab} V^{ab} + e_{3}^{} V^{ab} W_{ab} + e_{4}^{} e^{-\varphi}W_{ab} W^{ab}\nn\\[0.2cm]
&&+ e_{5}^{} \nabla_{a}\bar{\phi} \nabla^{a}\bar{\phi} + e_{6}^{} \nabla_{a}\varphi \nabla^{a}\bar{\phi} + e_{7}^{} \nabla_{a}\varphi \nabla^{a}\varphi+ e_{8}^{} \nabla_{a}\nabla^{a}\bar{\phi} + e_{9}^{} \nabla_{a}\nabla^{a}\varphi , \nn\\[0.4cm]
  \Delta g_{a}^{(1)}&\!\!\!\!\!=\!\!\!\!\!&d_{1}^{} e^{\varphi/2} \bar{H}_{abc} V^{bc} +d_{2}^{} e^{-\varphi/2}\bar{H}_{abc} W^{bc} + d_{3}^{} e^{\varphi/2} V_{ab} \nabla^{b}\bar{\phi} + d_{4}^{} e^{ \varphi/2} W_{ab} \nabla^{b}\bar{\phi}\nn\\[0.2cm]
&&+ d_{5}^{} e^{\varphi/2} \nabla_{b}V_{a}{}^{b} + d_{6}^{} e^{-\varphi/2} \nabla_{b}W_{a}{}^{b}+ d_{7}^{} e^{\varphi/2} V_{ab} \nabla^{b}\varphi + d_{8}^{} e^{-\varphi/2} W_{ab} \nabla^{b}\varphi, \nn\\[0.4cm]
   \Delta b_{a}^{(1)}&\!\!\!\!\!=\!\!\!\!\!&f_{1}^{} e^{\varphi/2} \bar{H}_{abc} V^{bc} +f_{2}^{} e^{-\varphi/2}\bar{H}_{abc} W^{bc} + f_{3}^{} e^{\varphi/2} V_{ab} \nabla^{b}\bar{\phi} + f_{4}^{} e^{ \varphi/2} W_{ab} \nabla^{b}\bar{\phi}\nn\\[0.2cm]
&&+ f_{5}^{} e^{\varphi/2} \nabla_{b}V_{a}{}^{b} + f_{6}^{} e^{-\varphi/2} \nabla_{b}W_{a}{}^{b}+ f_{7}^{} e^{\varphi/2} V_{ab} \nabla^{b}\varphi + f_{8}^{} e^{-\varphi/2} W_{ab} \nabla^{b}\varphi,\labell{dbH1n}\\[0.2cm]
   \bar{B}_{ab}^{(1)}&\!\!\!\!\!=\!\!\!\!\!&g_{1}^{} (V_{b}{}^{c} W_{ac} \!-\!  V_{a}{}^{c} W_{bc}) \!+\! g_{2}^{} \nabla_{c}\bar{H}_{ab}{}^{c} + g_{3}^{} \bar{H}_{abc} \nabla^{c}\bar{\phi}\! +\! g_{4}^{} \bar{H}_{abc} \nabla^{c}\varphi \!+\! g_{5}^{} (\nabla_{a}\varphi \nabla_{b}\bar{\phi} -  \nabla_{a}\bar{\phi} \nabla_{b}\varphi), \nn
\eeqa
where \(c_i, e_i, b_i, d_i, f_i\), and \(g_i\) are 57 arbitrary parameters.

To obtain the form of the above all-contraction structures, we assign each field a letter. Each letter \(L\) is a sum of letters \(L_1 + L_2 + \cdots\), where the lower indices indicate the number of derivatives. Up to two-derivative order, this assignment is as follows:
\beqa
&&\varphi: ~~~ A=A_1+A_2\,,\nn\\
&&\bphi: ~~~ B=B_1+B_2\,,\nn\\
&&\bar{g}_{ab}: ~~ C=C_1+C_2\,,\nn\\
&&g_a: ~~ D=D_1+D_2\,,\nn\\
&&b_a: ~~ E=E_1+E_2\,,\nn\\
&&\bar{B}_{ab}: ~~ F=F_1+F_2\,.\nn
\labell{AllC1}
\eeqa
Then, to find all possible combinations that make two-derivative structures such as \((A_1)^2\), \(B_1 \cdot F_1\), etc., one needs to expand the polynomial \((1 + A + \cdots + F)^2\), where the power two stands for two-derivative structures (higher orders require higher powers). This will include one- and two-derivative structures and has terms up to four-derivative order. Choosing only the two-derivative terms of this expansion results in 20 general structures. Now one can assign to each derivative letter \(L_i\) its own gauge-covariant field with \(i\) derivatives. For any field \(\psi\), this assignment is \(L_i \to (\nabla_{p_1} \cdots \nabla_{p_i}) \psi\). We then obtain all possible contractions of the indices \(p_1, \ldots, p_i\) with the indices of \(\psi\) to form scalar structures for \(\varphi^{(1)}\) and \(\bar{\phi}^{(1)}\), vector structures for \(\Delta g_a^{(1)}\) and \(\Delta b_a^{(1)}\), symmetric rank-two tensors for \(\bar{g}_{ab}^{(1)}\), and antisymmetric rank-two tensors for \(\bar{B}_{ab}^{(1)}\) corrections. These corrections contain both even- and odd-parity terms, and their explicit form is given in \reef{dbH1n}.

However, not all the arbitrary parameters in \reef{dbH1n} are independent. When they are used to transform the leading-order action \reef{S00} under the Buscher rules with the corrections \reef{dbH1n}, one finds that some of them may be related by total derivative terms and others may be related by the appropriate Bianchi identities. To find the independent parameters, we use the constraint that when they are inserted into the homogeneous part of the constraint \reef{Tn}, all parameters must vanish. In other words, at two-derivative order they should satisfy the following constraint:

\beqa
\displaystyle S^{(0,1)}_{(1)}(\psi_0) \sim 0\,.
\labell{homo}
\eeqa
where \(S^{(0,1)}_{(1)}(\psi_0)\) is the transformation of \(S^{(0,1)}_{(1)}(\psi)\) under the Buscher rules, and \(S^{(0,1)}_{(1)}(\psi)\) is the Taylor expansion of the reduction \reef{S00} at first order using the corrections \reef{dbH1n}. Note that the symbol \(\sim\) indicates equality up to total derivative terms and the use of Bianchi identities.

In order to solve the above equation, one must add all total derivative terms to it and impose all Bianchi identities as well. The construction of all total derivative terms at four-derivative order is similar to the construction of the corrections to the Buscher rules that we clarified before. It introduces 57 parameters from constructing a current \(J^a\) from all contractions at three-derivative order. The total derivative terms are then \(\partial_a (e^{-2\bar{\phi}} J^a)\). To impose the Bianchi identities \(dW = 0\), \(dV = 0\), we write the field strengths of \(W\) and \(V\) in terms of their corresponding gauge fields in the terms that have derivatives of \(W\) or \(V\). To impose the Bianchi identity \(d\bar{H} + \frac{3}{2} V \wedge W = 0\), we contract the left-hand side of this identity, or its derivatives, with base-space fields with arbitrary parameters to produce a scalar. This introduces 8 parameters.

Adding the total derivative terms and the Bianchi identity terms to the above equation, one finds that the sum of 106 independent terms, whose coefficients include all the above parameters, must vanish. This produces 106 algebraic equations for the parameters. The ones in which we are interested are 50 relations among the 57 parameters in \reef{dbH1n}. They are
\beqa
&&b_3 = 0, b_4 = 10 b_1 - b_2, b_7 = 7 b_1, b_8 = -128 b_1, b_9 = 
 10 b_1 - 2 b_2, c_1 = 0, c_{10} = 64 b_1, \nn\\
&&c_{11} = 0, c_{13} = 0,c_{14} = 4 b_1, c_{15} = -288 b_1 + 2 b_5, c_{16} = -64 b_1, c_{17} = 
 40 b_{1} - 8 b_{2} + 2 b_{6}, \nn\\
&& c_{18} = -(c_{12}/2), c_2 = (4 b_1)/3, c_3 = 0, c_4 = 4 b_1 + c_{12}/4, c_5 = 0, c_6 = 0, c_7 = 0,  \nn\\
&& c_8= 
 4 b_1- c_{12}/4, c_9 = 0, d_1 = 0, d_3 = 4 d_2, d_4 = 144 b_1 - b_5, d_5 = -2 d_2, d_6 = 0,  \nn\\
&& d_7 = -2 d_2, d_8 = -20 b_1 + 4 b_2 - 
  b_6, e_1 = (10 b_1)/3 - (2 b_2)/3 + c_{12}/8,  \nn\\
&& e_2 = 
 10 b_1 - 2 b_2 + (5 c_{12})/8, e_3 = 0, e_4 = 10 b_1 - 2 b_2 + (5 c_{12})/8,e_5 = 
 160 b_1 - 32 b_2 + 18 c_{12}, \nn\\
&&   e_6 = 
 144 b_1 - b_5, e_7 = -10 b_1 + 2 b_2 - b_6 + (7 c_{12})/8,  e_8 = -160 b_1 + 32 b_2 - 16 c_{12}, e_9 = 0,  \nn\\
&& f_1 = -d_2, f_2 = 0, f_3 =
  144 b_1 - b_5, f_4 = -4 d_2, f_5 = 0, f_6 = 2 d_2, f_7 = -20 b_1 + 4 b_2 - 
  b_6, \nn\\
&& f_8 = -2 d_2, g_1 = 0, g_2 = 0, g_3 = -144 b_1 + b_5, g_4 = 20 b_1 - 4 b_2 + b_6 \,.
\labell{setequ}
\eeqa
The other equations are relations between the parameters of the total derivative terms and the parameters of the Bianchi identities, expressed in terms of the unfixed parameters of \reef{dbH1n}, in which we are not interested. Hence, there are 7 redundant parameters in \reef{dbH1n}. To find the independent parameters, we should choose 7 parameters to be zero. However, we are not allowed to choose any arbitrary 7 of the 57 parameters to be zero. One must choose the 7 parameters such that, when they are set to zero, the above equations still produce 50 relations.

We choose the following parameters to be zero:
\beqa
b_1=0, b_6=0, e_1=0, e_6=0, e_8=0, f_6=0, g_5=0
\labell{arbitrary-choice}
\eeqa
This choice will not affect the number of equations in \reef{setequ}, which means that this choice is allowed. We impose this choice for the free parameters in \reef{dbH1n} to reduce the number of terms and find the following minimal basis for the corrections:
\beqa
   \bar{g}_{ab}^{(1)}&=& c_{1}^{} \bar{H}_{a}{}^{cd} \bar{H}_{bcd} + c_{2}^{} \bar{H}_{cde} \bar{H}^{cde} \bar{g}_{ab} + c_{3}^{} e^{\varphi} V_{a}{}^{c} V_{bc} + c_{4}^{} e^{\varphi} \bar{g}_{ab} V_{cd} V^{cd} \nn\\[0.2cm]
&&
 + c_{5}^{} (V_{b}{}^{c} W_{ac} + V_{a}{}^{c} W_{bc}) + c_{6}^{} \bar{g}_{ab} V^{cd} W_{cd} + c_{7}^{} e^{-\varphi} W_{a}{}^{c} W_{bc} \nn\\[0.2cm]
&&
+ c_{8}^{} e^{-\varphi} \bar{g}_{ab} W_{cd} W^{cd}+ c_{9}^{} \nabla_{a}\bar{\phi} \nabla_{b}\bar{\phi} + c_{10}^{} \bg_{ab} \nabla_{c}\bar{\phi} \nabla^{c}\bar{\phi} + c_{11}^{} (\nabla_{a}\varphi \nabla_{b}\bar{\phi} + \nabla_{a}\bar{\phi} \nabla_{b}\varphi)  \nn\\[0.2cm]
&&+ c_{12}^{} \bar{g}_{ab} \nabla_{c}\varphi \nabla^{c}\bar{\phi}+ c_{13}^{} \nabla_{a}\varphi \nabla_{b}\varphi  + c_{14}^{} \bar{g}_{ab} \nabla_{c}\varphi \nabla^{c}\varphi+ c_{15}^{} \nabla_{b}\nabla_{a}\bar{\phi} \nn\\[0.2cm]
&&
+ c_{17}^{} \nabla_{b}\nabla_{a}\varphi + c_{16}^{} \bar{g}_{ab} \nabla_{c}\nabla^{c}\bar{\phi} + c_{18}^{} \bar{g}_{ab} \nabla_{c}\nabla^{c}\varphi \,, \nn\\[0.4cm]
 \bphi^{(1)}&=& b_{2}^{} ~e^{\varphi} V_{ab} V^{ab} + b_{3}^{} V^{ab} W_{ab} +b_{4}^{} e^{-\varphi}W_{ab} W^{ab}\nn\\[0.2cm]
&&+ b_{5}^{} \nabla_{a}\bar{\phi} \nabla^{a}\bar{\phi} + b_{7}^{} \nabla_{a}\varphi \nabla^{a}\varphi+ b_{8}^{} \nabla_{a}\nabla^{a}\bar{\phi} + b_{9}^{} \nabla_{a}\nabla^{a}\varphi , \nn\\[0.4cm]
 \vp^{(1)}&=& e_{2}^{} ~e^{\varphi} V_{ab} V^{ab} + e_{3}^{} V^{ab} W_{ab} + e_{4}^{} e^{-\varphi}W_{ab} W^{ab}\nn\\[0.2cm]
&&+ e_{5}^{} \nabla_{a}\bar{\phi} \nabla^{a}\bar{\phi}+ e_{7}^{} \nabla_{a}\varphi \nabla^{a}\varphi+ e_{9}^{} \nabla_{a}\nabla^{a}\varphi , \nn\\[0.4cm]
  \Delta g_{a}^{(1)}&=&d_{1}^{} e^{\varphi/2} \bar{H}_{abc} V^{bc} +d_{2}^{} e^{-\varphi/2}\bar{H}_{abc} W^{bc} + d_{3}^{} e^{\varphi/2} V_{ab} \nabla^{b}\bar{\phi} + d_{4}^{} e^{ \varphi/2} W_{ab} \nabla^{b}\bar{\phi}\nn\\[0.2cm]
&&+ d_{5}^{} e^{\varphi/2} \nabla_{b}V_{a}{}^{b} + d_{6}^{} e^{-\varphi/2} \nabla_{b}W_{a}{}^{b}+ d_{7}^{} e^{\varphi/2} V_{ab} \nabla^{b}\varphi + d_{8}^{} e^{-\varphi/2} W_{ab} \nabla^{b}\varphi, \nn\\[0.4cm]
   \Delta b_{a}^{(1)}&=&f_{1}^{} e^{\varphi/2} \bar{H}_{abc} V^{bc} +f_{2}^{} e^{-\varphi/2}\bar{H}_{abc} W^{bc} + f_{3}^{} e^{\varphi/2} V_{ab} \nabla^{b}\bar{\phi} + f_{4}^{} e^{ \varphi/2} W_{ab} \nabla^{b}\bar{\phi}\nn\\[0.2cm]
&&+ f_{5}^{} e^{\varphi/2} \nabla_{b}V_{a}{}^{b} + f_{7}^{} e^{\varphi/2} V_{ab} \nabla^{b}\varphi + f_{8}^{} e^{-\varphi/2} W_{ab} \nabla^{b}\varphi,\labell{dbH1n2}\\[0.2cm]
   \bar{B}_{ab}^{(1)}&=&g_{1}^{} (V_{b}{}^{c} W_{ac} -  V_{a}{}^{c} W_{bc}) + g_{2}^{} \nabla_{c}\bar{H}_{ab}{}^{c} + g_{3}^{} \bar{H}_{abc} \nabla^{c}\bar{\phi} + g_{4}^{} \bar{H}_{abc} \nabla^{c}\varphi , \nn
\eeqa
Using the above independent corrections to the Buscher rules in solving equation \reef{T1} will fix all the parameters of the corrections to those in \reef{dbH1}.

One can proceed with this machinery to higher orders and find independent corrections to the Buscher rules, but we omit explicitly showing them here due to the large number of terms. On the other hand, working with the 57 most general parameters in the corrections to the Buscher rules and setting the 7 unfixed parameters to zero at the end of the calculation, or working with the 50 independent parameters in the corrections to the Buscher rules, does not reduce the complexity of the calculations, because in the latter case we must first find the 7 parameters appropriately, as we explained above, and then set them to zero. We perform the calculation in this paper using the most general corrections to the Buscher rules and set the remaining arbitrary parameters in the corrections to the Buscher rules to zero at the end, after solving the constraint \reef{Tn}. At order \(\alpha'^3\), there are 111,788 parameters in the corrections to the Buscher rules, 139,700 parameters in the total derivative terms, 53,201 parameters in imposing the Bianchi identities, as well as 1,349 parameters in the effective action in the minimal basis. There are 276,659 algebraic equations among these parameters. They fix all 1,349 parameters up to one parameter, which appears in \reef{S3}.

\section{Additional couplings in  Metsaev–Tseytlin scheme\label{AppA}}
\vskip 0.5 cm
 The complete couplings  in the MT scheme for the structures \([H^2\nabla H^2 R]_{85}\), \([H^4\nabla H^2]_{99}\), \([H^4 R\nabla H]_{41}\), and \([H^2\nabla H^3]_{31}\) are too voluminous for the main body of the paper; they are collected in full in this appendix. In the following, we explicitly list the \([H^2\nabla H^2 R]_{85}\) couplings:
\beqa
&&- \frac{27}{512} H^{\alpha \beta \gamma } H^{\delta \epsilon \varepsilon } R_{\varepsilon \lambda \theta \mu } \nabla_{\epsilon }H_{\gamma }{}^{\lambda \mu } \nabla^{\theta }H_{\alpha \beta \delta }- \frac{1}{32} H^{\alpha \beta \gamma } H^{\delta \epsilon \varepsilon } R_{\epsilon \lambda \varepsilon \mu } \nabla_{\theta }H_{\gamma }{}^{\lambda \mu } \nabla^{\theta }H_{\alpha \beta \delta }\nn\\
&&- \frac{25}{512} H_{\alpha }{}^{\delta \epsilon } H^{\alpha \beta \gamma } R_{\epsilon \lambda \theta \mu } \nabla_{\varepsilon }H_{\delta }{}^{\lambda \mu } \nabla^{\theta }H_{\beta \gamma }{}^{\varepsilon }- \frac{19}{256} H_{\alpha }{}^{\delta \epsilon } H^{\alpha \beta \gamma } R_{\epsilon \lambda \varepsilon \mu } \nabla_{\theta }H_{\delta }{}^{\lambda \mu } \nabla^{\theta }H_{\beta \gamma }{}^{\varepsilon }\nn\\
&&- \frac{7}{64} H_{\alpha }{}^{\delta \epsilon } H^{\alpha \beta \gamma } R_{\delta \lambda \epsilon \mu } \nabla_{\theta }H_{\varepsilon }{}^{\lambda \mu } \nabla^{\theta }H_{\beta \gamma }{}^{\varepsilon }+\frac{1}{128} H_{\alpha }{}^{\delta \epsilon } H^{\alpha \beta \gamma } R_{\epsilon \lambda \varepsilon \mu } \nabla_{\theta }H_{\gamma }{}^{\lambda \mu } \nabla^{\theta }H_{\beta \delta }{}^{\varepsilon }\nn\\
&&- \frac{1}{64} H_{\alpha \beta }{}^{\delta } H^{\alpha \beta \gamma } R_{\varepsilon \lambda \theta \mu } \nabla_{\delta }H_{\epsilon }{}^{\lambda \mu } \nabla^{\theta }H_{\gamma }{}^{\epsilon \varepsilon }- \frac{25}{512} H_{\alpha \beta }{}^{\delta } H^{\alpha \beta \gamma } R_{\delta \lambda \varepsilon \mu } \nabla_{\theta }H_{\epsilon }{}^{\lambda \mu } \nabla^{\theta }H_{\gamma }{}^{\epsilon \varepsilon }\nn\\
&&- \frac{19}{64} H^{\alpha \beta \gamma } H^{\delta \epsilon \varepsilon } R_{\beta \varepsilon \gamma \mu } \nabla_{\delta }H_{\alpha }{}^{\theta \lambda } \nabla_{\lambda }H_{\epsilon \theta }{}^{\mu }+\frac{23}{256} H^{\alpha \beta \gamma } H^{\delta \epsilon \varepsilon } R_{\epsilon \lambda \varepsilon \mu } \nabla_{\delta }H_{\gamma \theta }{}^{\mu } \nabla^{\lambda }H_{\alpha \beta }{}^{\theta }\nn\\
&&- \frac{25}{256} H^{\alpha \beta \gamma } H^{\delta \epsilon \varepsilon } R_{\epsilon \theta \varepsilon \mu } \nabla_{\delta }H_{\gamma \lambda }{}^{\mu } \nabla^{\lambda }H_{\alpha \beta }{}^{\theta }+\frac{53}{512} H^{\alpha \beta \gamma } H^{\delta \epsilon \varepsilon } R_{\gamma \mu \epsilon \varepsilon } \nabla_{\theta }H_{\delta \lambda }{}^{\mu } \nabla^{\lambda }H_{\alpha \beta }{}^{\theta }\nn\\
&&- \frac{37}{512} H^{\alpha \beta \gamma } H^{\delta \epsilon \varepsilon } R_{\epsilon \theta \varepsilon \mu } \nabla_{\lambda }H_{\gamma \delta }{}^{\mu } \nabla^{\lambda }H_{\alpha \beta }{}^{\theta }+\frac{31}{1024} H^{\alpha \beta \gamma } H^{\delta \epsilon \varepsilon } R_{\gamma \varepsilon \theta \mu } \nabla_{\lambda }H_{\delta \epsilon }{}^{\mu } \nabla^{\lambda }H_{\alpha \beta }{}^{\theta }\nn\\
&&- \frac{29}{1024} H^{\alpha \beta \gamma } H^{\delta \epsilon \varepsilon } R_{\gamma \mu \varepsilon \theta } \nabla_{\lambda }H_{\delta \epsilon }{}^{\mu } \nabla^{\lambda }H_{\alpha \beta }{}^{\theta }- \frac{137}{1024} H^{\alpha \beta \gamma } H^{\delta \epsilon \varepsilon } R_{\gamma \mu \epsilon \varepsilon } \nabla_{\lambda }H_{\delta \theta }{}^{\mu } \nabla^{\lambda }H_{\alpha \beta }{}^{\theta }\nn\\
&&- \frac{7}{64} H^{\alpha \beta \gamma } H^{\delta \epsilon \varepsilon } R_{\epsilon \theta \varepsilon \mu } \nabla_{\gamma }H_{\beta \lambda }{}^{\mu } \nabla^{\lambda }H_{\alpha \delta }{}^{\theta }- \frac{1}{128} H^{\alpha \beta \gamma } H^{\delta \epsilon \varepsilon } R_{\gamma \mu \varepsilon \lambda } \nabla_{\epsilon }H_{\beta \theta }{}^{\mu } \nabla^{\lambda }H_{\alpha \delta }{}^{\theta }\nn\\
&&+\frac{9}{32} H^{\alpha \beta \gamma } H^{\delta \epsilon \varepsilon } R_{\gamma \epsilon \varepsilon \mu } \nabla_{\theta }H_{\beta \lambda }{}^{\mu } \nabla^{\lambda }H_{\alpha \delta }{}^{\theta }+\frac{9}{64} H^{\alpha \beta \gamma } H^{\delta \epsilon \varepsilon } R_{\gamma \theta \varepsilon \mu } \nabla_{\lambda }H_{\beta \epsilon }{}^{\mu } \nabla^{\lambda }H_{\alpha \delta }{}^{\theta }\nn\\
&&+\frac{37}{256} H^{\alpha \beta \gamma } H^{\delta \epsilon \varepsilon } R_{\gamma \mu \epsilon \varepsilon } \nabla_{\lambda }H_{\beta \theta }{}^{\mu } \nabla^{\lambda }H_{\alpha \delta }{}^{\theta }+\frac{1}{128} H_{\alpha }{}^{\delta \epsilon } H^{\alpha \beta \gamma } R_{\delta \theta \epsilon \mu } \nabla_{\varepsilon }H_{\gamma \lambda }{}^{\mu } \nabla^{\lambda }H_{\beta }{}^{\varepsilon \theta }\nn\\
&&- \frac{75}{256} H_{\alpha }{}^{\delta \epsilon } H^{\alpha \beta \gamma } R_{\gamma \theta \epsilon \mu } \nabla_{\lambda }H_{\delta \varepsilon }{}^{\mu } \nabla^{\lambda }H_{\beta }{}^{\varepsilon \theta }+\frac{39}{256} H_{\alpha }{}^{\delta \epsilon } H^{\alpha \beta \gamma } R_{\gamma \mu \epsilon \theta } \nabla_{\lambda }H_{\delta \varepsilon }{}^{\mu } \nabla^{\lambda }H_{\beta }{}^{\varepsilon \theta }\nn\\
&&- \frac{33}{512} H_{\alpha }{}^{\delta \epsilon } H^{\alpha \beta \gamma } R_{\gamma \mu \delta \epsilon } \nabla_{\lambda }H_{\varepsilon \theta }{}^{\mu } \nabla^{\lambda }H_{\beta }{}^{\varepsilon \theta }+\frac{13}{1024} H_{\alpha \beta }{}^{\delta } H^{\alpha \beta \gamma } R_{\gamma \theta \delta \mu } \nabla_{\lambda }H_{\epsilon \varepsilon }{}^{\mu } \nabla^{\lambda }H^{\epsilon \varepsilon \theta }\nn\\
&&+\frac{13}{128} H^{\alpha \beta \gamma } H^{\delta \epsilon \varepsilon } R_{\beta \epsilon \gamma \varepsilon } \nabla_{\lambda }H_{\delta \theta \mu } \nabla^{\mu }H_{\alpha }{}^{\theta \lambda }- \frac{5}{256} H^{\alpha \beta \gamma } H^{\delta \epsilon \varepsilon } R_{\beta \epsilon \gamma \varepsilon } \nabla_{\mu }H_{\delta \theta \lambda } \nabla^{\mu }H_{\alpha }{}^{\theta \lambda }\nn\\
&&- \frac{9}{64} H^{\alpha \beta \gamma } H^{\delta \epsilon \varepsilon } R_{\gamma \theta \varepsilon \mu } \nabla^{\lambda }H_{\alpha \delta }{}^{\theta } \nabla^{\mu }H_{\beta \epsilon \lambda }- \frac{23}{128} H^{\alpha \beta \gamma } H^{\delta \epsilon \varepsilon } R_{\gamma \mu \epsilon \varepsilon } \nabla^{\lambda }H_{\alpha \delta }{}^{\theta } \nabla^{\mu }H_{\beta \theta \lambda }\nn\\
&&+\frac{3}{32} H^{\alpha \beta \gamma } H^{\delta \epsilon \varepsilon } R_{\epsilon \lambda \varepsilon \mu } \nabla^{\lambda }H_{\alpha \beta }{}^{\theta } \nabla^{\mu }H_{\gamma \delta \theta }- \frac{19}{512} H^{\alpha \beta \gamma } H^{\delta \epsilon \varepsilon } R_{\epsilon \theta \varepsilon \mu } \nabla^{\lambda }H_{\alpha \beta }{}^{\theta } \nabla^{\mu }H_{\gamma \delta \lambda }\nn\\
&&- \frac{79}{512} H^{\alpha \beta \gamma } H^{\delta \epsilon \varepsilon } R_{\varepsilon \theta \lambda \mu } \nabla_{\delta }H_{\alpha \beta }{}^{\theta } \nabla^{\mu }H_{\gamma \epsilon }{}^{\lambda }+\frac{15}{128} H^{\alpha \beta \gamma } H^{\delta \epsilon \varepsilon } R_{\varepsilon \lambda \theta \mu } \nabla_{\delta }H_{\alpha \beta }{}^{\theta } \nabla^{\mu }H_{\gamma \epsilon }{}^{\lambda }\nn\\
&&+\frac{85}{512} H^{\alpha \beta \gamma } H^{\delta \epsilon \varepsilon } R_{\varepsilon \lambda \theta \mu } \nabla^{\theta }H_{\alpha \beta \delta } \nabla^{\mu }H_{\gamma \epsilon }{}^{\lambda }- \frac{5}{32} H^{\alpha \beta \gamma } H^{\delta \epsilon \varepsilon } R_{\varepsilon \mu \theta \lambda } \nabla^{\theta }H_{\alpha \beta \delta } \nabla^{\mu }H_{\gamma \epsilon }{}^{\lambda }\nn\\
&&- \frac{5}{256} H_{\alpha }{}^{\delta \epsilon } H^{\alpha \beta \gamma } R_{\varepsilon \theta \lambda \mu } \nabla^{\theta }H_{\beta \delta }{}^{\varepsilon } \nabla^{\mu }H_{\gamma \epsilon }{}^{\lambda }- \frac{23}{512} H_{\alpha }{}^{\delta \epsilon } H^{\alpha \beta \gamma } R_{\delta \epsilon \lambda \mu } \nabla^{\lambda }H_{\beta }{}^{\varepsilon \theta } \nabla^{\mu }H_{\gamma \varepsilon \theta }\nn\\
&&+\frac{5}{32} H_{\alpha }{}^{\delta \epsilon } H^{\alpha \beta \gamma } R_{\delta \theta \epsilon \mu } \nabla^{\lambda }H_{\beta }{}^{\varepsilon \theta } \nabla^{\mu }H_{\gamma \varepsilon \lambda }+\frac{5}{128} H_{\alpha }{}^{\delta \epsilon } H^{\alpha \beta \gamma } R_{\epsilon \theta \lambda \mu } \nabla^{\theta }H_{\beta \delta }{}^{\varepsilon } \nabla^{\mu }H_{\gamma \varepsilon }{}^{\lambda }\nn\\
&&+\frac{15}{256} H^{\alpha \beta \gamma } H^{\delta \epsilon \varepsilon } R_{\epsilon \lambda \varepsilon \mu } \nabla_{\delta }H_{\alpha \beta }{}^{\theta } \nabla^{\mu }H_{\gamma \theta }{}^{\lambda }+\frac{5}{512} H^{\alpha \beta \gamma } H^{\delta \epsilon \varepsilon } R_{\epsilon \lambda \varepsilon \mu } \nabla^{\theta }H_{\alpha \beta \delta } \nabla^{\mu }H_{\gamma \theta }{}^{\lambda }\nn\\
&&- \frac{1}{256} H^{\alpha \beta \gamma } H^{\delta \epsilon \varepsilon } R_{\gamma \lambda \varepsilon \mu } \nabla^{\lambda }H_{\alpha \beta }{}^{\theta } \nabla^{\mu }H_{\delta \epsilon \theta }+\frac{25}{1024} H^{\alpha \beta \gamma } H^{\delta \epsilon \varepsilon } R_{\gamma \mu \varepsilon \lambda } \nabla^{\lambda }H_{\alpha \beta }{}^{\theta } \nabla^{\mu }H_{\delta \epsilon \theta }\nn\\
&&- \frac{31}{1024} H^{\alpha \beta \gamma } H^{\delta \epsilon \varepsilon } R_{\gamma \theta \varepsilon \mu } \nabla^{\lambda }H_{\alpha \beta }{}^{\theta } \nabla^{\mu }H_{\delta \epsilon \lambda }+\frac{31}{1024} H^{\alpha \beta \gamma } H^{\delta \epsilon \varepsilon } R_{\gamma \mu \varepsilon \theta } \nabla^{\lambda }H_{\alpha \beta }{}^{\theta } \nabla^{\mu }H_{\delta \epsilon \lambda }\nn\\
&&- \frac{11}{1024} H^{\alpha \beta \gamma } H^{\delta \epsilon \varepsilon } R_{\varepsilon \lambda \theta \mu } \nabla^{\theta }H_{\alpha \beta \gamma } \nabla^{\mu }H_{\delta \epsilon }{}^{\lambda }+\frac{29}{3072} H^{\alpha \beta \gamma } H^{\delta \epsilon \varepsilon } R_{\varepsilon \mu \theta \lambda } \nabla^{\theta }H_{\alpha \beta \gamma } \nabla^{\mu }H_{\delta \epsilon }{}^{\lambda }\nn\\
&&+\frac{7}{1024} H_{\alpha }{}^{\delta \epsilon } H^{\alpha \beta \gamma } R_{\varepsilon \theta \lambda \mu } \nabla^{\theta }H_{\beta \gamma }{}^{\varepsilon } \nabla^{\mu }H_{\delta \epsilon }{}^{\lambda }+\frac{1}{128} H_{\alpha \beta }{}^{\delta } H^{\alpha \beta \gamma } R_{\varepsilon \mu \theta \lambda } \nabla^{\theta }H_{\gamma }{}^{\epsilon \varepsilon } \nabla^{\mu }H_{\delta \epsilon }{}^{\lambda }\nn\\
&&- \frac{19}{256} H_{\alpha }{}^{\delta \epsilon } H^{\alpha \beta \gamma } R_{\gamma \mu \epsilon \lambda } \nabla^{\lambda }H_{\beta }{}^{\varepsilon \theta } \nabla^{\mu }H_{\delta \varepsilon \theta }+\frac{37}{128} H_{\alpha }{}^{\delta \epsilon } H^{\alpha \beta \gamma } R_{\gamma \epsilon \theta \mu } \nabla^{\lambda }H_{\beta }{}^{\varepsilon \theta } \nabla^{\mu }H_{\delta \varepsilon \lambda }\nn\\
&&+\frac{37}{128} H_{\alpha }{}^{\delta \epsilon } H^{\alpha \beta \gamma } R_{\gamma \mu \epsilon \theta } \nabla^{\lambda }H_{\beta }{}^{\varepsilon \theta } \nabla^{\mu }H_{\delta \varepsilon \lambda }+\frac{15}{512} H_{\alpha }{}^{\delta \epsilon } H^{\alpha \beta \gamma } R_{\epsilon \theta \lambda \mu } \nabla^{\theta }H_{\beta \gamma }{}^{\varepsilon } \nabla^{\mu }H_{\delta \varepsilon }{}^{\lambda }\nn\\
&&- \frac{3}{512} H_{\alpha }{}^{\delta \epsilon } H^{\alpha \beta \gamma } R_{\epsilon \lambda \theta \mu } \nabla^{\theta }H_{\beta \gamma }{}^{\varepsilon } \nabla^{\mu }H_{\delta \varepsilon }{}^{\lambda }+\frac{11}{1024} H^{\alpha \beta \gamma } H^{\delta \epsilon \varepsilon } R_{\gamma \mu \epsilon \varepsilon } \nabla^{\lambda }H_{\alpha \beta }{}^{\theta } \nabla^{\mu }H_{\delta \theta \lambda }\nn\\
&&+\frac{5}{1536} H^{\alpha \beta \gamma } H^{\delta \epsilon \varepsilon } R_{\epsilon \lambda \varepsilon \mu } \nabla^{\theta }H_{\alpha \beta \gamma } \nabla^{\mu }H_{\delta \theta }{}^{\lambda }- \frac{9}{256} H_{\alpha }{}^{\delta \epsilon } H^{\alpha \beta \gamma } R_{\epsilon \lambda \varepsilon \mu } \nabla^{\theta }H_{\beta \gamma }{}^{\varepsilon } \nabla^{\mu }H_{\delta \theta }{}^{\lambda }\nn\\
&&- \frac{1}{256} H_{\alpha }{}^{\delta \epsilon } H^{\alpha \beta \gamma } R_{\epsilon \mu \varepsilon \lambda } \nabla^{\theta }H_{\beta \gamma }{}^{\varepsilon } \nabla^{\mu }H_{\delta \theta }{}^{\lambda }- \frac{1}{128} H_{\alpha \beta }{}^{\delta } H^{\alpha \beta \gamma } R_{\epsilon \lambda \varepsilon \mu } \nabla^{\theta }H_{\gamma }{}^{\epsilon \varepsilon } \nabla^{\mu }H_{\delta \theta }{}^{\lambda }\nn\\
&&- \frac{25}{1024} H_{\alpha \beta }{}^{\delta } H^{\alpha \beta \gamma } R_{\gamma \theta \delta \mu } \nabla^{\lambda }H^{\epsilon \varepsilon \theta } \nabla^{\mu }H_{\epsilon \varepsilon \lambda }+\frac{1}{256} H_{\alpha \beta }{}^{\delta } H^{\alpha \beta \gamma } R_{\delta \mu \theta \lambda } \nabla_{\gamma }H^{\epsilon \varepsilon \theta } \nabla^{\mu }H_{\epsilon \varepsilon }{}^{\lambda }\nn\\
&&+\frac{25}{1024} H^{\alpha \beta \gamma } H^{\delta \epsilon \varepsilon } R_{\gamma \lambda \theta \mu } \nabla_{\delta }H_{\alpha \beta }{}^{\theta } \nabla^{\mu }H_{\epsilon \varepsilon }{}^{\lambda }+\frac{7}{256} H^{\alpha \beta \gamma } H^{\delta \epsilon \varepsilon } R_{\gamma \mu \theta \lambda } \nabla_{\delta }H_{\alpha \beta }{}^{\theta } \nabla^{\mu }H_{\epsilon \varepsilon }{}^{\lambda }\nn\\
&&+\frac{9}{64} H_{\alpha }{}^{\delta \epsilon } H^{\alpha \beta \gamma } R_{\gamma \theta \lambda \mu } \nabla_{\delta }H_{\beta }{}^{\varepsilon \theta } \nabla^{\mu }H_{\epsilon \varepsilon }{}^{\lambda }+\frac{13}{1024} H^{\alpha \beta \gamma } H^{\delta \epsilon \varepsilon } R_{\gamma \lambda \theta \mu } \nabla^{\theta }H_{\alpha \beta \delta } \nabla^{\mu }H_{\epsilon \varepsilon }{}^{\lambda }\nn\\
&&- \frac{39}{512} H^{\alpha \beta \gamma } H^{\delta \epsilon \varepsilon } R_{\gamma \mu \theta \lambda } \nabla^{\theta }H_{\alpha \beta \delta } \nabla^{\mu }H_{\epsilon \varepsilon }{}^{\lambda }+\frac{53}{1024} H_{\alpha \beta }{}^{\delta } H^{\alpha \beta \gamma } R_{\delta \theta \lambda \mu } \nabla^{\theta }H_{\gamma }{}^{\epsilon \varepsilon } \nabla^{\mu }H_{\epsilon \varepsilon }{}^{\lambda }\nn\\
&&+\frac{21}{64} H^{\alpha \beta \gamma } H^{\delta \epsilon \varepsilon } R_{\beta \varepsilon \gamma \mu } \nabla_{\delta }H_{\alpha }{}^{\theta \lambda } \nabla^{\mu }H_{\epsilon \theta \lambda }- \frac{15}{128} H^{\alpha \beta \gamma } H^{\delta \epsilon \varepsilon } R_{\gamma \mu \varepsilon \lambda } \nabla_{\beta }H_{\alpha \delta }{}^{\theta } \nabla^{\mu }H_{\epsilon \theta }{}^{\lambda }\nn\\
&&+\frac{5}{512} H^{\alpha \beta \gamma } H^{\delta \epsilon \varepsilon } R_{\gamma \varepsilon \lambda \mu } \nabla_{\delta }H_{\alpha \beta }{}^{\theta } \nabla^{\mu }H_{\epsilon \theta }{}^{\lambda }+\frac{3}{256} H^{\alpha \beta \gamma } H^{\delta \epsilon \varepsilon } R_{\gamma \mu \varepsilon \lambda } \nabla_{\delta }H_{\alpha \beta }{}^{\theta } \nabla^{\mu }H_{\epsilon \theta }{}^{\lambda }\nn\\
&&- \frac{11}{128} H^{\alpha \beta \gamma } H^{\delta \epsilon \varepsilon } R_{\gamma \lambda \varepsilon \mu } \nabla^{\theta }H_{\alpha \beta \delta } \nabla^{\mu }H_{\epsilon \theta }{}^{\lambda }- \frac{13}{512} H_{\alpha \beta }{}^{\delta } H^{\alpha \beta \gamma } R_{\delta \lambda \varepsilon \mu } \nabla^{\theta }H_{\gamma }{}^{\epsilon \varepsilon } \nabla^{\mu }H_{\epsilon \theta }{}^{\lambda }\nn\\
&&- \frac{1}{256} H_{\alpha \beta }{}^{\delta } H^{\alpha \beta \gamma } R_{\delta \mu \varepsilon \lambda } \nabla^{\theta }H_{\gamma }{}^{\epsilon \varepsilon } \nabla^{\mu }H_{\epsilon \theta }{}^{\lambda }- \frac{3}{1024} H^{\alpha \beta \gamma } H^{\delta \epsilon \varepsilon } R_{\varepsilon \mu \theta \lambda } \nabla_{\delta }H_{\alpha \beta \gamma } \nabla^{\mu }H_{\epsilon }{}^{\theta \lambda }\nn\\
&&- \frac{9}{512} H_{\alpha }{}^{\delta \epsilon } H^{\alpha \beta \gamma } R_{\varepsilon \mu \theta \lambda } \nabla_{\delta }H_{\beta \gamma }{}^{\varepsilon } \nabla^{\mu }H_{\epsilon }{}^{\theta \lambda }- \frac{5}{128} H_{\alpha \beta }{}^{\delta } H^{\alpha \beta \gamma } R_{\varepsilon \theta \lambda \mu } \nabla_{\delta }H_{\gamma }{}^{\epsilon \varepsilon } \nabla^{\mu }H_{\epsilon }{}^{\theta \lambda }\nn\\
&&- \frac{1}{512} H_{\alpha }{}^{\delta \epsilon } H^{\alpha \beta \gamma } R_{\varepsilon \mu \theta \lambda } \nabla^{\varepsilon }H_{\beta \gamma \delta } \nabla^{\mu }H_{\epsilon }{}^{\theta \lambda }+\frac{25}{512} H_{\alpha }{}^{\delta \epsilon } H^{\alpha \beta \gamma } R_{\gamma \mu \delta \epsilon } \nabla^{\lambda }H_{\beta }{}^{\varepsilon \theta } \nabla^{\mu }H_{\varepsilon \theta \lambda }\nn\\
&&+\frac{19}{256} H_{\alpha }{}^{\delta \epsilon } H^{\alpha \beta \gamma } R_{\gamma \mu \epsilon \lambda } \nabla_{\delta }H_{\beta }{}^{\varepsilon \theta } \nabla^{\mu }H_{\varepsilon \theta }{}^{\lambda }+\frac{39}{512} H_{\alpha }{}^{\delta \epsilon } H^{\alpha \beta \gamma } R_{\delta \lambda \epsilon \mu } \nabla^{\theta }H_{\beta \gamma }{}^{\varepsilon } \nabla^{\mu }H_{\varepsilon \theta }{}^{\lambda }\nn\\
&&- \frac{29}{1024} H_{\alpha }{}^{\delta \epsilon } H^{\alpha \beta \gamma } R_{\epsilon \mu \theta \lambda } \nabla_{\delta }H_{\beta \gamma }{}^{\varepsilon } \nabla^{\mu }H_{\varepsilon }{}^{\theta \lambda }- \frac{115}{1024} H^{\alpha \beta \gamma } H^{\delta \epsilon \varepsilon } R_{\gamma \mu \theta \lambda } \nabla_{\epsilon }H_{\alpha \beta \delta } \nabla^{\mu }H_{\varepsilon }{}^{\theta \lambda }\nn\\
&&- \frac{1}{256} H_{\alpha }{}^{\delta \epsilon } H^{\alpha \beta \gamma } R_{\beta \delta \gamma \epsilon } \nabla_{\mu }H_{\varepsilon \theta \lambda } \nabla^{\mu }H^{\varepsilon \theta \lambda }
\eeqa
The couplings with structure $[H^4\nabla H^2]_{99}$ are the following:
\beqa
&&- \frac{5}{512} H_{\alpha }{}^{\delta \epsilon } H^{\alpha \beta \gamma } H_{\beta }{}^{\varepsilon \theta } H^{\lambda \mu \nu } \nabla_{\delta }H_{\gamma \lambda \mu } \nabla_{\theta }H_{\epsilon \varepsilon \nu }+\frac{3}{256} H_{\alpha \beta }{}^{\delta } H^{\alpha \beta \gamma } H_{\gamma }{}^{\epsilon \varepsilon } H^{\theta \lambda \mu } \nabla_{\delta }H_{\lambda \mu \nu } \nabla_{\theta }H_{\epsilon \varepsilon }{}^{\nu }\nn\\
&&- \frac{43}{2048} H_{\alpha }{}^{\delta \epsilon } H^{\alpha \beta \gamma } H_{\beta }{}^{\varepsilon \theta } H^{\lambda \mu \nu } \nabla_{\varepsilon }H_{\gamma \delta \lambda } \nabla_{\theta }H_{\epsilon \mu \nu }+\frac{61}{4096} H_{\alpha \beta }{}^{\delta } H^{\alpha \beta \gamma } H_{\gamma }{}^{\epsilon \varepsilon } H_{\epsilon }{}^{\theta \lambda } \nabla_{\delta }H_{\lambda \mu \nu } \nabla_{\theta }H_{\varepsilon }{}^{\mu \nu }\nn\\
&&+\frac{9}{1024} H_{\alpha }{}^{\delta \epsilon } H^{\alpha \beta \gamma } H_{\beta }{}^{\varepsilon \theta } H^{\lambda \mu \nu } \nabla_{\theta }H_{\epsilon \mu \nu } \nabla_{\lambda }H_{\gamma \delta \varepsilon }+\frac{123}{16384} H_{\alpha \beta }{}^{\delta } H^{\alpha \beta \gamma } H^{\epsilon \varepsilon \theta } H^{\lambda \mu \nu } \nabla_{\delta }H_{\theta \mu \nu } \nabla_{\lambda }H_{\gamma \epsilon \varepsilon }\nn\\
&&- \frac{13}{1024} H_{\alpha \beta }{}^{\delta } H^{\alpha \beta \gamma } H_{\epsilon }{}^{\lambda \mu } H^{\epsilon \varepsilon \theta } \nabla_{\theta }H_{\delta \mu \nu } \nabla_{\lambda }H_{\gamma \varepsilon }{}^{\nu }- \frac{9}{512} H_{\alpha }{}^{\delta \epsilon } H^{\alpha \beta \gamma } H_{\beta }{}^{\varepsilon \theta } H_{\delta }{}^{\lambda \mu } \nabla_{\theta }H_{\epsilon \mu \nu } \nabla_{\lambda }H_{\gamma \varepsilon }{}^{\nu }\nn\\
&&+\frac{155}{49152} H_{\alpha \beta }{}^{\delta } H^{\alpha \beta \gamma } H_{\epsilon \varepsilon }{}^{\lambda } H^{\epsilon \varepsilon \theta } \nabla_{\theta }H_{\gamma }{}^{\mu \nu } \nabla_{\lambda }H_{\delta \mu \nu }- \frac{55}{4096} H_{\alpha }{}^{\delta \epsilon } H^{\alpha \beta \gamma } H_{\varepsilon }{}^{\mu \nu } H^{\varepsilon \theta \lambda } \nabla_{\theta }H_{\beta \gamma \delta } \nabla_{\lambda }H_{\epsilon \mu \nu }\nn\\
&&+\frac{13}{1024} H_{\alpha }{}^{\delta \epsilon } H^{\alpha \beta \gamma } H_{\beta \delta }{}^{\varepsilon } H_{\gamma }{}^{\theta \lambda } \nabla_{\theta }H_{\epsilon }{}^{\mu \nu } \nabla_{\lambda }H_{\varepsilon \mu \nu }+\frac{91}{4096} H_{\alpha \beta }{}^{\delta } H^{\alpha \beta \gamma } H_{\gamma }{}^{\epsilon \varepsilon } H_{\delta }{}^{\theta \lambda } \nabla_{\theta }H_{\epsilon }{}^{\mu \nu } \nabla_{\lambda }H_{\varepsilon \mu \nu }\nn\\
&&- \frac{59}{49152} H_{\alpha \beta }{}^{\delta } H^{\alpha \beta \gamma } H_{\epsilon \varepsilon }{}^{\lambda } H^{\epsilon \varepsilon \theta } \nabla_{\delta }H_{\gamma }{}^{\mu \nu } \nabla_{\lambda }H_{\theta \mu \nu }+\frac{509}{24576} H_{\alpha }{}^{\delta \epsilon } H^{\alpha \beta \gamma } H_{\varepsilon }{}^{\mu \nu } H^{\varepsilon \theta \lambda } \nabla_{\lambda }H_{\delta \epsilon \nu } \nabla_{\mu }H_{\beta \gamma \theta }\nn\\
&&- \frac{9}{1024} H_{\alpha }{}^{\delta \epsilon } H^{\alpha \beta \gamma } H_{\varepsilon }{}^{\mu \nu } H^{\varepsilon \theta \lambda } \nabla_{\lambda }H_{\gamma \epsilon \nu } \nabla_{\mu }H_{\beta \delta \theta }+\frac{581}{3072} H_{\alpha }{}^{\delta \epsilon } H^{\alpha \beta \gamma } H_{\beta }{}^{\varepsilon \theta } H^{\lambda \mu \nu } \nabla_{\theta }H_{\epsilon \varepsilon \nu } \nabla_{\mu }H_{\gamma \delta \lambda }\nn\\
&&- \frac{581}{6144} H_{\alpha }{}^{\delta \epsilon } H^{\alpha \beta \gamma } H_{\beta }{}^{\varepsilon \theta } H_{\delta }{}^{\lambda \mu } \nabla_{\theta }H_{\epsilon \varepsilon \nu } \nabla_{\mu }H_{\gamma \lambda }{}^{\nu }- \frac{21}{2048} H_{\alpha \beta }{}^{\delta } H^{\alpha \beta \gamma } H_{\epsilon }{}^{\lambda \mu } H^{\epsilon \varepsilon \theta } \nabla_{\lambda }H_{\gamma \varepsilon }{}^{\nu } \nabla_{\mu }H_{\delta \theta \nu }\nn\\
&&+\frac{37}{768} H_{\alpha \beta }{}^{\delta } H^{\alpha \beta \gamma } H_{\gamma }{}^{\epsilon \varepsilon } H^{\theta \lambda \mu } \nabla_{\lambda }H_{\delta \theta }{}^{\nu } \nabla_{\mu }H_{\epsilon \varepsilon \nu }- \frac{13}{6144} H_{\alpha }{}^{\delta \epsilon } H^{\alpha \beta \gamma } H_{\beta }{}^{\varepsilon \theta } H_{\delta }{}^{\lambda \mu } \nabla_{\theta }H_{\gamma \varepsilon }{}^{\nu } \nabla_{\mu }H_{\epsilon \lambda \nu }\nn\\
&&+\frac{5}{2048} H_{\alpha \beta }{}^{\delta } H^{\alpha \beta \gamma } H_{\gamma }{}^{\epsilon \varepsilon } H^{\theta \lambda \mu } \nabla_{\epsilon }H_{\delta \theta }{}^{\nu } \nabla_{\mu }H_{\varepsilon \lambda \nu }- \frac{5}{1024} H_{\alpha }{}^{\delta \epsilon } H^{\alpha \beta \gamma } H_{\beta \delta }{}^{\varepsilon } H^{\theta \lambda \mu } \nabla_{\theta }H_{\gamma \epsilon }{}^{\nu } \nabla_{\mu }H_{\varepsilon \lambda \nu }\nn\\
&&- \frac{31}{2048} H_{\alpha \beta }{}^{\delta } H^{\alpha \beta \gamma } H_{\gamma }{}^{\epsilon \varepsilon } H^{\theta \lambda \mu } \nabla_{\theta }H_{\delta \epsilon }{}^{\nu } \nabla_{\mu }H_{\varepsilon \lambda \nu }+\frac{27}{1024} H_{\alpha \beta }{}^{\delta } H^{\alpha \beta \gamma } H_{\epsilon }{}^{\lambda \mu } H^{\epsilon \varepsilon \theta } \nabla_{\delta }H_{\gamma \varepsilon }{}^{\nu } \nabla_{\mu }H_{\theta \lambda \nu }\nn\\
&&+\frac{1}{128} H_{\alpha }{}^{\delta \epsilon } H^{\alpha \beta \gamma } H_{\beta }{}^{\varepsilon \theta } H_{\delta }{}^{\lambda \mu } \nabla_{\epsilon }H_{\gamma \varepsilon }{}^{\nu } \nabla_{\mu }H_{\theta \lambda \nu }- \frac{51}{1024} H_{\alpha }{}^{\delta \epsilon } H^{\alpha \beta \gamma } H_{\beta }{}^{\varepsilon \theta } H_{\delta }{}^{\lambda \mu } \nabla_{\varepsilon }H_{\gamma \epsilon }{}^{\nu } \nabla_{\mu }H_{\theta \lambda \nu }\nn\\
&&- \frac{113}{3072} H_{\alpha }{}^{\delta \epsilon } H^{\alpha \beta \gamma } H_{\beta }{}^{\varepsilon \theta } H_{\gamma }{}^{\lambda \mu } \nabla_{\varepsilon }H_{\delta \epsilon }{}^{\nu } \nabla_{\mu }H_{\theta \lambda \nu }+\frac{9}{1024} H_{\alpha }{}^{\delta \epsilon } H^{\alpha \beta \gamma } H_{\varepsilon }{}^{\mu \nu } H^{\varepsilon \theta \lambda } \nabla_{\mu }H_{\beta \delta \theta } \nabla_{\nu }H_{\gamma \epsilon \lambda }\nn\\
&&- \frac{1}{1024} H_{\alpha }{}^{\delta \epsilon } H^{\alpha \beta \gamma } H_{\beta }{}^{\varepsilon \theta } H^{\lambda \mu \nu } \nabla_{\theta }H_{\delta \epsilon \varepsilon } \nabla_{\nu }H_{\gamma \lambda \mu }- \frac{379}{12288} H_{\alpha }{}^{\delta \epsilon } H^{\alpha \beta \gamma } H_{\varepsilon }{}^{\mu \nu } H^{\varepsilon \theta \lambda } \nabla_{\mu }H_{\beta \gamma \theta } \nabla_{\nu }H_{\delta \epsilon \lambda }\nn\\
&&- \frac{133}{24576} H_{\alpha }{}^{\delta \epsilon } H^{\alpha \beta \gamma } H_{\varepsilon }{}^{\mu \nu } H^{\varepsilon \theta \lambda } \nabla_{\lambda }H_{\beta \gamma \theta } \nabla_{\nu }H_{\delta \epsilon \mu }- \frac{1}{512} H_{\alpha \beta }{}^{\delta } H^{\alpha \beta \gamma } H^{\epsilon \varepsilon \theta } H^{\lambda \mu \nu } \nabla_{\varepsilon }H_{\gamma \epsilon \lambda } \nabla_{\nu }H_{\delta \theta \mu }\nn\\
&&- \frac{613}{24576} H_{\alpha \beta }{}^{\delta } H^{\alpha \beta \gamma } H^{\epsilon \varepsilon \theta } H^{\lambda \mu \nu } \nabla_{\lambda }H_{\gamma \epsilon \varepsilon } \nabla_{\nu }H_{\delta \theta \mu }- \frac{59}{49152} H_{\alpha \beta }{}^{\delta } H^{\alpha \beta \gamma } H^{\epsilon \varepsilon \theta } H^{\lambda \mu \nu } \nabla_{\theta }H_{\gamma \epsilon \varepsilon } \nabla_{\nu }H_{\delta \lambda \mu }\nn\\
&&- \frac{611}{6144} H_{\alpha }{}^{\delta \epsilon } H^{\alpha \beta \gamma } H_{\beta }{}^{\varepsilon \theta } H^{\lambda \mu \nu } \nabla_{\mu }H_{\gamma \delta \lambda } \nabla_{\nu }H_{\epsilon \varepsilon \theta }+\frac{15}{512} H_{\alpha }{}^{\delta \epsilon } H^{\alpha \beta \gamma } H_{\varepsilon }{}^{\mu \nu } H^{\varepsilon \theta \lambda } \nabla_{\delta }H_{\beta \gamma \theta } \nabla_{\nu }H_{\epsilon \lambda \mu }\nn\\
&&- \frac{1}{1024} H_{\alpha }{}^{\delta \epsilon } H^{\alpha \beta \gamma } H_{\varepsilon }{}^{\mu \nu } H^{\varepsilon \theta \lambda } \nabla_{\theta }H_{\beta \gamma \delta } \nabla_{\nu }H_{\epsilon \lambda \mu }- \frac{11}{512} H_{\alpha \beta }{}^{\delta } H^{\alpha \beta \gamma } H^{\epsilon \varepsilon \theta } H^{\lambda \mu \nu } \nabla_{\delta }H_{\gamma \epsilon \lambda } \nabla_{\nu }H_{\varepsilon \theta \mu }\nn\\
&&+\frac{157}{6144} H_{\alpha }{}^{\delta \epsilon } H^{\alpha \beta \gamma } H_{\beta }{}^{\varepsilon \theta } H^{\lambda \mu \nu } \nabla_{\epsilon }H_{\gamma \delta \lambda } \nabla_{\nu }H_{\varepsilon \theta \mu }- \frac{125}{6144} H_{\alpha }{}^{\delta \epsilon } H^{\alpha \beta \gamma } H_{\beta }{}^{\varepsilon \theta } H^{\lambda \mu \nu } \nabla_{\lambda }H_{\gamma \delta \epsilon } \nabla_{\nu }H_{\varepsilon \theta \mu }\nn\\
&&+\frac{35}{6144} H_{\alpha \beta }{}^{\delta } H^{\alpha \beta \gamma } H_{\gamma }{}^{\epsilon \varepsilon } H^{\theta \lambda \mu } \nabla_{\epsilon }H_{\delta \theta }{}^{\nu } \nabla_{\nu }H_{\varepsilon \lambda \mu }- \frac{41}{4096} H_{\alpha }{}^{\delta \epsilon } H^{\alpha \beta \gamma } H_{\beta \delta }{}^{\varepsilon } H^{\theta \lambda \mu } \nabla_{\theta }H_{\gamma \epsilon }{}^{\nu } \nabla_{\nu }H_{\varepsilon \lambda \mu }\nn\\
&&+\frac{17}{1024} H_{\alpha \beta }{}^{\delta } H^{\alpha \beta \gamma } H_{\gamma }{}^{\epsilon \varepsilon } H^{\theta \lambda \mu } \nabla_{\theta }H_{\delta \epsilon }{}^{\nu } \nabla_{\nu }H_{\varepsilon \lambda \mu }- \frac{59}{24576} H_{\alpha \beta }{}^{\delta } H^{\alpha \beta \gamma } H^{\epsilon \varepsilon \theta } H^{\lambda \mu \nu } \nabla_{\delta }H_{\gamma \epsilon \varepsilon } \nabla_{\nu }H_{\theta \lambda \mu }\nn\\
&&- \frac{7}{512} H_{\alpha \beta }{}^{\delta } H^{\alpha \beta \gamma } H_{\epsilon }{}^{\lambda \mu } H^{\epsilon \varepsilon \theta } \nabla_{\delta }H_{\gamma \varepsilon }{}^{\nu } \nabla_{\nu }H_{\theta \lambda \mu }- \frac{15}{1024} H_{\alpha }{}^{\delta \epsilon } H^{\alpha \beta \gamma } H_{\beta }{}^{\varepsilon \theta } H^{\lambda \mu \nu } \nabla_{\epsilon }H_{\gamma \delta \varepsilon } \nabla_{\nu }H_{\theta \lambda \mu }\nn\\
&&- \frac{31}{2048} H_{\alpha }{}^{\delta \epsilon } H^{\alpha \beta \gamma } H_{\beta }{}^{\varepsilon \theta } H_{\delta }{}^{\lambda \mu } \nabla_{\epsilon }H_{\gamma \varepsilon }{}^{\nu } \nabla_{\nu }H_{\theta \lambda \mu }+\frac{7}{1024} H_{\alpha }{}^{\delta \epsilon } H^{\alpha \beta \gamma } H_{\beta }{}^{\varepsilon \theta } H^{\lambda \mu \nu } \nabla_{\varepsilon }H_{\gamma \delta \epsilon } \nabla_{\nu }H_{\theta \lambda \mu }\nn\\
&&+\frac{41}{1024} H_{\alpha }{}^{\delta \epsilon } H^{\alpha \beta \gamma } H_{\beta }{}^{\varepsilon \theta } H_{\delta }{}^{\lambda \mu } \nabla_{\varepsilon }H_{\gamma \epsilon }{}^{\nu } \nabla_{\nu }H_{\theta \lambda \mu }- \frac{13}{6144} H_{\alpha \beta }{}^{\delta } H^{\alpha \beta \gamma } H_{\gamma }{}^{\epsilon \varepsilon } H^{\theta \lambda \mu } \nabla_{\varepsilon }H_{\delta \epsilon }{}^{\nu } \nabla_{\nu }H_{\theta \lambda \mu }\nn\\
&&+\frac{599}{12288} H_{\alpha \beta }{}^{\delta } H^{\alpha \beta \gamma } H_{\gamma }{}^{\epsilon \varepsilon } H_{\epsilon }{}^{\theta \lambda } \nabla_{\varepsilon }H_{\delta }{}^{\mu \nu } \nabla_{\nu }H_{\theta \lambda \mu }+\frac{25}{512} H_{\alpha }{}^{\delta \epsilon } H^{\alpha \beta \gamma } H_{\beta }{}^{\varepsilon \theta } H_{\delta }{}^{\lambda \mu } \nabla_{\mu }H_{\theta \lambda \nu } \nabla^{\nu }H_{\gamma \epsilon \varepsilon }\nn\\
&&- \frac{29}{1024} H_{\alpha }{}^{\delta \epsilon } H^{\alpha \beta \gamma } H_{\beta }{}^{\varepsilon \theta } H_{\delta }{}^{\lambda \mu } \nabla_{\nu }H_{\theta \lambda \mu } \nabla^{\nu }H_{\gamma \epsilon \varepsilon }- \frac{3}{1024} H_{\alpha }{}^{\delta \epsilon } H^{\alpha \beta \gamma } H_{\beta \delta }{}^{\varepsilon } H^{\theta \lambda \mu } \nabla_{\mu }H_{\varepsilon \lambda \nu } \nabla^{\nu }H_{\gamma \epsilon \theta }\nn\\
&&+\frac{25}{4096} H_{\alpha }{}^{\delta \epsilon } H^{\alpha \beta \gamma } H_{\beta \delta }{}^{\varepsilon } H^{\theta \lambda \mu } \nabla_{\nu }H_{\varepsilon \lambda \mu } \nabla^{\nu }H_{\gamma \epsilon \theta }+\frac{1}{1024} H_{\alpha }{}^{\delta \epsilon } H^{\alpha \beta \gamma } H_{\beta }{}^{\varepsilon \theta } H_{\delta \varepsilon }{}^{\lambda } \nabla_{\mu }H_{\theta \lambda \nu } \nabla^{\nu }H_{\gamma \epsilon }{}^{\mu }\nn\\
&&+\frac{1}{1024} H_{\alpha }{}^{\delta \epsilon } H^{\alpha \beta \gamma } H_{\beta }{}^{\varepsilon \theta } H_{\delta \varepsilon }{}^{\lambda } \nabla_{\nu }H_{\theta \lambda \mu } \nabla^{\nu }H_{\gamma \epsilon }{}^{\mu }- \frac{281}{12288} H_{\alpha \beta }{}^{\delta } H^{\alpha \beta \gamma } H_{\epsilon }{}^{\lambda \mu } H^{\epsilon \varepsilon \theta } \nabla_{\mu }H_{\delta \lambda \nu } \nabla^{\nu }H_{\gamma \varepsilon \theta }\nn\\
&&+\frac{19}{6144} H_{\alpha }{}^{\delta \epsilon } H^{\alpha \beta \gamma } H_{\beta }{}^{\varepsilon \theta } H_{\delta }{}^{\lambda \mu } \nabla_{\mu }H_{\epsilon \lambda \nu } \nabla^{\nu }H_{\gamma \varepsilon \theta }+\frac{107}{49152} H_{\alpha \beta }{}^{\delta } H^{\alpha \beta \gamma } H_{\epsilon }{}^{\lambda \mu } H^{\epsilon \varepsilon \theta } \nabla_{\nu }H_{\delta \lambda \mu } \nabla^{\nu }H_{\gamma \varepsilon \theta }\nn\\
&&- \frac{37}{24576} H_{\alpha }{}^{\delta \epsilon } H^{\alpha \beta \gamma } H_{\beta }{}^{\varepsilon \theta } H_{\delta }{}^{\lambda \mu } \nabla_{\nu }H_{\epsilon \lambda \mu } \nabla^{\nu }H_{\gamma \varepsilon \theta }+\frac{5}{256} H_{\alpha }{}^{\delta \epsilon } H^{\alpha \beta \gamma } H_{\beta }{}^{\varepsilon \theta } H_{\delta }{}^{\lambda \mu } \nabla_{\epsilon }H_{\theta \mu \nu } \nabla^{\nu }H_{\gamma \varepsilon \lambda }\nn\\
&&+\frac{37}{2048} H_{\alpha \beta }{}^{\delta } H^{\alpha \beta \gamma } H_{\epsilon }{}^{\lambda \mu } H^{\epsilon \varepsilon \theta } \nabla_{\mu }H_{\delta \theta \nu } \nabla^{\nu }H_{\gamma \varepsilon \lambda }+\frac{5}{256} H_{\alpha }{}^{\delta \epsilon } H^{\alpha \beta \gamma } H_{\beta }{}^{\varepsilon \theta } H_{\delta }{}^{\lambda \mu } \nabla_{\mu }H_{\epsilon \theta \nu } \nabla^{\nu }H_{\gamma \varepsilon \lambda }\nn\\
&&- \frac{9}{1024} H_{\alpha \beta }{}^{\delta } H^{\alpha \beta \gamma } H_{\epsilon }{}^{\lambda \mu } H^{\epsilon \varepsilon \theta } \nabla_{\nu }H_{\delta \theta \mu } \nabla^{\nu }H_{\gamma \varepsilon \lambda }- \frac{11}{512} H_{\alpha }{}^{\delta \epsilon } H^{\alpha \beta \gamma } H_{\beta }{}^{\varepsilon \theta } H_{\delta }{}^{\lambda \mu } \nabla_{\nu }H_{\epsilon \theta \mu } \nabla^{\nu }H_{\gamma \varepsilon \lambda }\nn\\
&&+\frac{215}{16384} H_{\alpha \beta }{}^{\delta } H^{\alpha \beta \gamma } H_{\epsilon \varepsilon }{}^{\lambda } H^{\epsilon \varepsilon \theta } \nabla_{\mu }H_{\delta \lambda \nu } \nabla^{\nu }H_{\gamma \theta }{}^{\mu }- \frac{613}{49152} H_{\alpha \beta }{}^{\delta } H^{\alpha \beta \gamma } H_{\epsilon \varepsilon }{}^{\lambda } H^{\epsilon \varepsilon \theta } \nabla_{\nu }H_{\delta \lambda \mu } \nabla^{\nu }H_{\gamma \theta }{}^{\mu }\nn\\
&&+\frac{65}{512} H_{\alpha }{}^{\delta \epsilon } H^{\alpha \beta \gamma } H_{\beta }{}^{\varepsilon \theta } H_{\delta }{}^{\lambda \mu } \nabla_{\theta }H_{\epsilon \varepsilon \nu } \nabla^{\nu }H_{\gamma \lambda \mu }- \frac{1009}{24576} H_{\alpha }{}^{\delta \epsilon } H^{\alpha \beta \gamma } H_{\beta }{}^{\varepsilon \theta } H_{\delta }{}^{\lambda \mu } \nabla_{\nu }H_{\epsilon \varepsilon \theta } \nabla^{\nu }H_{\gamma \lambda \mu }\nn\\
&&+\frac{1}{2048} H_{\alpha }{}^{\delta \epsilon } H^{\alpha \beta \gamma } H_{\beta }{}^{\varepsilon \theta } H_{\gamma }{}^{\lambda \mu } \nabla_{\theta }H_{\lambda \mu \nu } \nabla^{\nu }H_{\delta \epsilon \varepsilon }- \frac{7}{2048} H_{\alpha }{}^{\delta \epsilon } H^{\alpha \beta \gamma } H_{\beta }{}^{\varepsilon \theta } H_{\gamma }{}^{\lambda \mu } \nabla_{\mu }H_{\theta \lambda \nu } \nabla^{\nu }H_{\delta \epsilon \varepsilon }\nn\\
&&- \frac{1}{1024} H_{\alpha \beta }{}^{\delta } H^{\alpha \beta \gamma } H_{\gamma }{}^{\epsilon \varepsilon } H^{\theta \lambda \mu } \nabla_{\nu }H_{\theta \lambda \mu } \nabla^{\nu }H_{\delta \epsilon \varepsilon }+\frac{475}{12288} H_{\alpha \beta }{}^{\delta } H^{\alpha \beta \gamma } H_{\gamma }{}^{\epsilon \varepsilon } H^{\theta \lambda \mu } \nabla_{\varepsilon }H_{\lambda \mu \nu } \nabla^{\nu }H_{\delta \epsilon \theta }\nn\\
&&- \frac{91}{12288} H_{\alpha \beta }{}^{\delta } H^{\alpha \beta \gamma } H_{\gamma }{}^{\epsilon \varepsilon } H^{\theta \lambda \mu } \nabla_{\nu }H_{\varepsilon \lambda \mu } \nabla^{\nu }H_{\delta \epsilon \theta }- \frac{403}{12288} H_{\alpha \beta }{}^{\delta } H^{\alpha \beta \gamma } H_{\gamma }{}^{\epsilon \varepsilon } H_{\epsilon }{}^{\theta \lambda } \nabla_{\mu }H_{\theta \lambda \nu } \nabla^{\nu }H_{\delta \varepsilon }{}^{\mu }\nn\\
&&+\frac{115}{12288} H_{\alpha \beta }{}^{\delta } H^{\alpha \beta \gamma } H_{\gamma }{}^{\epsilon \varepsilon } H_{\epsilon }{}^{\theta \lambda } \nabla_{\nu }H_{\theta \lambda \mu } \nabla^{\nu }H_{\delta \varepsilon }{}^{\mu }- \frac{45}{2048} H_{\alpha \beta }{}^{\delta } H^{\alpha \beta \gamma } H_{\gamma }{}^{\epsilon \varepsilon } H^{\theta \lambda \mu } \nabla_{\nu }H_{\epsilon \varepsilon \mu } \nabla^{\nu }H_{\delta \theta \lambda }\nn\\
&&+\frac{11}{512} H_{\alpha \beta }{}^{\delta } H^{\alpha \beta \gamma } H_{\gamma }{}^{\epsilon \varepsilon } H_{\epsilon }{}^{\theta \lambda } \nabla_{\varepsilon }H_{\lambda \mu \nu } \nabla^{\nu }H_{\delta \theta }{}^{\mu }- \frac{7}{512} H_{\alpha \beta }{}^{\delta } H^{\alpha \beta \gamma } H_{\gamma }{}^{\epsilon \varepsilon } H_{\epsilon }{}^{\theta \lambda } \nabla_{\mu }H_{\varepsilon \lambda \nu } \nabla^{\nu }H_{\delta \theta }{}^{\mu }\nn\\
&&+\frac{9}{512} H_{\alpha \beta }{}^{\delta } H^{\alpha \beta \gamma } H_{\gamma }{}^{\epsilon \varepsilon } H_{\epsilon }{}^{\theta \lambda } \nabla_{\nu }H_{\varepsilon \lambda \mu } \nabla^{\nu }H_{\delta \theta }{}^{\mu }- \frac{469}{24576} H_{\alpha \beta }{}^{\delta } H^{\alpha \beta \gamma } H_{\gamma }{}^{\epsilon \varepsilon } H_{\epsilon \varepsilon }{}^{\theta } \nabla_{\mu }H_{\theta \lambda \nu } \nabla^{\nu }H_{\delta }{}^{\lambda \mu }\nn\\
&&- \frac{155}{49152} H_{\alpha \beta }{}^{\delta } H^{\alpha \beta \gamma } H_{\gamma }{}^{\epsilon \varepsilon } H_{\epsilon \varepsilon }{}^{\theta } \nabla_{\nu }H_{\theta \lambda \mu } \nabla^{\nu }H_{\delta }{}^{\lambda \mu }+\frac{27}{2048} H_{\alpha \beta }{}^{\delta } H^{\alpha \beta \gamma } H_{\gamma }{}^{\epsilon \varepsilon } H^{\theta \lambda \mu } \nabla_{\delta }H_{\lambda \mu \nu } \nabla^{\nu }H_{\epsilon \varepsilon \theta }\nn\\
&&- \frac{19}{4096} H_{\alpha }{}^{\delta \epsilon } H^{\alpha \beta \gamma } H_{\beta \delta }{}^{\varepsilon } H_{\gamma }{}^{\theta \lambda } \nabla_{\mu }H_{\theta \lambda \nu } \nabla^{\nu }H_{\epsilon \varepsilon }{}^{\mu }- \frac{647}{49152} H_{\alpha \beta }{}^{\delta } H^{\alpha \beta \gamma } H_{\gamma }{}^{\epsilon \varepsilon } H_{\delta }{}^{\theta \lambda } \nabla_{\mu }H_{\theta \lambda \nu } \nabla^{\nu }H_{\epsilon \varepsilon }{}^{\mu }\nn\\
&&+\frac{19}{4096} H_{\alpha }{}^{\delta \epsilon } H^{\alpha \beta \gamma } H_{\beta \delta }{}^{\varepsilon } H_{\gamma }{}^{\theta \lambda } \nabla_{\nu }H_{\theta \lambda \mu } \nabla^{\nu }H_{\epsilon \varepsilon }{}^{\mu }+\frac{647}{49152} H_{\alpha \beta }{}^{\delta } H^{\alpha \beta \gamma } H_{\gamma }{}^{\epsilon \varepsilon } H_{\delta }{}^{\theta \lambda } \nabla_{\nu }H_{\theta \lambda \mu } \nabla^{\nu }H_{\epsilon \varepsilon }{}^{\mu }\nn\\
&&+\frac{5}{256} H_{\alpha }{}^{\delta \epsilon } H^{\alpha \beta \gamma } H_{\beta \delta }{}^{\varepsilon } H_{\gamma }{}^{\theta \lambda } \nabla_{\varepsilon }H_{\lambda \mu \nu } \nabla^{\nu }H_{\epsilon \theta }{}^{\mu }+\frac{9}{512} H_{\alpha }{}^{\delta \epsilon } H^{\alpha \beta \gamma } H_{\beta \delta }{}^{\varepsilon } H_{\gamma }{}^{\theta \lambda } \nabla_{\mu }H_{\varepsilon \lambda \nu } \nabla^{\nu }H_{\epsilon \theta }{}^{\mu }\nn\\&&+\frac{103}{4096} H_{\alpha \beta }{}^{\delta } H^{\alpha \beta \gamma } H_{\gamma }{}^{\epsilon \varepsilon } H_{\delta }{}^{\theta \lambda } \nabla_{\mu }H_{\varepsilon \lambda \nu } \nabla^{\nu }H_{\epsilon \theta }{}^{\mu }- \frac{7}{512} H_{\alpha }{}^{\delta \epsilon } H^{\alpha \beta \gamma } H_{\beta \delta }{}^{\varepsilon } H_{\gamma }{}^{\theta \lambda } \nabla_{\nu }H_{\varepsilon \lambda \mu } \nabla^{\nu }H_{\epsilon \theta }{}^{\mu }\nn\\
&&- \frac{107}{4096} H_{\alpha \beta }{}^{\delta } H^{\alpha \beta \gamma } H_{\gamma }{}^{\epsilon \varepsilon } H_{\delta }{}^{\theta \lambda } \nabla_{\nu }H_{\varepsilon \lambda \mu } \nabla^{\nu }H_{\epsilon \theta }{}^{\mu }+\frac{31}{2048} H_{\alpha \beta }{}^{\delta } H^{\alpha \beta \gamma } H_{\gamma }{}^{\epsilon \varepsilon } H_{\epsilon }{}^{\theta \lambda } \nabla_{\delta }H_{\lambda \mu \nu } \nabla^{\nu }H_{\varepsilon \theta }{}^{\mu }\nn\\
&&+\frac{5}{1024} H_{\alpha }{}^{\delta \epsilon } H^{\alpha \beta \gamma } H_{\beta \delta }{}^{\varepsilon } H_{\gamma \epsilon }{}^{\theta } \nabla_{\theta }H_{\lambda \mu \nu } \nabla^{\nu }H_{\varepsilon }{}^{\lambda \mu }- \frac{1}{256} H_{\alpha }{}^{\delta \epsilon } H^{\alpha \beta \gamma } H_{\beta \delta }{}^{\varepsilon } H_{\gamma \epsilon }{}^{\theta } \nabla_{\mu }H_{\theta \lambda \nu } \nabla^{\nu }H_{\varepsilon }{}^{\lambda \mu }\nn\\
&&- \frac{361}{6144} H_{\alpha \beta }{}^{\delta } H^{\alpha \beta \gamma } H_{\gamma }{}^{\epsilon \varepsilon } H_{\delta \epsilon }{}^{\theta } \nabla_{\mu }H_{\theta \lambda \nu } \nabla^{\nu }H_{\varepsilon }{}^{\lambda \mu }+\frac{283}{12288} H_{\alpha \beta }{}^{\delta } H^{\alpha \beta \gamma } H_{\gamma }{}^{\epsilon \varepsilon } H_{\delta \epsilon }{}^{\theta } \nabla_{\nu }H_{\theta \lambda \mu } \nabla^{\nu }H_{\varepsilon }{}^{\lambda \mu }\nn\\
&&+\frac{1}{2048} H_{\alpha \beta }{}^{\delta } H^{\alpha \beta \gamma } H_{\gamma }{}^{\epsilon \varepsilon } H_{\delta \epsilon \varepsilon } \nabla_{\nu }H_{\theta \lambda \mu } \nabla^{\nu }H^{\theta \lambda \mu }
\eeqa
The couplings with structure $[H^4 R\nabla H]_{41}$ are the following:
\beqa
&&\frac{31}{512} H_{\alpha }{}^{\delta \epsilon } H^{\alpha \beta \gamma } H_{\beta }{}^{\varepsilon \lambda } H_{\delta }{}^{\mu \nu } R_{\epsilon \mu \nu \rho } \nabla_{\gamma }H_{\varepsilon \lambda }{}^{\rho }+\frac{29}{512} H_{\alpha \beta }{}^{\delta } H^{\alpha \beta \gamma } H_{\epsilon }{}^{\mu \nu } H^{\epsilon \varepsilon \lambda } R_{\lambda \mu \nu \rho } \nabla_{\delta }H_{\gamma \varepsilon }{}^{\rho }\nn\\
&&- \frac{3}{1024} H_{\alpha \beta }{}^{\delta } H^{\alpha \beta \gamma } H_{\gamma }{}^{\epsilon \varepsilon } H_{\epsilon }{}^{\lambda \mu } R_{\lambda \mu \nu \rho } \nabla_{\delta }H_{\varepsilon }{}^{\nu \rho }- \frac{59}{512} H_{\alpha \beta }{}^{\delta } H^{\alpha \beta \gamma } H_{\gamma }{}^{\epsilon \varepsilon } H_{\epsilon }{}^{\lambda \mu } R_{\varepsilon \mu \nu \rho } \nabla_{\delta }H_{\lambda }{}^{\nu \rho }\nn\\
&&- \frac{1}{128} H_{\alpha }{}^{\delta \epsilon } H^{\alpha \beta \gamma } H_{\beta }{}^{\varepsilon \lambda } H_{\delta }{}^{\mu \nu } R_{\lambda \rho \mu \nu } \nabla_{\epsilon }H_{\gamma \varepsilon }{}^{\rho }- \frac{29}{512} H_{\alpha \beta }{}^{\delta } H^{\alpha \beta \gamma } H_{\gamma }{}^{\epsilon \varepsilon } H^{\lambda \mu \nu } R_{\varepsilon \rho \mu \nu } \nabla_{\epsilon }H_{\delta \lambda }{}^{\rho }\nn\\
&&- \frac{1}{256} H_{\alpha }{}^{\delta \epsilon } H^{\alpha \beta \gamma } H_{\beta }{}^{\varepsilon \lambda } H_{\delta }{}^{\mu \nu } R_{\gamma \rho \mu \nu } \nabla_{\epsilon }H_{\varepsilon \lambda }{}^{\rho }+\frac{13}{512} H_{\alpha \beta }{}^{\delta } H^{\alpha \beta \gamma } H_{\gamma }{}^{\epsilon \varepsilon } H^{\lambda \mu \nu } R_{\delta \varepsilon \nu \rho } \nabla_{\epsilon }H_{\lambda \mu }{}^{\rho }\nn\\
&&+\frac{13}{512} H_{\alpha \beta }{}^{\delta } H^{\alpha \beta \gamma } H_{\gamma }{}^{\epsilon \varepsilon } H^{\lambda \mu \nu } R_{\delta \rho \varepsilon \nu } \nabla_{\epsilon }H_{\lambda \mu }{}^{\rho }+\frac{33}{512} H_{\alpha \beta }{}^{\delta } H^{\alpha \beta \gamma } H^{\epsilon \varepsilon \lambda } H^{\mu \nu \rho } R_{\delta \nu \lambda \rho } \nabla_{\varepsilon }H_{\gamma \epsilon \mu }\nn\\
&&+\frac{13}{256} H_{\alpha }{}^{\delta \epsilon } H^{\alpha \beta \gamma } H_{\beta }{}^{\varepsilon \lambda } H^{\mu \nu \rho } R_{\gamma \lambda \nu \rho } \nabla_{\varepsilon }H_{\delta \epsilon \mu }+\frac{1}{64} H_{\alpha }{}^{\delta \epsilon } H^{\alpha \beta \gamma } H_{\beta }{}^{\varepsilon \lambda } H_{\gamma }{}^{\mu \nu } R_{\lambda \mu \nu \rho } \nabla_{\varepsilon }H_{\delta \epsilon }{}^{\rho }\nn\\
&&- \frac{55}{1024} H_{\alpha \beta }{}^{\delta } H^{\alpha \beta \gamma } H_{\gamma }{}^{\epsilon \varepsilon } H_{\epsilon }{}^{\lambda \mu } R_{\lambda \mu \nu \rho } \nabla_{\varepsilon }H_{\delta }{}^{\nu \rho }+\frac{33}{1024} H_{\alpha \beta }{}^{\delta } H^{\alpha \beta \gamma } H_{\gamma }{}^{\epsilon \varepsilon } H^{\lambda \mu \nu } R_{\varepsilon \rho \mu \nu } \nabla_{\lambda }H_{\delta \epsilon }{}^{\rho }\nn\\
&&+\frac{87}{1024} H_{\alpha \beta }{}^{\delta } H^{\alpha \beta \gamma } H_{\gamma }{}^{\epsilon \varepsilon } H_{\epsilon }{}^{\lambda \mu } R_{\varepsilon \mu \nu \rho } \nabla_{\lambda }H_{\delta }{}^{\nu \rho }- \frac{1}{128} H_{\alpha }{}^{\delta \epsilon } H^{\alpha \beta \gamma } H_{\beta \delta }{}^{\varepsilon } H_{\gamma }{}^{\lambda \mu } R_{\varepsilon \mu \nu \rho } \nabla_{\lambda }H_{\epsilon }{}^{\nu \rho }\nn\\
&&+\frac{31}{1024} H_{\alpha \beta }{}^{\delta } H^{\alpha \beta \gamma } H_{\gamma }{}^{\epsilon \varepsilon } H_{\delta }{}^{\lambda \mu } R_{\varepsilon \mu \nu \rho } \nabla_{\lambda }H_{\epsilon }{}^{\nu \rho }+\frac{35}{256} H_{\alpha }{}^{\delta \epsilon } H^{\alpha \beta \gamma } H_{\beta }{}^{\varepsilon \lambda } H_{\delta }{}^{\mu \nu } R_{\gamma \epsilon \nu \rho } \nabla_{\lambda }H_{\varepsilon \mu }{}^{\rho }\nn\\
&&+\frac{1}{128} H_{\alpha \beta }{}^{\delta } H^{\alpha \beta \gamma } H_{\gamma }{}^{\epsilon \varepsilon } H_{\epsilon }{}^{\lambda \mu } R_{\delta \mu \nu \rho } \nabla_{\lambda }H_{\varepsilon }{}^{\nu \rho }- \frac{77}{512} H_{\alpha \beta }{}^{\delta } H^{\alpha \beta \gamma } H^{\epsilon \varepsilon \lambda } H^{\mu \nu \rho } R_{\delta \nu \lambda \rho } \nabla_{\mu }H_{\gamma \epsilon \varepsilon }\nn\\
&&+\frac{33}{512} H_{\alpha \beta }{}^{\delta } H^{\alpha \beta \gamma } H_{\epsilon }{}^{\mu \nu } H^{\epsilon \varepsilon \lambda } R_{\delta \nu \lambda \rho } \nabla_{\mu }H_{\gamma \varepsilon }{}^{\rho }- \frac{1}{64} H_{\alpha }{}^{\delta \epsilon } H^{\alpha \beta \gamma } H_{\beta \delta }{}^{\varepsilon } H^{\lambda \mu \nu } R_{\epsilon \nu \varepsilon \rho } \nabla_{\mu }H_{\gamma \lambda }{}^{\rho }\nn\\
&&+\frac{3}{512} H_{\alpha }{}^{\delta \epsilon } H^{\alpha \beta \gamma } H_{\beta }{}^{\varepsilon \lambda } H^{\mu \nu \rho } R_{\gamma \nu \lambda \rho } \nabla_{\mu }H_{\delta \epsilon \varepsilon }- \frac{91}{512} H_{\alpha \beta }{}^{\delta } H^{\alpha \beta \gamma } H_{\gamma }{}^{\epsilon \varepsilon } H^{\lambda \mu \nu } R_{\epsilon \nu \varepsilon \rho } \nabla_{\mu }H_{\delta \lambda }{}^{\rho }\nn\\
&&- \frac{17}{1024} H_{\alpha \beta }{}^{\delta } H^{\alpha \beta \gamma } H_{\gamma }{}^{\epsilon \varepsilon } H^{\lambda \mu \nu } R_{\delta \nu \varepsilon \rho } \nabla_{\mu }H_{\epsilon \lambda }{}^{\rho }+\frac{5}{1024} H_{\alpha \beta }{}^{\delta } H^{\alpha \beta \gamma } H_{\gamma }{}^{\epsilon \varepsilon } H^{\lambda \mu \nu } R_{\delta \rho \varepsilon \nu } \nabla_{\mu }H_{\epsilon \lambda }{}^{\rho }\nn\\
&&- \frac{35}{512} H_{\alpha }{}^{\delta \epsilon } H^{\alpha \beta \gamma } H_{\varepsilon }{}^{\nu \rho } H^{\varepsilon \lambda \mu } R_{\delta \mu \epsilon \rho } \nabla_{\nu }H_{\beta \gamma \lambda }- \frac{1}{64} H_{\alpha }{}^{\delta \epsilon } H^{\alpha \beta \gamma } H_{\beta }{}^{\varepsilon \lambda } H_{\delta }{}^{\mu \nu } R_{\epsilon \rho \varepsilon \lambda } \nabla_{\nu }H_{\gamma \mu }{}^{\rho }\nn\\
&&+\frac{3}{128} H_{\alpha }{}^{\delta \epsilon } H^{\alpha \beta \gamma } H_{\beta }{}^{\varepsilon \lambda } H^{\mu \nu \rho } R_{\gamma \epsilon \lambda \rho } \nabla_{\nu }H_{\delta \varepsilon \mu }- \frac{1}{128} H_{\alpha }{}^{\delta \epsilon } H^{\alpha \beta \gamma } H_{\beta \delta }{}^{\varepsilon } H^{\lambda \mu \nu } R_{\epsilon \nu \varepsilon \rho } \nabla^{\rho }H_{\gamma \lambda \mu }\nn\\
&&- \frac{75}{512} H_{\alpha \beta }{}^{\delta } H^{\alpha \beta \gamma } H_{\epsilon \varepsilon }{}^{\mu } H^{\epsilon \varepsilon \lambda } R_{\delta \nu \mu \rho } \nabla^{\rho }H_{\gamma \lambda }{}^{\nu }+\frac{31}{512} H_{\alpha \beta }{}^{\delta } H^{\alpha \beta \gamma } H_{\gamma }{}^{\epsilon \varepsilon } H^{\lambda \mu \nu } R_{\varepsilon \mu \nu \rho } \nabla^{\rho }H_{\delta \epsilon \lambda }\nn\\
&&- \frac{29}{256} H_{\alpha \beta }{}^{\delta } H^{\alpha \beta \gamma } H_{\gamma }{}^{\epsilon \varepsilon } H^{\lambda \mu \nu } R_{\epsilon \nu \varepsilon \rho } \nabla^{\rho }H_{\delta \lambda \mu }+\frac{1}{256} H_{\alpha \beta }{}^{\delta } H^{\alpha \beta \gamma } H_{\gamma }{}^{\epsilon \varepsilon } H_{\epsilon }{}^{\lambda \mu } R_{\varepsilon \nu \mu \rho } \nabla^{\rho }H_{\delta \lambda }{}^{\nu }\nn\\
&&- \frac{29}{512} H_{\alpha \beta }{}^{\delta } H^{\alpha \beta \gamma } H_{\gamma }{}^{\epsilon \varepsilon } H_{\epsilon \varepsilon }{}^{\lambda } R_{\lambda \rho \mu \nu } \nabla^{\rho }H_{\delta }{}^{\mu \nu }- \frac{31}{512} H_{\alpha \beta }{}^{\delta } H^{\alpha \beta \gamma } H_{\gamma }{}^{\epsilon \varepsilon } H^{\lambda \mu \nu } R_{\delta \rho \mu \nu } \nabla^{\rho }H_{\epsilon \varepsilon \lambda }\nn\\
&&- \frac{37}{512} H_{\alpha }{}^{\delta \epsilon } H^{\alpha \beta \gamma } H_{\beta }{}^{\varepsilon \lambda } H_{\delta }{}^{\mu \nu } R_{\gamma \rho \epsilon \nu } \nabla^{\rho }H_{\varepsilon \lambda \mu }- \frac{1}{256} H_{\alpha \beta }{}^{\delta } H^{\alpha \beta \gamma } H_{\gamma }{}^{\epsilon \varepsilon } H_{\epsilon }{}^{\lambda \mu } R_{\delta \mu \nu \rho } \nabla^{\rho }H_{\varepsilon \lambda }{}^{\nu }\nn\\
&&- \frac{1}{128} H_{\alpha }{}^{\delta \epsilon } H^{\alpha \beta \gamma } H_{\beta \delta }{}^{\varepsilon } H_{\gamma \epsilon }{}^{\lambda } R_{\lambda \mu \nu \rho } \nabla^{\rho }H_{\varepsilon }{}^{\mu \nu }- \frac{7}{256} H_{\alpha \beta }{}^{\delta } H^{\alpha \beta \gamma } H_{\gamma }{}^{\epsilon \varepsilon } H_{\delta \epsilon }{}^{\lambda } R_{\lambda \mu \nu \rho } \nabla^{\rho }H_{\varepsilon }{}^{\mu \nu }\nn\\
&&- \frac{25}{2048} H_{\alpha \beta }{}^{\delta } H^{\alpha \beta \gamma } H_{\gamma }{}^{\epsilon \varepsilon } H_{\epsilon }{}^{\lambda \mu } R_{\delta \varepsilon \nu \rho } \nabla^{\rho }H_{\lambda \mu }{}^{\nu }
\eeqa
The couplings with structure  $[H^2\nabla H^3]_{31}$ are the following:
\beqa
	&&- \frac{153}{256} H^{\alpha \beta \gamma } H^{\delta \epsilon \varepsilon } \nabla_{\beta }H_{\alpha }{}^{\lambda \mu } \nabla_{\delta }H_{\gamma \lambda }{}^{\nu } \nabla_{\varepsilon }H_{\epsilon \mu \nu }+\frac{7}{256} H^{\alpha \beta \gamma } H^{\delta \epsilon \varepsilon } \nabla_{\delta }H_{\alpha \beta }{}^{\lambda } \nabla_{\epsilon }H_{\gamma }{}^{\mu \nu } \nabla_{\varepsilon }H_{\lambda \mu \nu }\nn\\
	&&+\frac{3}{256} H^{\alpha \beta \gamma } H^{\delta \epsilon \varepsilon } \nabla_{\beta }H_{\alpha \delta }{}^{\lambda } \nabla_{\gamma }H_{\epsilon }{}^{\mu \nu } \nabla_{\lambda }H_{\varepsilon \mu \nu }+\frac{1}{512} H^{\alpha \beta \gamma } H^{\delta \epsilon \varepsilon } \nabla_{\gamma }H_{\epsilon }{}^{\mu \nu } \nabla_{\varepsilon }H_{\lambda \mu \nu } \nabla^{\lambda }H_{\alpha \beta \delta }\nn\\
	&&+\frac{29}{1024} H_{\alpha }{}^{\delta \epsilon } H^{\alpha \beta \gamma } \nabla_{\varepsilon }H_{\delta }{}^{\mu \nu } \nabla_{\lambda }H_{\epsilon \mu \nu } \nabla^{\lambda }H_{\beta \gamma }{}^{\varepsilon }- \frac{1}{128} H_{\alpha }{}^{\delta \epsilon } H^{\alpha \beta \gamma } \nabla_{\epsilon }H_{\varepsilon \mu \nu } \nabla_{\lambda }H_{\gamma }{}^{\mu \nu } \nabla^{\lambda }H_{\beta \delta }{}^{\varepsilon }\nn\\
	&&+\frac{5}{512} H^{\alpha \beta \gamma } H^{\delta \epsilon \varepsilon } \nabla_{\delta }H_{\gamma \lambda }{}^{\nu } \nabla_{\varepsilon }H_{\epsilon \mu \nu } \nabla^{\mu }H_{\alpha \beta }{}^{\lambda }- \frac{7}{128} H^{\alpha \beta \gamma } H^{\delta \epsilon \varepsilon } \nabla_{\epsilon }H_{\gamma \delta }{}^{\nu } \nabla_{\varepsilon }H_{\lambda \mu \nu } \nabla^{\mu }H_{\alpha \beta }{}^{\lambda }\nn\\
	&&- \frac{79}{256} H^{\alpha \beta \gamma } H^{\delta \epsilon \varepsilon } \nabla_{\varepsilon }H_{\epsilon \mu \nu } \nabla_{\lambda }H_{\gamma \delta }{}^{\nu } \nabla^{\mu }H_{\alpha \beta }{}^{\lambda }+\frac{65}{1024} H^{\alpha \beta \gamma } H^{\delta \epsilon \varepsilon } \nabla_{\lambda }H_{\gamma }{}^{\mu \nu } \nabla^{\lambda }H_{\alpha \beta \delta } \nabla_{\nu }H_{\epsilon \varepsilon \mu }\nn\\
	&&+\frac{7}{2048} H^{\alpha \beta \gamma } H^{\delta \epsilon \varepsilon } \nabla_{\delta }H_{\gamma \lambda }{}^{\nu } \nabla^{\mu }H_{\alpha \beta }{}^{\lambda } \nabla_{\nu }H_{\epsilon \varepsilon \mu }+\frac{35}{512} H^{\alpha \beta \gamma } H^{\delta \epsilon \varepsilon } \nabla_{\epsilon }H_{\gamma }{}^{\mu \nu } \nabla^{\lambda }H_{\alpha \beta \delta } \nabla_{\nu }H_{\varepsilon \lambda \mu }\nn\\
	&&- \frac{309}{2048} H^{\alpha \beta \gamma } H^{\delta \epsilon \varepsilon } \nabla^{\mu }H_{\alpha \beta }{}^{\lambda } \nabla_{\nu }H_{\epsilon \varepsilon \mu } \nabla^{\nu }H_{\gamma \delta \lambda }+\frac{25}{512} H^{\alpha \beta \gamma } H^{\delta \epsilon \varepsilon } \nabla_{\delta }H_{\alpha \beta }{}^{\lambda } \nabla_{\varepsilon }H_{\lambda \mu \nu } \nabla^{\nu }H_{\gamma \epsilon }{}^{\mu }\nn\\
	&&- \frac{1}{256} H^{\alpha \beta \gamma } H^{\delta \epsilon \varepsilon } \nabla_{\delta }H_{\alpha \beta }{}^{\lambda } \nabla_{\mu }H_{\varepsilon \lambda \nu } \nabla^{\nu }H_{\gamma \epsilon }{}^{\mu }+\frac{1}{128} H^{\alpha \beta \gamma } H^{\delta \epsilon \varepsilon } \nabla^{\lambda }H_{\alpha \beta \delta } \nabla_{\mu }H_{\varepsilon \lambda \nu } \nabla^{\nu }H_{\gamma \epsilon }{}^{\mu }\nn\\
	&&+\frac{3}{512} H^{\alpha \beta \gamma } H^{\delta \epsilon \varepsilon } \nabla_{\beta }H_{\alpha \delta }{}^{\lambda } \nabla_{\varepsilon }H_{\epsilon \mu \nu } \nabla^{\nu }H_{\gamma \lambda }{}^{\mu }- \frac{213}{1024} H^{\alpha \beta \gamma } H^{\delta \epsilon \varepsilon } \nabla_{\varepsilon }H_{\epsilon \mu \nu } \nabla^{\lambda }H_{\alpha \beta \delta } \nabla^{\nu }H_{\gamma \lambda }{}^{\mu }\nn\\
	&&- \frac{1}{64} H_{\alpha }{}^{\delta \epsilon } H^{\alpha \beta \gamma } \nabla_{\epsilon }H_{\varepsilon \mu \nu } \nabla^{\lambda }H_{\beta \delta }{}^{\varepsilon } \nabla^{\nu }H_{\gamma \lambda }{}^{\mu }+\frac{29}{256} H^{\alpha \beta \gamma } H^{\delta \epsilon \varepsilon } \nabla^{\lambda }H_{\alpha \beta \delta } \nabla_{\mu }H_{\epsilon \varepsilon \nu } \nabla^{\nu }H_{\gamma \lambda }{}^{\mu }\nn\\
	&&- \frac{37}{3072} H^{\alpha \beta \gamma } H^{\delta \epsilon \varepsilon } \nabla_{\lambda }H_{\varepsilon \mu \nu } \nabla^{\lambda }H_{\alpha \beta \gamma } \nabla^{\nu }H_{\delta \epsilon }{}^{\mu }+\frac{1}{128} H^{\alpha \beta \gamma } H^{\delta \epsilon \varepsilon } \nabla_{\gamma }H_{\alpha \beta }{}^{\lambda } \nabla_{\nu }H_{\varepsilon \lambda \mu } \nabla^{\nu }H_{\delta \epsilon }{}^{\mu }\nn\\
	&&+\frac{31}{1024} H_{\alpha }{}^{\delta \epsilon } H^{\alpha \beta \gamma } \nabla^{\lambda }H_{\beta \gamma }{}^{\varepsilon } \nabla_{\nu }H_{\varepsilon \lambda \mu } \nabla^{\nu }H_{\delta \epsilon }{}^{\mu }+\frac{41}{256} H_{\alpha }{}^{\delta \epsilon } H^{\alpha \beta \gamma } \nabla_{\gamma }H_{\beta }{}^{\varepsilon \lambda } \nabla_{\lambda }H_{\epsilon \mu \nu } \nabla^{\nu }H_{\delta \varepsilon }{}^{\mu }\nn\\
	&&+\frac{17}{128} H_{\alpha }{}^{\delta \epsilon } H^{\alpha \beta \gamma } \nabla_{\epsilon }H_{\lambda \mu \nu } \nabla^{\lambda }H_{\beta \gamma }{}^{\varepsilon } \nabla^{\nu }H_{\delta \varepsilon }{}^{\mu }+\frac{15}{512} H^{\alpha \beta \gamma } H^{\delta \epsilon \varepsilon } \nabla_{\varepsilon }H_{\epsilon \mu \nu } \nabla^{\lambda }H_{\alpha \beta \gamma } \nabla^{\nu }H_{\delta \lambda }{}^{\mu }\nn\\
	&&- \frac{105}{512} H_{\alpha }{}^{\delta \epsilon } H^{\alpha \beta \gamma } \nabla_{\varepsilon }H_{\epsilon \mu \nu } \nabla^{\lambda }H_{\beta \gamma }{}^{\varepsilon } \nabla^{\nu }H_{\delta \lambda }{}^{\mu }- \frac{89}{1024} H_{\alpha \beta }{}^{\delta } H^{\alpha \beta \gamma } \nabla_{\delta }H_{\lambda \mu \nu } \nabla^{\lambda }H_{\gamma }{}^{\epsilon \varepsilon } \nabla^{\nu }H_{\epsilon \varepsilon }{}^{\mu }\nn\\
	&&+\frac{1}{8} H_{\alpha }{}^{\delta \epsilon } H^{\alpha \beta \gamma } \nabla_{\delta }H_{\beta \gamma }{}^{\varepsilon } \nabla_{\mu }H_{\varepsilon \lambda \nu } \nabla^{\nu }H_{\epsilon }{}^{\lambda \mu }- \frac{5}{256} H_{\alpha }{}^{\delta \epsilon } H^{\alpha \beta \gamma } \nabla_{\gamma }H_{\beta \delta }{}^{\varepsilon } \nabla_{\nu }H_{\varepsilon \lambda \mu } \nabla^{\nu }H_{\epsilon }{}^{\lambda \mu }\nn\\
	&&- \frac{5}{512} H_{\alpha }{}^{\delta \epsilon } H^{\alpha \beta \gamma } \nabla^{\varepsilon }H_{\beta \gamma \delta } \nabla_{\nu }H_{\varepsilon \lambda \mu } \nabla^{\nu }H_{\epsilon }{}^{\lambda \mu }
\eeqa
\vskip 0.5 cm
\section{Additional couplings in Meissner scheme \label{AppB}}
\vskip 0.5 cm
 The full expressions for the \([H^2\nabla H^2 R]_{96}\), \([H^4\nabla H^2]_{93}\), \([H^2 R^2 \nabla H]_{31}\), and \([H^4 R\nabla H]_{52}\) couplings in the Meissner scheme are too lengthy to include in the main text and are therefore relegated to this appendix. Below, we list the \([H^2\nabla H^2 R]_{96}\) couplings:
\beqa
&&- \frac{3}{256} H^{\alpha \beta \gamma } H^{\delta \epsilon \varepsilon } R_{\gamma \eta \varepsilon \theta } \nabla_{\delta }H_{\alpha \beta }{}^{\zeta } \nabla_{\zeta }H_{\epsilon }{}^{\eta \theta }+\frac{3}{256} H^{\alpha \beta \gamma } H^{\delta \epsilon \varepsilon } R_{\varepsilon \eta \zeta \theta } \nabla_{\epsilon }H_{\gamma }{}^{\eta \theta } \nabla^{\zeta }H_{\alpha \beta \delta }\nn\\
&&- \frac{3}{128} H^{\alpha \beta \gamma } H^{\delta \epsilon \varepsilon } R_{\epsilon \eta \varepsilon \theta } \nabla_{\zeta }H_{\gamma }{}^{\eta \theta } \nabla^{\zeta }H_{\alpha \beta \delta }- \frac{1}{128} H_{\alpha }{}^{\delta \epsilon } H^{\alpha \beta \gamma } R_{\epsilon \eta \zeta \theta } \nabla_{\varepsilon }H_{\delta }{}^{\eta \theta } \nabla^{\zeta }H_{\beta \gamma }{}^{\varepsilon }\nn\\
&&+\frac{9}{256} H_{\alpha }{}^{\delta \epsilon } H^{\alpha \beta \gamma } R_{\epsilon \eta \varepsilon \theta } \nabla_{\zeta }H_{\delta }{}^{\eta \theta } \nabla^{\zeta }H_{\beta \gamma }{}^{\varepsilon }- \frac{1}{256} H_{\alpha }{}^{\delta \epsilon } H^{\alpha \beta \gamma } R_{\delta \eta \epsilon \theta } \nabla_{\zeta }H_{\varepsilon }{}^{\eta \theta } \nabla^{\zeta }H_{\beta \gamma }{}^{\varepsilon }\nn\\
&&+\frac{3}{128} H_{\alpha }{}^{\delta \epsilon } H^{\alpha \beta \gamma } R_{\epsilon \eta \zeta \theta } \nabla_{\varepsilon }H_{\gamma }{}^{\eta \theta } \nabla^{\zeta }H_{\beta \delta }{}^{\varepsilon }+\frac{15}{256} H_{\alpha }{}^{\delta \epsilon } H^{\alpha \beta \gamma } R_{\gamma \eta \epsilon \theta } \nabla_{\zeta }H_{\varepsilon }{}^{\eta \theta } \nabla^{\zeta }H_{\beta \delta }{}^{\varepsilon }\nn\\
&&- \frac{5}{256} H_{\alpha \beta }{}^{\delta } H^{\alpha \beta \gamma } R_{\epsilon \eta \varepsilon \theta } \nabla_{\zeta }H_{\delta }{}^{\eta \theta } \nabla^{\zeta }H_{\gamma }{}^{\epsilon \varepsilon }- \frac{5}{256} H_{\alpha \beta }{}^{\delta } H^{\alpha \beta \gamma } R_{\delta \eta \varepsilon \theta } \nabla_{\zeta }H_{\epsilon }{}^{\eta \theta } \nabla^{\zeta }H_{\gamma }{}^{\epsilon \varepsilon }\nn\\
&&+\frac{3}{32} H^{\alpha \beta \gamma } H^{\delta \epsilon \varepsilon } R_{\beta \varepsilon \gamma \theta } \nabla_{\delta }H_{\alpha }{}^{\zeta \eta } \nabla_{\eta }H_{\epsilon \zeta }{}^{\theta }+\frac{11}{256} H^{\alpha \beta \gamma } H^{\delta \epsilon \varepsilon } R_{\epsilon \eta \varepsilon \theta } \nabla_{\delta }H_{\gamma \zeta }{}^{\theta } \nabla^{\eta }H_{\alpha \beta }{}^{\zeta }\nn\\
&&- \frac{25}{256} H^{\alpha \beta \gamma } H^{\delta \epsilon \varepsilon } R_{\epsilon \zeta \varepsilon \theta } \nabla_{\delta }H_{\gamma \eta }{}^{\theta } \nabla^{\eta }H_{\alpha \beta }{}^{\zeta }+\frac{17}{256} H^{\alpha \beta \gamma } H^{\delta \epsilon \varepsilon } R_{\gamma \theta \epsilon \varepsilon } \nabla_{\zeta }H_{\delta \eta }{}^{\theta } \nabla^{\eta }H_{\alpha \beta }{}^{\zeta }\nn\\
&&+\frac{9}{256} H^{\alpha \beta \gamma } H^{\delta \epsilon \varepsilon } R_{\epsilon \zeta \varepsilon \theta } \nabla_{\eta }H_{\gamma \delta }{}^{\theta } \nabla^{\eta }H_{\alpha \beta }{}^{\zeta }- \frac{3}{512} H^{\alpha \beta \gamma } H^{\delta \epsilon \varepsilon } R_{\gamma \zeta \varepsilon \theta } \nabla_{\eta }H_{\delta \epsilon }{}^{\theta } \nabla^{\eta }H_{\alpha \beta }{}^{\zeta }\nn\\
&&+\frac{7}{512} H^{\alpha \beta \gamma } H^{\delta \epsilon \varepsilon } R_{\gamma \theta \varepsilon \zeta } \nabla_{\eta }H_{\delta \epsilon }{}^{\theta } \nabla^{\eta }H_{\alpha \beta }{}^{\zeta }- \frac{31}{512} H^{\alpha \beta \gamma } H^{\delta \epsilon \varepsilon } R_{\gamma \theta \epsilon \varepsilon } \nabla_{\eta }H_{\delta \zeta }{}^{\theta } \nabla^{\eta }H_{\alpha \beta }{}^{\zeta }\nn\\
&&+\frac{7}{128} H^{\alpha \beta \gamma } H^{\delta \epsilon \varepsilon } R_{\epsilon \zeta \varepsilon \theta } \nabla_{\gamma }H_{\beta \eta }{}^{\theta } \nabla^{\eta }H_{\alpha \delta }{}^{\zeta }+\frac{3}{32} H^{\alpha \beta \gamma } H^{\delta \epsilon \varepsilon } R_{\gamma \theta \varepsilon \eta } \nabla_{\epsilon }H_{\beta \zeta }{}^{\theta } \nabla^{\eta }H_{\alpha \delta }{}^{\zeta }\nn\\
&&- \frac{3}{64} H^{\alpha \beta \gamma } H^{\delta \epsilon \varepsilon } R_{\gamma \theta \epsilon \varepsilon } \nabla_{\zeta }H_{\beta \eta }{}^{\theta } \nabla^{\eta }H_{\alpha \delta }{}^{\zeta }+\frac{3}{64} H^{\alpha \beta \gamma } H^{\delta \epsilon \varepsilon } R_{\gamma \zeta \varepsilon \theta } \nabla_{\eta }H_{\beta \epsilon }{}^{\theta } \nabla^{\eta }H_{\alpha \delta }{}^{\zeta }\nn\\
&&+\frac{13}{128} H^{\alpha \beta \gamma } H^{\delta \epsilon \varepsilon } R_{\gamma \theta \epsilon \varepsilon } \nabla_{\eta }H_{\beta \zeta }{}^{\theta } \nabla^{\eta }H_{\alpha \delta }{}^{\zeta }- \frac{27}{256} H_{\alpha }{}^{\delta \epsilon } H^{\alpha \beta \gamma } R_{\delta \zeta \epsilon \theta } \nabla_{\varepsilon }H_{\gamma \eta }{}^{\theta } \nabla^{\eta }H_{\beta }{}^{\varepsilon \zeta }\nn\\
&&+\frac{13}{128} H_{\alpha }{}^{\delta \epsilon } H^{\alpha \beta \gamma } R_{\delta \zeta \epsilon \theta } \nabla_{\eta }H_{\gamma \varepsilon }{}^{\theta } \nabla^{\eta }H_{\beta }{}^{\varepsilon \zeta }- \frac{27}{256} H_{\alpha }{}^{\delta \epsilon } H^{\alpha \beta \gamma } R_{\gamma \zeta \epsilon \theta } \nabla_{\eta }H_{\delta \varepsilon }{}^{\theta } \nabla^{\eta }H_{\beta }{}^{\varepsilon \zeta }\nn\\
&&+\frac{15}{256} H_{\alpha }{}^{\delta \epsilon } H^{\alpha \beta \gamma } R_{\gamma \theta \epsilon \zeta } \nabla_{\eta }H_{\delta \varepsilon }{}^{\theta } \nabla^{\eta }H_{\beta }{}^{\varepsilon \zeta }+\frac{1}{256} H_{\alpha }{}^{\delta \epsilon } H^{\alpha \beta \gamma } R_{\gamma \theta \delta \epsilon } \nabla_{\eta }H_{\varepsilon \zeta }{}^{\theta } \nabla^{\eta }H_{\beta }{}^{\varepsilon \zeta }\nn\\
&&- \frac{1}{256} H_{\alpha \beta }{}^{\delta } H^{\alpha \beta \gamma } R_{\gamma \zeta \delta \theta } \nabla_{\eta }H_{\epsilon \varepsilon }{}^{\theta } \nabla^{\eta }H^{\epsilon \varepsilon \zeta }- \frac{1}{32} H^{\alpha \beta \gamma } H^{\delta \epsilon \varepsilon } R_{\beta \epsilon \gamma \varepsilon } \nabla_{\eta }H_{\delta \zeta \theta } \nabla^{\theta }H_{\alpha }{}^{\zeta \eta }\nn\\
&&+\frac{7}{512} H^{\alpha \beta \gamma } H^{\delta \epsilon \varepsilon } R_{\beta \epsilon \gamma \varepsilon } \nabla_{\theta }H_{\delta \zeta \eta } \nabla^{\theta }H_{\alpha }{}^{\zeta \eta }- \frac{3}{64} H^{\alpha \beta \gamma } H^{\delta \epsilon \varepsilon } R_{\gamma \zeta \varepsilon \theta } \nabla^{\eta }H_{\alpha \delta }{}^{\zeta } \nabla^{\theta }H_{\beta \epsilon \eta }\nn\\
&&- \frac{1}{256} H^{\alpha \beta \gamma } H^{\delta \epsilon \varepsilon } R_{\gamma \theta \epsilon \varepsilon } \nabla^{\eta }H_{\alpha \delta }{}^{\zeta } \nabla^{\theta }H_{\beta \zeta \eta }+\frac{11}{512} H^{\alpha \beta \gamma } H^{\delta \epsilon \varepsilon } R_{\epsilon \eta \varepsilon \theta } \nabla^{\eta }H_{\alpha \beta }{}^{\zeta } \nabla^{\theta }H_{\gamma \delta \zeta }\nn\\
&&- \frac{29}{512} H^{\alpha \beta \gamma } H^{\delta \epsilon \varepsilon } R_{\epsilon \zeta \varepsilon \theta } \nabla^{\eta }H_{\alpha \beta }{}^{\zeta } \nabla^{\theta }H_{\gamma \delta \eta }- \frac{53}{512} H^{\alpha \beta \gamma } H^{\delta \epsilon \varepsilon } R_{\varepsilon \eta \zeta \theta } \nabla_{\delta }H_{\alpha \beta }{}^{\zeta } \nabla^{\theta }H_{\gamma \epsilon }{}^{\eta }\nn\\
&&+\frac{21}{256} H^{\alpha \beta \gamma } H^{\delta \epsilon \varepsilon } R_{\varepsilon \theta \zeta \eta } \nabla_{\delta }H_{\alpha \beta }{}^{\zeta } \nabla^{\theta }H_{\gamma \epsilon }{}^{\eta }+\frac{7}{128} H^{\alpha \beta \gamma } H^{\delta \epsilon \varepsilon } R_{\varepsilon \eta \zeta \theta } \nabla^{\zeta }H_{\alpha \beta \delta } \nabla^{\theta }H_{\gamma \epsilon }{}^{\eta }\nn\\
&&- \frac{31}{512} H^{\alpha \beta \gamma } H^{\delta \epsilon \varepsilon } R_{\varepsilon \theta \zeta \eta } \nabla^{\zeta }H_{\alpha \beta \delta } \nabla^{\theta }H_{\gamma \epsilon }{}^{\eta }- \frac{7}{512} H_{\alpha }{}^{\delta \epsilon } H^{\alpha \beta \gamma } R_{\varepsilon \eta \zeta \theta } \nabla^{\zeta }H_{\beta \delta }{}^{\varepsilon } \nabla^{\theta }H_{\gamma \epsilon }{}^{\eta }\nn\\
&&+\frac{1}{512} H_{\alpha }{}^{\delta \epsilon } H^{\alpha \beta \gamma } R_{\varepsilon \theta \zeta \eta } \nabla^{\zeta }H_{\beta \delta }{}^{\varepsilon } \nabla^{\theta }H_{\gamma \epsilon }{}^{\eta }+\frac{3}{128} H_{\alpha }{}^{\delta \epsilon } H^{\alpha \beta \gamma } R_{\delta \eta \epsilon \theta } \nabla^{\eta }H_{\beta }{}^{\varepsilon \zeta } \nabla^{\theta }H_{\gamma \varepsilon \zeta }\nn\\
&&- \frac{9}{64} H_{\alpha }{}^{\delta \epsilon } H^{\alpha \beta \gamma } R_{\delta \zeta \epsilon \theta } \nabla^{\eta }H_{\beta }{}^{\varepsilon \zeta } \nabla^{\theta }H_{\gamma \varepsilon \eta }+\frac{13}{256} H_{\alpha }{}^{\delta \epsilon } H^{\alpha \beta \gamma } R_{\epsilon \eta \zeta \theta } \nabla^{\zeta }H_{\beta \delta }{}^{\varepsilon } \nabla^{\theta }H_{\gamma \varepsilon }{}^{\eta }\nn\\
&&- \frac{7}{256} H_{\alpha }{}^{\delta \epsilon } H^{\alpha \beta \gamma } R_{\epsilon \theta \zeta \eta } \nabla^{\zeta }H_{\beta \delta }{}^{\varepsilon } \nabla^{\theta }H_{\gamma \varepsilon }{}^{\eta }+\frac{1}{128} H^{\alpha \beta \gamma } H^{\delta \epsilon \varepsilon } R_{\epsilon \eta \varepsilon \theta } \nabla_{\delta }H_{\alpha \beta }{}^{\zeta } \nabla^{\theta }H_{\gamma \zeta }{}^{\eta }\nn\\
&&- \frac{51}{512} H^{\alpha \beta \gamma } H^{\delta \epsilon \varepsilon } R_{\epsilon \eta \varepsilon \theta } \nabla^{\zeta }H_{\alpha \beta \delta } \nabla^{\theta }H_{\gamma \zeta }{}^{\eta }- \frac{5}{256} H_{\alpha }{}^{\delta \epsilon } H^{\alpha \beta \gamma } R_{\epsilon \eta \varepsilon \theta } \nabla^{\zeta }H_{\beta \delta }{}^{\varepsilon } \nabla^{\theta }H_{\gamma \zeta }{}^{\eta }\nn\\
&&+\frac{5}{256} H_{\alpha }{}^{\delta \epsilon } H^{\alpha \beta \gamma } R_{\epsilon \theta \varepsilon \eta } \nabla^{\zeta }H_{\beta \delta }{}^{\varepsilon } \nabla^{\theta }H_{\gamma \zeta }{}^{\eta }+\frac{1}{128} H^{\alpha \beta \gamma } H^{\delta \epsilon \varepsilon } R_{\gamma \eta \varepsilon \theta } \nabla^{\eta }H_{\alpha \beta }{}^{\zeta } \nabla^{\theta }H_{\delta \epsilon \zeta }\nn\\
&&- \frac{1}{128} H^{\alpha \beta \gamma } H^{\delta \epsilon \varepsilon } R_{\gamma \theta \varepsilon \eta } \nabla^{\eta }H_{\alpha \beta }{}^{\zeta } \nabla^{\theta }H_{\delta \epsilon \zeta }+\frac{3}{512} H^{\alpha \beta \gamma } H^{\delta \epsilon \varepsilon } R_{\gamma \zeta \varepsilon \theta } \nabla^{\eta }H_{\alpha \beta }{}^{\zeta } \nabla^{\theta }H_{\delta \epsilon \eta }\nn\\
&&+\frac{1}{512} H^{\alpha \beta \gamma } H^{\delta \epsilon \varepsilon } R_{\gamma \theta \varepsilon \zeta } \nabla^{\eta }H_{\alpha \beta }{}^{\zeta } \nabla^{\theta }H_{\delta \epsilon \eta }+\frac{1}{3072} H^{\alpha \beta \gamma } H^{\delta \epsilon \varepsilon } R_{\varepsilon \eta \zeta \theta } \nabla^{\zeta }H_{\alpha \beta \gamma } \nabla^{\theta }H_{\delta \epsilon }{}^{\eta }\nn\\
&&+\frac{7}{3072} H^{\alpha \beta \gamma } H^{\delta \epsilon \varepsilon } R_{\varepsilon \theta \zeta \eta } \nabla^{\zeta }H_{\alpha \beta \gamma } \nabla^{\theta }H_{\delta \epsilon }{}^{\eta }- \frac{9}{1024} H_{\alpha }{}^{\delta \epsilon } H^{\alpha \beta \gamma } R_{\varepsilon \eta \zeta \theta } \nabla^{\zeta }H_{\beta \gamma }{}^{\varepsilon } \nabla^{\theta }H_{\delta \epsilon }{}^{\eta }\nn\\
&&- \frac{3}{1024} H_{\alpha }{}^{\delta \epsilon } H^{\alpha \beta \gamma } R_{\varepsilon \theta \zeta \eta } \nabla^{\zeta }H_{\beta \gamma }{}^{\varepsilon } \nabla^{\theta }H_{\delta \epsilon }{}^{\eta }- \frac{3}{512} H_{\alpha \beta }{}^{\delta } H^{\alpha \beta \gamma } R_{\varepsilon \eta \zeta \theta } \nabla^{\zeta }H_{\gamma }{}^{\epsilon \varepsilon } \nabla^{\theta }H_{\delta \epsilon }{}^{\eta }\nn\\
&&+\frac{3}{512} H_{\alpha \beta }{}^{\delta } H^{\alpha \beta \gamma } R_{\varepsilon \theta \zeta \eta } \nabla^{\zeta }H_{\gamma }{}^{\epsilon \varepsilon } \nabla^{\theta }H_{\delta \epsilon }{}^{\eta }+\frac{3}{128} H_{\alpha }{}^{\delta \epsilon } H^{\alpha \beta \gamma } R_{\gamma \theta \epsilon \eta } \nabla^{\eta }H_{\beta }{}^{\varepsilon \zeta } \nabla^{\theta }H_{\delta \varepsilon \zeta }\nn\\
&&+\frac{13}{128} H_{\alpha }{}^{\delta \epsilon } H^{\alpha \beta \gamma } R_{\gamma \zeta \epsilon \theta } \nabla^{\eta }H_{\beta }{}^{\varepsilon \zeta } \nabla^{\theta }H_{\delta \varepsilon \eta }- \frac{13}{128} H_{\alpha }{}^{\delta \epsilon } H^{\alpha \beta \gamma } R_{\gamma \theta \epsilon \zeta } \nabla^{\eta }H_{\beta }{}^{\varepsilon \zeta } \nabla^{\theta }H_{\delta \varepsilon \eta }\nn\\
&&+\frac{3}{64} H_{\alpha }{}^{\delta \epsilon } H^{\alpha \beta \gamma } R_{\epsilon \eta \zeta \theta } \nabla^{\zeta }H_{\beta \gamma }{}^{\varepsilon } \nabla^{\theta }H_{\delta \varepsilon }{}^{\eta }- \frac{19}{512} H_{\alpha }{}^{\delta \epsilon } H^{\alpha \beta \gamma } R_{\epsilon \theta \zeta \eta } \nabla^{\zeta }H_{\beta \gamma }{}^{\varepsilon } \nabla^{\theta }H_{\delta \varepsilon }{}^{\eta }\nn\\
&&+\frac{3}{1024} H^{\alpha \beta \gamma } H^{\delta \epsilon \varepsilon } R_{\gamma \theta \epsilon \varepsilon } \nabla^{\eta }H_{\alpha \beta }{}^{\zeta } \nabla^{\theta }H_{\delta \zeta \eta }+\frac{1}{1536} H^{\alpha \beta \gamma } H^{\delta \epsilon \varepsilon } R_{\epsilon \eta \varepsilon \theta } \nabla^{\zeta }H_{\alpha \beta \gamma } \nabla^{\theta }H_{\delta \zeta }{}^{\eta }\nn\\
&&- \frac{1}{16} H_{\alpha }{}^{\delta \epsilon } H^{\alpha \beta \gamma } R_{\epsilon \eta \varepsilon \theta } \nabla^{\zeta }H_{\beta \gamma }{}^{\varepsilon } \nabla^{\theta }H_{\delta \zeta }{}^{\eta }+\frac{17}{512} H_{\alpha }{}^{\delta \epsilon } H^{\alpha \beta \gamma } R_{\epsilon \theta \varepsilon \eta } \nabla^{\zeta }H_{\beta \gamma }{}^{\varepsilon } \nabla^{\theta }H_{\delta \zeta }{}^{\eta }\nn\\
&&- \frac{23}{512} H_{\alpha \beta }{}^{\delta } H^{\alpha \beta \gamma } R_{\epsilon \eta \varepsilon \theta } \nabla^{\zeta }H_{\gamma }{}^{\epsilon \varepsilon } \nabla^{\theta }H_{\delta \zeta }{}^{\eta }+\frac{1}{256} H_{\alpha \beta }{}^{\delta } H^{\alpha \beta \gamma } R_{\gamma \zeta \delta \theta } \nabla^{\eta }H^{\epsilon \varepsilon \zeta } \nabla^{\theta }H_{\epsilon \varepsilon \eta }\nn\\
&&- \frac{1}{256} H^{\alpha \beta \gamma } H^{\delta \epsilon \varepsilon } R_{\gamma \eta \zeta \theta } \nabla_{\delta }H_{\alpha \beta }{}^{\zeta } \nabla^{\theta }H_{\epsilon \varepsilon }{}^{\eta }+\frac{5}{512} H^{\alpha \beta \gamma } H^{\delta \epsilon \varepsilon } R_{\gamma \theta \zeta \eta } \nabla_{\delta }H_{\alpha \beta }{}^{\zeta } \nabla^{\theta }H_{\epsilon \varepsilon }{}^{\eta }\nn\\
&&+\frac{3}{64} H_{\alpha }{}^{\delta \epsilon } H^{\alpha \beta \gamma } R_{\gamma \eta \zeta \theta } \nabla_{\delta }H_{\beta }{}^{\varepsilon \zeta } \nabla^{\theta }H_{\epsilon \varepsilon }{}^{\eta }- \frac{3}{64} H_{\alpha }{}^{\delta \epsilon } H^{\alpha \beta \gamma } R_{\gamma \theta \zeta \eta } \nabla_{\delta }H_{\beta }{}^{\varepsilon \zeta } \nabla^{\theta }H_{\epsilon \varepsilon }{}^{\eta }\nn\\
&&+\frac{15}{1024} H^{\alpha \beta \gamma } H^{\delta \epsilon \varepsilon } R_{\gamma \eta \zeta \theta } \nabla^{\zeta }H_{\alpha \beta \delta } \nabla^{\theta }H_{\epsilon \varepsilon }{}^{\eta }- \frac{9}{1024} H^{\alpha \beta \gamma } H^{\delta \epsilon \varepsilon } R_{\gamma \theta \zeta \eta } \nabla^{\zeta }H_{\alpha \beta \delta } \nabla^{\theta }H_{\epsilon \varepsilon }{}^{\eta }\nn\\
&&- \frac{1}{1024} H_{\alpha \beta }{}^{\delta } H^{\alpha \beta \gamma } R_{\delta \eta \zeta \theta } \nabla^{\zeta }H_{\gamma }{}^{\epsilon \varepsilon } \nabla^{\theta }H_{\epsilon \varepsilon }{}^{\eta }- \frac{3}{1024} H_{\alpha \beta }{}^{\delta } H^{\alpha \beta \gamma } R_{\delta \theta \zeta \eta } \nabla^{\zeta }H_{\gamma }{}^{\epsilon \varepsilon } \nabla^{\theta }H_{\epsilon \varepsilon }{}^{\eta }\nn\\
&&- \frac{11}{256} H^{\alpha \beta \gamma } H^{\delta \epsilon \varepsilon } R_{\beta \varepsilon \gamma \theta } \nabla_{\delta }H_{\alpha }{}^{\zeta \eta } \nabla^{\theta }H_{\epsilon \zeta \eta }+\frac{5}{128} H^{\alpha \beta \gamma } H^{\delta \epsilon \varepsilon } R_{\gamma \eta \varepsilon \theta } \nabla_{\delta }H_{\alpha \beta }{}^{\zeta } \nabla^{\theta }H_{\epsilon \zeta }{}^{\eta }\nn\\
&&+\frac{3}{512} H^{\alpha \beta \gamma } H^{\delta \epsilon \varepsilon } R_{\gamma \theta \varepsilon \eta } \nabla_{\delta }H_{\alpha \beta }{}^{\zeta } \nabla^{\theta }H_{\epsilon \zeta }{}^{\eta }- \frac{27}{512} H^{\alpha \beta \gamma } H^{\delta \epsilon \varepsilon } R_{\gamma \eta \varepsilon \theta } \nabla^{\zeta }H_{\alpha \beta \delta } \nabla^{\theta }H_{\epsilon \zeta }{}^{\eta }\nn\\
&&- \frac{1}{64} H^{\alpha \beta \gamma } H^{\delta \epsilon \varepsilon } R_{\gamma \theta \varepsilon \eta } \nabla^{\zeta }H_{\alpha \beta \delta } \nabla^{\theta }H_{\epsilon \zeta }{}^{\eta }- \frac{23}{512} H_{\alpha \beta }{}^{\delta } H^{\alpha \beta \gamma } R_{\delta \eta \varepsilon \theta } \nabla^{\zeta }H_{\gamma }{}^{\epsilon \varepsilon } \nabla^{\theta }H_{\epsilon \zeta }{}^{\eta }\nn\\
&&+\frac{9}{512} H_{\alpha \beta }{}^{\delta } H^{\alpha \beta \gamma } R_{\delta \theta \varepsilon \eta } \nabla^{\zeta }H_{\gamma }{}^{\epsilon \varepsilon } \nabla^{\theta }H_{\epsilon \zeta }{}^{\eta }+\frac{1}{256} H^{\alpha \beta \gamma } H^{\delta \epsilon \varepsilon } R_{\varepsilon \theta \zeta \eta } \nabla_{\delta }H_{\alpha \beta \gamma } \nabla^{\theta }H_{\epsilon }{}^{\zeta \eta }\nn\\
&&- \frac{3}{1024} H_{\alpha }{}^{\delta \epsilon } H^{\alpha \beta \gamma } R_{\varepsilon \theta \zeta \eta } \nabla_{\delta }H_{\beta \gamma }{}^{\varepsilon } \nabla^{\theta }H_{\epsilon }{}^{\zeta \eta }+\frac{3}{1024} H_{\alpha \beta }{}^{\delta } H^{\alpha \beta \gamma } R_{\varepsilon \theta \zeta \eta } \nabla_{\delta }H_{\gamma }{}^{\epsilon \varepsilon } \nabla^{\theta }H_{\epsilon }{}^{\zeta \eta }\nn\\
&&- \frac{5}{512} H_{\alpha }{}^{\delta \epsilon } H^{\alpha \beta \gamma } R_{\gamma \theta \delta \epsilon } \nabla^{\eta }H_{\beta }{}^{\varepsilon \zeta } \nabla^{\theta }H_{\varepsilon \zeta \eta }- \frac{3}{128} H_{\alpha }{}^{\delta \epsilon } H^{\alpha \beta \gamma } R_{\gamma \theta \epsilon \eta } \nabla_{\delta }H_{\beta }{}^{\varepsilon \zeta } \nabla^{\theta }H_{\varepsilon \zeta }{}^{\eta }\nn\\
&&+\frac{17}{512} H_{\alpha }{}^{\delta \epsilon } H^{\alpha \beta \gamma } R_{\delta \eta \epsilon \theta } \nabla^{\zeta }H_{\beta \gamma }{}^{\varepsilon } \nabla^{\theta }H_{\varepsilon \zeta }{}^{\eta }- \frac{5}{256} H_{\alpha }{}^{\delta \epsilon } H^{\alpha \beta \gamma } R_{\gamma \eta \epsilon \theta } \nabla^{\zeta }H_{\beta \delta }{}^{\varepsilon } \nabla^{\theta }H_{\varepsilon \zeta }{}^{\eta }\nn\\
&&- \frac{5}{512} H_{\alpha }{}^{\delta \epsilon } H^{\alpha \beta \gamma } R_{\epsilon \theta \zeta \eta } \nabla_{\delta }H_{\beta \gamma }{}^{\varepsilon } \nabla^{\theta }H_{\varepsilon }{}^{\zeta \eta }- \frac{35}{1024} H^{\alpha \beta \gamma } H^{\delta \epsilon \varepsilon } R_{\gamma \theta \zeta \eta } \nabla_{\epsilon }H_{\alpha \beta \delta } \nabla^{\theta }H_{\varepsilon }{}^{\zeta \eta }\nn\\
&&+\frac{11}{1024} H_{\alpha }{}^{\delta \epsilon } H^{\alpha \beta \gamma } R_{\epsilon \theta \zeta \eta } \nabla^{\varepsilon }H_{\beta \gamma \delta } \nabla^{\theta }H_{\varepsilon }{}^{\zeta \eta }+\frac{7}{3072} H_{\alpha }{}^{\delta \epsilon } H^{\alpha \beta \gamma } R_{\beta \delta \gamma \epsilon } \nabla_{\theta }H_{\varepsilon \zeta \eta } \nabla^{\theta }H^{\varepsilon \zeta \eta }\,.
\eeqa
The couplings with the structure  $[H^4\nabla H^2]_{93}$ are the following: 
\beqa
&&\frac{95}{512} H_{\alpha }{}^{\delta \epsilon } H^{\alpha \beta \gamma } H_{\beta }{}^{\varepsilon \zeta } H_{\delta }{}^{\eta \theta } \nabla_{\zeta }H_{\epsilon \theta \iota } \nabla_{\eta }H_{\gamma \varepsilon }{}^{\iota }+\frac{39}{2048} H_{\alpha }{}^{\delta \epsilon } H^{\alpha \beta \gamma } H_{\varepsilon }{}^{\theta \iota } H^{\varepsilon \zeta \eta } \nabla_{\zeta }H_{\beta \gamma \delta } \nabla_{\eta }H_{\epsilon \theta \iota }\nn\\
&&+\frac{19}{1024} H_{\alpha \beta }{}^{\delta } H^{\alpha \beta \gamma } H_{\gamma }{}^{\epsilon \varepsilon } H_{\epsilon }{}^{\zeta \eta } \nabla_{\zeta }H_{\delta }{}^{\theta \iota } \nabla_{\eta }H_{\varepsilon \theta \iota }- \frac{93}{1024} H_{\alpha }{}^{\delta \epsilon } H^{\alpha \beta \gamma } H_{\beta \delta }{}^{\varepsilon } H_{\gamma }{}^{\zeta \eta } \nabla_{\zeta }H_{\epsilon }{}^{\theta \iota } \nabla_{\eta }H_{\varepsilon \theta \iota }\nn\\
&&- \frac{39}{2048} H_{\alpha \beta }{}^{\delta } H^{\alpha \beta \gamma } H_{\gamma }{}^{\epsilon \varepsilon } H_{\delta }{}^{\zeta \eta } \nabla_{\zeta }H_{\epsilon }{}^{\theta \iota } \nabla_{\eta }H_{\varepsilon \theta \iota }+\frac{95}{1024} H_{\alpha }{}^{\delta \epsilon } H^{\alpha \beta \gamma } H_{\varepsilon }{}^{\theta \iota } H^{\varepsilon \zeta \eta } \nabla_{\eta }H_{\gamma \epsilon \iota } \nabla_{\theta }H_{\beta \delta \zeta }\nn\\
&&- \frac{63}{256} H_{\alpha }{}^{\delta \epsilon } H^{\alpha \beta \gamma } H_{\beta }{}^{\varepsilon \zeta } H^{\eta \theta \iota } \nabla_{\zeta }H_{\epsilon \varepsilon \iota } \nabla_{\theta }H_{\gamma \delta \eta }+\frac{93}{1024} H_{\alpha }{}^{\delta \epsilon } H^{\alpha \beta \gamma } H_{\beta }{}^{\varepsilon \zeta } H_{\delta }{}^{\eta \theta } \nabla_{\zeta }H_{\epsilon \varepsilon \iota } \nabla_{\theta }H_{\gamma \eta }{}^{\iota }\nn\\
&&+\frac{37}{1024} H_{\alpha \beta }{}^{\delta } H^{\alpha \beta \gamma } H_{\epsilon }{}^{\eta \theta } H^{\epsilon \varepsilon \zeta } \nabla_{\eta }H_{\gamma \varepsilon }{}^{\iota } \nabla_{\theta }H_{\delta \zeta \iota }+\frac{17}{1024} H_{\alpha \beta }{}^{\delta } H^{\alpha \beta \gamma } H_{\epsilon }{}^{\eta \theta } H^{\epsilon \varepsilon \zeta } \nabla_{\zeta }H_{\gamma \varepsilon }{}^{\iota } \nabla_{\theta }H_{\delta \eta \iota }\nn\\
&&- \frac{1}{512} H_{\alpha \beta }{}^{\delta } H^{\alpha \beta \gamma } H_{\gamma }{}^{\epsilon \varepsilon } H^{\zeta \eta \theta } \nabla_{\eta }H_{\delta \zeta }{}^{\iota } \nabla_{\theta }H_{\epsilon \varepsilon \iota }- \frac{25}{128} H_{\alpha }{}^{\delta \epsilon } H^{\alpha \beta \gamma } H_{\beta }{}^{\varepsilon \zeta } H_{\delta }{}^{\eta \theta } \nabla_{\eta }H_{\gamma \varepsilon }{}^{\iota } \nabla_{\theta }H_{\epsilon \zeta \iota }\nn\\
&&+\frac{27}{512} H_{\alpha }{}^{\delta \epsilon } H^{\alpha \beta \gamma } H_{\beta }{}^{\varepsilon \zeta } H_{\delta }{}^{\eta \theta } \nabla_{\zeta }H_{\gamma \varepsilon }{}^{\iota } \nabla_{\theta }H_{\epsilon \eta \iota }+\frac{11}{512} H_{\alpha \beta }{}^{\delta } H^{\alpha \beta \gamma } H_{\gamma }{}^{\epsilon \varepsilon } H^{\zeta \eta \theta } \nabla_{\epsilon }H_{\delta \zeta }{}^{\iota } \nabla_{\theta }H_{\varepsilon \eta \iota }\nn\\
&&+\frac{1}{1024} H_{\alpha }{}^{\delta \epsilon } H^{\alpha \beta \gamma } H_{\beta \delta }{}^{\varepsilon } H^{\zeta \eta \theta } \nabla_{\zeta }H_{\gamma \epsilon }{}^{\iota } \nabla_{\theta }H_{\varepsilon \eta \iota }- \frac{1}{256} H_{\alpha \beta }{}^{\delta } H^{\alpha \beta \gamma } H_{\gamma }{}^{\epsilon \varepsilon } H^{\zeta \eta \theta } \nabla_{\zeta }H_{\delta \epsilon }{}^{\iota } \nabla_{\theta }H_{\varepsilon \eta \iota }\nn\\
&&- \frac{13}{512} H_{\alpha \beta }{}^{\delta } H^{\alpha \beta \gamma } H_{\epsilon }{}^{\eta \theta } H^{\epsilon \varepsilon \zeta } \nabla_{\delta }H_{\gamma \varepsilon }{}^{\iota } \nabla_{\theta }H_{\zeta \eta \iota }+\frac{41}{512} H_{\alpha }{}^{\delta \epsilon } H^{\alpha \beta \gamma } H_{\beta }{}^{\varepsilon \zeta } H_{\delta }{}^{\eta \theta } \nabla_{\epsilon }H_{\gamma \varepsilon }{}^{\iota } \nabla_{\theta }H_{\zeta \eta \iota }\nn\\
&&+\frac{37}{512} H_{\alpha }{}^{\delta \epsilon } H^{\alpha \beta \gamma } H_{\beta }{}^{\varepsilon \zeta } H_{\delta }{}^{\eta \theta } \nabla_{\varepsilon }H_{\gamma \epsilon }{}^{\iota } \nabla_{\theta }H_{\zeta \eta \iota }+\frac{61}{2048} H_{\alpha }{}^{\delta \epsilon } H^{\alpha \beta \gamma } H_{\beta }{}^{\varepsilon \zeta } H_{\gamma }{}^{\eta \theta } \nabla_{\varepsilon }H_{\delta \epsilon }{}^{\iota } \nabla_{\theta }H_{\zeta \eta \iota }\nn\\
&&+\frac{5}{1024} H_{\alpha }{}^{\delta \epsilon } H^{\alpha \beta \gamma } H_{\varepsilon }{}^{\theta \iota } H^{\varepsilon \zeta \eta } \nabla_{\theta }H_{\beta \delta \zeta } \nabla_{\iota }H_{\gamma \epsilon \eta }+\frac{171}{2048} H_{\alpha }{}^{\delta \epsilon } H^{\alpha \beta \gamma } H_{\varepsilon }{}^{\theta \iota } H^{\varepsilon \zeta \eta } \nabla_{\eta }H_{\beta \delta \zeta } \nabla_{\iota }H_{\gamma \epsilon \theta }\nn\\
&&+\frac{13}{2048} H_{\alpha }{}^{\delta \epsilon } H^{\alpha \beta \gamma } H_{\beta }{}^{\varepsilon \zeta } H^{\eta \theta \iota } \nabla_{\zeta }H_{\delta \epsilon \varepsilon } \nabla_{\iota }H_{\gamma \eta \theta }+\frac{1}{128} H_{\alpha }{}^{\delta \epsilon } H^{\alpha \beta \gamma } H_{\varepsilon }{}^{\theta \iota } H^{\varepsilon \zeta \eta } \nabla_{\theta }H_{\beta \gamma \zeta } \nabla_{\iota }H_{\delta \epsilon \eta }\nn\\
&&- \frac{33}{2048} H_{\alpha }{}^{\delta \epsilon } H^{\alpha \beta \gamma } H_{\varepsilon }{}^{\theta \iota } H^{\varepsilon \zeta \eta } \nabla_{\eta }H_{\beta \gamma \zeta } \nabla_{\iota }H_{\delta \epsilon \theta }- \frac{17}{1024} H_{\alpha \beta }{}^{\delta } H^{\alpha \beta \gamma } H^{\epsilon \varepsilon \zeta } H^{\eta \theta \iota } \nabla_{\varepsilon }H_{\gamma \epsilon \eta } \nabla_{\iota }H_{\delta \zeta \theta }\nn\\
&&+\frac{23}{1024} H_{\alpha \beta }{}^{\delta } H^{\alpha \beta \gamma } H^{\epsilon \varepsilon \zeta } H^{\eta \theta \iota } \nabla_{\eta }H_{\gamma \epsilon \varepsilon } \nabla_{\iota }H_{\delta \zeta \theta }- \frac{17}{1024} H_{\alpha }{}^{\delta \epsilon } H^{\alpha \beta \gamma } H_{\beta }{}^{\varepsilon \zeta } H^{\eta \theta \iota } \nabla_{\delta }H_{\gamma \eta \theta } \nabla_{\iota }H_{\epsilon \varepsilon \zeta }\nn\\
&&+\frac{109}{1024} H_{\alpha }{}^{\delta \epsilon } H^{\alpha \beta \gamma } H_{\beta }{}^{\varepsilon \zeta } H^{\eta \theta \iota } \nabla_{\theta }H_{\gamma \delta \eta } \nabla_{\iota }H_{\epsilon \varepsilon \zeta }- \frac{37}{512} H_{\alpha }{}^{\delta \epsilon } H^{\alpha \beta \gamma } H_{\beta }{}^{\varepsilon \zeta } H^{\eta \theta \iota } \nabla_{\varepsilon }H_{\gamma \delta \eta } \nabla_{\iota }H_{\epsilon \zeta \theta }\nn\\
&&- \frac{29}{512} H_{\alpha }{}^{\delta \epsilon } H^{\alpha \beta \gamma } H_{\varepsilon }{}^{\theta \iota } H^{\varepsilon \zeta \eta } \nabla_{\delta }H_{\beta \gamma \zeta } \nabla_{\iota }H_{\epsilon \eta \theta }+\frac{1}{1024} H_{\alpha }{}^{\delta \epsilon } H^{\alpha \beta \gamma } H_{\varepsilon }{}^{\theta \iota } H^{\varepsilon \zeta \eta } \nabla_{\zeta }H_{\beta \gamma \delta } \nabla_{\iota }H_{\epsilon \eta \theta }\nn\\
&&+\frac{13}{1024} H_{\alpha \beta }{}^{\delta } H^{\alpha \beta \gamma } H^{\epsilon \varepsilon \zeta } H^{\eta \theta \iota } \nabla_{\delta }H_{\gamma \epsilon \eta } \nabla_{\iota }H_{\varepsilon \zeta \theta }+\frac{35}{256} H_{\alpha }{}^{\delta \epsilon } H^{\alpha \beta \gamma } H_{\beta }{}^{\varepsilon \zeta } H^{\eta \theta \iota } \nabla_{\epsilon }H_{\gamma \delta \eta } \nabla_{\iota }H_{\varepsilon \zeta \theta }\nn\\
&&+\frac{3}{2048} H_{\alpha }{}^{\delta \epsilon } H^{\alpha \beta \gamma } H_{\beta }{}^{\varepsilon \zeta } H^{\eta \theta \iota } \nabla_{\eta }H_{\gamma \delta \epsilon } \nabla_{\iota }H_{\varepsilon \zeta \theta }+\frac{1}{1024} H_{\alpha \beta }{}^{\delta } H^{\alpha \beta \gamma } H_{\gamma }{}^{\epsilon \varepsilon } H^{\zeta \eta \theta } \nabla_{\zeta }H_{\delta \epsilon }{}^{\iota } \nabla_{\iota }H_{\varepsilon \eta \theta }\nn\\
&&- \frac{3}{2048} H_{\alpha \beta }{}^{\delta } H^{\alpha \beta \gamma } H_{\epsilon }{}^{\eta \theta } H^{\epsilon \varepsilon \zeta } \nabla_{\delta }H_{\gamma \varepsilon }{}^{\iota } \nabla_{\iota }H_{\zeta \eta \theta }- \frac{13}{1024} H_{\alpha }{}^{\delta \epsilon } H^{\alpha \beta \gamma } H_{\beta }{}^{\varepsilon \zeta } H^{\eta \theta \iota } \nabla_{\epsilon }H_{\gamma \delta \varepsilon } \nabla_{\iota }H_{\zeta \eta \theta }\nn\\
&&+\frac{9}{256} H_{\alpha }{}^{\delta \epsilon } H^{\alpha \beta \gamma } H_{\beta }{}^{\varepsilon \zeta } H_{\delta }{}^{\eta \theta } \nabla_{\epsilon }H_{\gamma \varepsilon }{}^{\iota } \nabla_{\iota }H_{\zeta \eta \theta }- \frac{1}{128} H_{\alpha }{}^{\delta \epsilon } H^{\alpha \beta \gamma } H_{\beta }{}^{\varepsilon \zeta } H^{\eta \theta \iota } \nabla_{\varepsilon }H_{\gamma \delta \epsilon } \nabla_{\iota }H_{\zeta \eta \theta }\nn\\
&&- \frac{55}{512} H_{\alpha }{}^{\delta \epsilon } H^{\alpha \beta \gamma } H_{\beta }{}^{\varepsilon \zeta } H_{\delta }{}^{\eta \theta } \nabla_{\varepsilon }H_{\gamma \epsilon }{}^{\iota } \nabla_{\iota }H_{\zeta \eta \theta }+\frac{1}{1536} H_{\alpha \beta }{}^{\delta } H^{\alpha \beta \gamma } H_{\gamma }{}^{\epsilon \varepsilon } H^{\zeta \eta \theta } \nabla_{\varepsilon }H_{\delta \epsilon }{}^{\iota } \nabla_{\iota }H_{\zeta \eta \theta }\nn\\
&&+\frac{5}{1024} H_{\alpha \beta }{}^{\delta } H^{\alpha \beta \gamma } H_{\gamma }{}^{\epsilon \varepsilon } H_{\epsilon }{}^{\zeta \eta } \nabla_{\varepsilon }H_{\delta }{}^{\theta \iota } \nabla_{\iota }H_{\zeta \eta \theta }- \frac{27}{128} H_{\alpha }{}^{\delta \epsilon } H^{\alpha \beta \gamma } H_{\beta }{}^{\varepsilon \zeta } H_{\delta }{}^{\eta \theta } \nabla_{\theta }H_{\zeta \eta \iota } \nabla^{\iota }H_{\gamma \epsilon \varepsilon }\nn\\
&&+\frac{55}{512} H_{\alpha }{}^{\delta \epsilon } H^{\alpha \beta \gamma } H_{\beta }{}^{\varepsilon \zeta } H_{\delta }{}^{\eta \theta } \nabla_{\iota }H_{\zeta \eta \theta } \nabla^{\iota }H_{\gamma \epsilon \varepsilon }+\frac{1}{9216} H_{\alpha }{}^{\delta \epsilon } H^{\alpha \beta \gamma } H_{\beta \delta }{}^{\varepsilon } H^{\zeta \eta \theta } \nabla_{\iota }H_{\zeta \eta \theta } \nabla^{\iota }H_{\gamma \epsilon \varepsilon }\nn\\
&&+\frac{19}{1024} H_{\alpha }{}^{\delta \epsilon } H^{\alpha \beta \gamma } H_{\beta \delta }{}^{\varepsilon } H^{\zeta \eta \theta } \nabla_{\theta }H_{\varepsilon \eta \iota } \nabla^{\iota }H_{\gamma \epsilon \zeta }- \frac{3}{1024} H_{\alpha }{}^{\delta \epsilon } H^{\alpha \beta \gamma } H_{\beta \delta }{}^{\varepsilon } H^{\zeta \eta \theta } \nabla_{\iota }H_{\varepsilon \eta \theta } \nabla^{\iota }H_{\gamma \epsilon \zeta }\nn\\
&&+\frac{1}{1024} H_{\alpha }{}^{\delta \epsilon } H^{\alpha \beta \gamma } H_{\beta }{}^{\varepsilon \zeta } H_{\delta \varepsilon }{}^{\eta } \nabla_{\theta }H_{\zeta \eta \iota } \nabla^{\iota }H_{\gamma \epsilon }{}^{\theta }- \frac{1}{1024} H_{\alpha }{}^{\delta \epsilon } H^{\alpha \beta \gamma } H_{\beta }{}^{\varepsilon \zeta } H_{\delta \varepsilon }{}^{\eta } \nabla_{\iota }H_{\zeta \eta \theta } \nabla^{\iota }H_{\gamma \epsilon }{}^{\theta }\nn\\
&&+\frac{3}{512} H_{\alpha \beta }{}^{\delta } H^{\alpha \beta \gamma } H_{\epsilon }{}^{\eta \theta } H^{\epsilon \varepsilon \zeta } \nabla_{\theta }H_{\delta \eta \iota } \nabla^{\iota }H_{\gamma \varepsilon \zeta }- \frac{23}{1024} H_{\alpha }{}^{\delta \epsilon } H^{\alpha \beta \gamma } H_{\beta }{}^{\varepsilon \zeta } H_{\delta }{}^{\eta \theta } \nabla_{\theta }H_{\epsilon \eta \iota } \nabla^{\iota }H_{\gamma \varepsilon \zeta }\nn\\
&&- \frac{3}{2048} H_{\alpha }{}^{\delta \epsilon } H^{\alpha \beta \gamma } H_{\beta }{}^{\varepsilon \zeta } H_{\delta }{}^{\eta \theta } \nabla_{\iota }H_{\epsilon \eta \theta } \nabla^{\iota }H_{\gamma \varepsilon \zeta }- \frac{47}{256} H_{\alpha }{}^{\delta \epsilon } H^{\alpha \beta \gamma } H_{\beta }{}^{\varepsilon \zeta } H_{\delta }{}^{\eta \theta } \nabla_{\zeta }H_{\epsilon \theta \iota } \nabla^{\iota }H_{\gamma \varepsilon \eta }\nn\\
&&- \frac{73}{512} H_{\alpha \beta }{}^{\delta } H^{\alpha \beta \gamma } H_{\epsilon }{}^{\eta \theta } H^{\epsilon \varepsilon \zeta } \nabla_{\theta }H_{\delta \zeta \iota } \nabla^{\iota }H_{\gamma \varepsilon \eta }+\frac{55}{1024} H_{\alpha \beta }{}^{\delta } H^{\alpha \beta \gamma } H_{\epsilon }{}^{\eta \theta } H^{\epsilon \varepsilon \zeta } \nabla_{\iota }H_{\delta \zeta \theta } \nabla^{\iota }H_{\gamma \varepsilon \eta }\nn\\
&&+\frac{1}{512} H_{\alpha }{}^{\delta \epsilon } H^{\alpha \beta \gamma } H_{\beta }{}^{\varepsilon \zeta } H_{\delta }{}^{\eta \theta } \nabla_{\iota }H_{\epsilon \zeta \theta } \nabla^{\iota }H_{\gamma \varepsilon \eta }+\frac{11}{128} H_{\alpha }{}^{\delta \epsilon } H^{\alpha \beta \gamma } H_{\beta \delta }{}^{\varepsilon } H^{\zeta \eta \theta } \nabla_{\varepsilon }H_{\epsilon \theta \iota } \nabla^{\iota }H_{\gamma \zeta \eta }\nn\\
&&- \frac{23}{2048} H_{\alpha \beta }{}^{\delta } H^{\alpha \beta \gamma } H_{\epsilon \varepsilon }{}^{\eta } H^{\epsilon \varepsilon \zeta } \nabla_{\theta }H_{\delta \eta \iota } \nabla^{\iota }H_{\gamma \zeta }{}^{\theta }+\frac{23}{2048} H_{\alpha \beta }{}^{\delta } H^{\alpha \beta \gamma } H_{\epsilon \varepsilon }{}^{\eta } H^{\epsilon \varepsilon \zeta } \nabla_{\iota }H_{\delta \eta \theta } \nabla^{\iota }H_{\gamma \zeta }{}^{\theta }\nn\\
&&- \frac{103}{1024} H_{\alpha }{}^{\delta \epsilon } H^{\alpha \beta \gamma } H_{\beta }{}^{\varepsilon \zeta } H_{\delta }{}^{\eta \theta } \nabla_{\zeta }H_{\epsilon \varepsilon \iota } \nabla^{\iota }H_{\gamma \eta \theta }+\frac{29}{1024} H_{\alpha }{}^{\delta \epsilon } H^{\alpha \beta \gamma } H_{\beta }{}^{\varepsilon \zeta } H_{\delta }{}^{\eta \theta } \nabla_{\iota }H_{\epsilon \varepsilon \zeta } \nabla^{\iota }H_{\gamma \eta \theta }\nn\\
&&+\frac{33}{2048} H_{\alpha }{}^{\delta \epsilon } H^{\alpha \beta \gamma } H_{\beta }{}^{\varepsilon \zeta } H_{\gamma }{}^{\eta \theta } \nabla_{\theta }H_{\zeta \eta \iota } \nabla^{\iota }H_{\delta \epsilon \varepsilon }- \frac{1}{128} H_{\alpha }{}^{\delta \epsilon } H^{\alpha \beta \gamma } H_{\beta }{}^{\varepsilon \zeta } H_{\gamma }{}^{\eta \theta } \nabla_{\iota }H_{\zeta \eta \theta } \nabla^{\iota }H_{\delta \epsilon \varepsilon }\nn\\
&&+\frac{1}{6144} H_{\alpha \beta }{}^{\delta } H^{\alpha \beta \gamma } H_{\gamma }{}^{\epsilon \varepsilon } H^{\zeta \eta \theta } \nabla_{\iota }H_{\zeta \eta \theta } \nabla^{\iota }H_{\delta \epsilon \varepsilon }+\frac{7}{512} H_{\alpha \beta }{}^{\delta } H^{\alpha \beta \gamma } H_{\gamma }{}^{\epsilon \varepsilon } H^{\zeta \eta \theta } \nabla_{\theta }H_{\varepsilon \eta \iota } \nabla^{\iota }H_{\delta \epsilon \zeta }\nn\\
&&- \frac{1}{256} H_{\alpha \beta }{}^{\delta } H^{\alpha \beta \gamma } H_{\gamma }{}^{\epsilon \varepsilon } H^{\zeta \eta \theta } \nabla_{\iota }H_{\varepsilon \eta \theta } \nabla^{\iota }H_{\delta \epsilon \zeta }+\frac{7}{1024} H_{\alpha \beta }{}^{\delta } H^{\alpha \beta \gamma } H_{\gamma }{}^{\epsilon \varepsilon } H_{\epsilon }{}^{\zeta \eta } \nabla_{\theta }H_{\zeta \eta \iota } \nabla^{\iota }H_{\delta \varepsilon }{}^{\theta }\nn\\
&&- \frac{1}{128} H_{\alpha \beta }{}^{\delta } H^{\alpha \beta \gamma } H_{\gamma }{}^{\epsilon \varepsilon } H_{\epsilon }{}^{\zeta \eta } \nabla_{\iota }H_{\zeta \eta \theta } \nabla^{\iota }H_{\delta \varepsilon }{}^{\theta }+\frac{1}{2048} H_{\alpha \beta }{}^{\delta } H^{\alpha \beta \gamma } H_{\gamma }{}^{\epsilon \varepsilon } H^{\zeta \eta \theta } \nabla_{\theta }H_{\epsilon \varepsilon \iota } \nabla^{\iota }H_{\delta \zeta \eta }\nn\\
&&- \frac{7}{2048} H_{\alpha \beta }{}^{\delta } H^{\alpha \beta \gamma } H_{\gamma }{}^{\epsilon \varepsilon } H^{\zeta \eta \theta } \nabla_{\iota }H_{\epsilon \varepsilon \theta } \nabla^{\iota }H_{\delta \zeta \eta }+\frac{11}{512} H_{\alpha \beta }{}^{\delta } H^{\alpha \beta \gamma } H_{\gamma }{}^{\epsilon \varepsilon } H_{\epsilon }{}^{\zeta \eta } \nabla_{\eta }H_{\varepsilon \theta \iota } \nabla^{\iota }H_{\delta \zeta }{}^{\theta }\nn\\
&&- \frac{3}{256} H_{\alpha \beta }{}^{\delta } H^{\alpha \beta \gamma } H_{\gamma }{}^{\epsilon \varepsilon } H_{\epsilon }{}^{\zeta \eta } \nabla_{\theta }H_{\varepsilon \eta \iota } \nabla^{\iota }H_{\delta \zeta }{}^{\theta }+\frac{3}{256} H_{\alpha \beta }{}^{\delta } H^{\alpha \beta \gamma } H_{\gamma }{}^{\epsilon \varepsilon } H_{\epsilon }{}^{\zeta \eta } \nabla_{\iota }H_{\varepsilon \eta \theta } \nabla^{\iota }H_{\delta \zeta }{}^{\theta }\nn\\
&&- \frac{1}{1024} H_{\alpha \beta }{}^{\delta } H^{\alpha \beta \gamma } H_{\gamma }{}^{\epsilon \varepsilon } H_{\epsilon \varepsilon }{}^{\zeta } \nabla_{\theta }H_{\zeta \eta \iota } \nabla^{\iota }H_{\delta }{}^{\eta \theta }+\frac{1}{512} H_{\alpha \beta }{}^{\delta } H^{\alpha \beta \gamma } H_{\gamma }{}^{\epsilon \varepsilon } H^{\zeta \eta \theta } \nabla_{\delta }H_{\eta \theta \iota } \nabla^{\iota }H_{\epsilon \varepsilon \zeta }\nn\\
&&+\frac{81}{2048} H_{\alpha }{}^{\delta \epsilon } H^{\alpha \beta \gamma } H_{\beta \delta }{}^{\varepsilon } H_{\gamma }{}^{\zeta \eta } \nabla_{\theta }H_{\zeta \eta \iota } \nabla^{\iota }H_{\epsilon \varepsilon }{}^{\theta }+\frac{9}{4096} H_{\alpha \beta }{}^{\delta } H^{\alpha \beta \gamma } H_{\gamma }{}^{\epsilon \varepsilon } H_{\delta }{}^{\zeta \eta } \nabla_{\theta }H_{\zeta \eta \iota } \nabla^{\iota }H_{\epsilon \varepsilon }{}^{\theta }\nn\\
&&- \frac{75}{2048} H_{\alpha }{}^{\delta \epsilon } H^{\alpha \beta \gamma } H_{\beta \delta }{}^{\varepsilon } H_{\gamma }{}^{\zeta \eta } \nabla_{\iota }H_{\zeta \eta \theta } \nabla^{\iota }H_{\epsilon \varepsilon }{}^{\theta }- \frac{5}{4096} H_{\alpha \beta }{}^{\delta } H^{\alpha \beta \gamma } H_{\gamma }{}^{\epsilon \varepsilon } H_{\delta }{}^{\zeta \eta } \nabla_{\iota }H_{\zeta \eta \theta } \nabla^{\iota }H_{\epsilon \varepsilon }{}^{\theta }\nn\\
&&- \frac{5}{1024} H_{\alpha \beta }{}^{\delta } H^{\alpha \beta \gamma } H_{\gamma }{}^{\epsilon \varepsilon } H^{\zeta \eta \theta } \nabla_{\delta }H_{\varepsilon \theta \iota } \nabla^{\iota }H_{\epsilon \zeta \eta }- \frac{47}{256} H_{\alpha }{}^{\delta \epsilon } H^{\alpha \beta \gamma } H_{\beta \delta }{}^{\varepsilon } H_{\gamma }{}^{\zeta \eta } \nabla_{\eta }H_{\varepsilon \theta \iota } \nabla^{\iota }H_{\epsilon \zeta }{}^{\theta }\nn\\
&&- \frac{1}{512} H_{\alpha }{}^{\delta \epsilon } H^{\alpha \beta \gamma } H_{\beta \delta }{}^{\varepsilon } H_{\gamma }{}^{\zeta \eta } \nabla_{\theta }H_{\varepsilon \eta \iota } \nabla^{\iota }H_{\epsilon \zeta }{}^{\theta }- \frac{3}{256} H_{\alpha \beta }{}^{\delta } H^{\alpha \beta \gamma } H_{\gamma }{}^{\epsilon \varepsilon } H_{\delta }{}^{\zeta \eta } \nabla_{\theta }H_{\varepsilon \eta \iota } \nabla^{\iota }H_{\epsilon \zeta }{}^{\theta }\nn\\
&&+\frac{1}{512} H_{\alpha }{}^{\delta \epsilon } H^{\alpha \beta \gamma } H_{\beta \delta }{}^{\varepsilon } H_{\gamma }{}^{\zeta \eta } \nabla_{\iota }H_{\varepsilon \eta \theta } \nabla^{\iota }H_{\epsilon \zeta }{}^{\theta }+\frac{3}{256} H_{\alpha \beta }{}^{\delta } H^{\alpha \beta \gamma } H_{\gamma }{}^{\epsilon \varepsilon } H_{\delta }{}^{\zeta \eta } \nabla_{\iota }H_{\varepsilon \eta \theta } \nabla^{\iota }H_{\epsilon \zeta }{}^{\theta }\nn\\
&&+\frac{25}{256} H_{\alpha \beta }{}^{\delta } H^{\alpha \beta \gamma } H_{\gamma }{}^{\epsilon \varepsilon } H_{\epsilon }{}^{\zeta \eta } \nabla_{\delta }H_{\eta \theta \iota } \nabla^{\iota }H_{\varepsilon \zeta }{}^{\theta }+\frac{53}{512} H_{\alpha }{}^{\delta \epsilon } H^{\alpha \beta \gamma } H_{\beta \delta }{}^{\varepsilon } H_{\gamma \epsilon }{}^{\zeta } \nabla_{\theta }H_{\zeta \eta \iota } \nabla^{\iota }H_{\varepsilon }{}^{\eta \theta }\nn\\
&&+\frac{39}{1024} H_{\alpha \beta }{}^{\delta } H^{\alpha \beta \gamma } H_{\gamma }{}^{\epsilon \varepsilon } H_{\delta \epsilon }{}^{\zeta } \nabla_{\theta }H_{\zeta \eta \iota } \nabla^{\iota }H_{\varepsilon }{}^{\eta \theta }- \frac{53}{1024} H_{\alpha }{}^{\delta \epsilon } H^{\alpha \beta \gamma } H_{\beta \delta }{}^{\varepsilon } H_{\gamma \epsilon }{}^{\zeta } \nabla_{\iota }H_{\zeta \eta \theta } \nabla^{\iota }H_{\varepsilon }{}^{\eta \theta }\nn\\
&&- \frac{19}{1024} H_{\alpha \beta }{}^{\delta } H^{\alpha \beta \gamma } H_{\gamma }{}^{\epsilon \varepsilon } H_{\delta \epsilon }{}^{\zeta } \nabla_{\iota }H_{\zeta \eta \theta } \nabla^{\iota }H_{\varepsilon }{}^{\eta \theta }- \frac{1}{36864} H_{\alpha }{}^{\delta \epsilon } H^{\alpha \beta \gamma } H_{\beta \delta }{}^{\varepsilon } H_{\gamma \epsilon \varepsilon } \nabla_{\iota }H_{\zeta \eta \theta } \nabla^{\iota }H^{\zeta \eta \theta }\nn\\
&&- \frac{1}{8192} H_{\alpha \beta }{}^{\delta } H^{\alpha \beta \gamma } H_{\gamma }{}^{\epsilon \varepsilon } H_{\delta \epsilon \varepsilon } \nabla_{\iota }H_{\zeta \eta \theta } \nabla^{\iota }H^{\zeta \eta \theta }\,.
\eeqa
The couplings with the structure  $[H^2 R^2\nabla H]_{31}$ are the following: 
\beqa
&&\frac{1}{64} H^{\alpha \beta \gamma } H^{\delta \epsilon \varepsilon } R_{\beta \delta \zeta \eta } R_{\gamma \theta \epsilon \varepsilon } \nabla_{\alpha }H^{\zeta \eta \theta }+\frac{1}{128} H^{\alpha \beta \gamma } H^{\delta \epsilon \varepsilon } R_{\beta \gamma \delta \zeta } R_{\epsilon \eta \varepsilon \theta } \nabla_{\alpha }H^{\zeta \eta \theta }\nn\\
&&- \frac{3}{64} H^{\alpha \beta \gamma } H^{\delta \epsilon \varepsilon } R_{\gamma }{}^{\eta }{}_{\zeta }{}^{\theta } R_{\epsilon \eta \varepsilon \theta } \nabla^{\zeta }H_{\alpha \beta \delta }+\frac{1}{32} H^{\alpha \beta \gamma } H^{\delta \epsilon \varepsilon } R_{\gamma }{}^{\eta }{}_{\epsilon }{}^{\theta } R_{\varepsilon \eta \zeta \theta } \nabla^{\zeta }H_{\alpha \beta \delta }\nn\\
&&+\frac{1}{64} H^{\alpha \beta \gamma } H^{\delta \epsilon \varepsilon } R_{\gamma }{}^{\eta }{}_{\epsilon }{}^{\theta } R_{\varepsilon \theta \zeta \eta } \nabla^{\zeta }H_{\alpha \beta \delta }- \frac{3}{256} H_{\alpha }{}^{\delta \epsilon } H^{\alpha \beta \gamma } R_{\delta }{}^{\eta }{}_{\varepsilon }{}^{\theta } R_{\epsilon \eta \zeta \theta } \nabla^{\zeta }H_{\beta \gamma }{}^{\varepsilon }\nn\\
&&+\frac{11}{256} H_{\alpha }{}^{\delta \epsilon } H^{\alpha \beta \gamma } R_{\delta }{}^{\eta }{}_{\varepsilon }{}^{\theta } R_{\epsilon \theta \zeta \eta } \nabla^{\zeta }H_{\beta \gamma }{}^{\varepsilon }- \frac{13}{256} H_{\alpha }{}^{\delta \epsilon } H^{\alpha \beta \gamma } R_{\delta }{}^{\eta }{}_{\epsilon }{}^{\theta } R_{\varepsilon \eta \zeta \theta } \nabla^{\zeta }H_{\beta \gamma }{}^{\varepsilon }\nn\\
&&- \frac{3}{64} H_{\alpha }{}^{\delta \epsilon } H^{\alpha \beta \gamma } R_{\gamma }{}^{\eta }{}_{\varepsilon }{}^{\theta } R_{\epsilon \eta \zeta \theta } \nabla^{\zeta }H_{\beta \delta }{}^{\varepsilon }+\frac{3}{64} H_{\alpha }{}^{\delta \epsilon } H^{\alpha \beta \gamma } R_{\gamma }{}^{\eta }{}_{\varepsilon }{}^{\theta } R_{\epsilon \theta \zeta \eta } \nabla^{\zeta }H_{\beta \delta }{}^{\varepsilon }\nn\\
&&+\frac{3}{64} H_{\alpha }{}^{\delta \epsilon } H^{\alpha \beta \gamma } R_{\gamma }{}^{\eta }{}_{\epsilon }{}^{\theta } R_{\varepsilon \eta \zeta \theta } \nabla^{\zeta }H_{\beta \delta }{}^{\varepsilon }- \frac{1}{32} H_{\alpha \beta }{}^{\delta } H^{\alpha \beta \gamma } R_{\delta }{}^{\eta }{}_{\zeta }{}^{\theta } R_{\epsilon \eta \varepsilon \theta } \nabla^{\zeta }H_{\gamma }{}^{\epsilon \varepsilon }\nn\\
&&+\frac{1}{32} H_{\alpha \beta }{}^{\delta } H^{\alpha \beta \gamma } R_{\delta }{}^{\eta }{}_{\epsilon }{}^{\theta } R_{\varepsilon \eta \zeta \theta } \nabla^{\zeta }H_{\gamma }{}^{\epsilon \varepsilon }- \frac{1}{64} H^{\alpha \beta \gamma } H^{\delta \epsilon \varepsilon } R_{\gamma \zeta \delta }{}^{\theta } R_{\epsilon \eta \varepsilon \theta } \nabla^{\eta }H_{\alpha \beta }{}^{\zeta }\nn\\
&&+\frac{1}{64} H^{\alpha \beta \gamma } H^{\delta \epsilon \varepsilon } R_{\gamma }{}^{\theta }{}_{\delta \zeta } R_{\epsilon \eta \varepsilon \theta } \nabla^{\eta }H_{\alpha \beta }{}^{\zeta }+\frac{1}{128} H^{\alpha \beta \gamma } H^{\delta \epsilon \varepsilon } R_{\gamma }{}^{\theta }{}_{\delta \epsilon } R_{\varepsilon \eta \zeta \theta } \nabla^{\eta }H_{\alpha \beta }{}^{\zeta }\nn\\
&&+\frac{1}{64} H^{\alpha \beta \gamma } H^{\delta \epsilon \varepsilon } R_{\beta \zeta \epsilon }{}^{\theta } R_{\gamma \eta \varepsilon \theta } \nabla^{\eta }H_{\alpha \delta }{}^{\zeta }- \frac{1}{64} H^{\alpha \beta \gamma } H^{\delta \epsilon \varepsilon } R_{\beta \zeta \epsilon }{}^{\theta } R_{\gamma \theta \varepsilon \eta } \nabla^{\eta }H_{\alpha \delta }{}^{\zeta }\nn\\
&&+\frac{1}{64} H^{\alpha \beta \gamma } H^{\delta \epsilon \varepsilon } R_{\beta \zeta \gamma }{}^{\theta } R_{\epsilon \eta \varepsilon \theta } \nabla^{\eta }H_{\alpha \delta }{}^{\zeta }+\frac{1}{32} H^{\alpha \beta \gamma } H^{\delta \epsilon \varepsilon } R_{\beta \epsilon \gamma }{}^{\theta } R_{\varepsilon \eta \zeta \theta } \nabla^{\eta }H_{\alpha \delta }{}^{\zeta }\nn\\
&&- \frac{1}{16} H^{\alpha \beta \gamma } H^{\delta \epsilon \varepsilon } R_{\beta \epsilon \gamma }{}^{\theta } R_{\varepsilon \theta \zeta \eta } \nabla^{\eta }H_{\alpha \delta }{}^{\zeta }+\frac{1}{32} H_{\alpha }{}^{\delta \epsilon } H^{\alpha \beta \gamma } R_{\gamma \eta \varepsilon }{}^{\theta } R_{\delta \zeta \epsilon \theta } \nabla^{\eta }H_{\beta }{}^{\varepsilon \zeta }\nn\\
&&- \frac{3}{32} H_{\alpha }{}^{\delta \epsilon } H^{\alpha \beta \gamma } R_{\gamma }{}^{\theta }{}_{\varepsilon \eta } R_{\delta \zeta \epsilon \theta } \nabla^{\eta }H_{\beta }{}^{\varepsilon \zeta }+\frac{1}{128} H_{\alpha }{}^{\delta \epsilon } H^{\alpha \beta \gamma } R_{\gamma }{}^{\theta }{}_{\varepsilon \zeta } R_{\delta \eta \epsilon \theta } \nabla^{\eta }H_{\beta }{}^{\varepsilon \zeta }\nn\\
&&- \frac{1}{64} H_{\alpha }{}^{\delta \epsilon } H^{\alpha \beta \gamma } R_{\gamma }{}^{\theta }{}_{\delta \varepsilon } R_{\epsilon \eta \zeta \theta } \nabla^{\eta }H_{\beta }{}^{\varepsilon \zeta }+\frac{1}{64} H_{\alpha }{}^{\delta \epsilon } H^{\alpha \beta \gamma } R_{\gamma }{}^{\theta }{}_{\delta \varepsilon } R_{\epsilon \theta \zeta \eta } \nabla^{\eta }H_{\beta }{}^{\varepsilon \zeta }\nn\\
&&- \frac{1}{128} H_{\alpha }{}^{\delta \epsilon } H^{\alpha \beta \gamma } R_{\gamma }{}^{\theta }{}_{\delta \epsilon } R_{\varepsilon \eta \zeta \theta } \nabla^{\eta }H_{\beta }{}^{\varepsilon \zeta }- \frac{1}{32} H^{\alpha \beta \gamma } H^{\delta \epsilon \varepsilon } R_{\beta \zeta \delta \eta } R_{\gamma \theta \epsilon \varepsilon } \nabla^{\theta }H_{\alpha }{}^{\zeta \eta }\nn\\
&&- \frac{1}{64} H^{\alpha \beta \gamma } H^{\delta \epsilon \varepsilon } R_{\beta \zeta \delta \epsilon } R_{\gamma \theta \varepsilon \eta } \nabla^{\theta }H_{\alpha }{}^{\zeta \eta }- \frac{3}{128} H^{\alpha \beta \gamma } H^{\delta \epsilon \varepsilon } R_{\beta \delta \gamma \epsilon } R_{\varepsilon \theta \zeta \eta } \nabla^{\theta }H_{\alpha }{}^{\zeta \eta }\nn\\
&&- \frac{1}{16} H_{\alpha }{}^{\delta \epsilon } H^{\alpha \beta \gamma } R_{\beta \delta \gamma \varepsilon } R_{\epsilon \theta \zeta \eta } \nabla^{\theta }H^{\varepsilon \zeta \eta }\,.
\eeqa
The couplings with the structure  $[H^4 R\nabla H]_{52}$ are the following: 
\beqa
&&\frac{1}{256} H_{\alpha }{}^{\delta \epsilon } H^{\alpha \beta \gamma } H_{\beta }{}^{\varepsilon \zeta } H_{\delta }{}^{\eta \theta } R_{\epsilon \theta \zeta \iota } \nabla_{\varepsilon }H_{\gamma \eta }{}^{\iota }+\frac{19}{512} H_{\alpha }{}^{\delta \epsilon } H^{\alpha \beta \gamma } H_{\beta }{}^{\varepsilon \zeta } H^{\eta \theta \iota } R_{\gamma \theta \zeta \iota } \nabla_{\varepsilon }H_{\delta \epsilon \eta }\nn\\
&&- \frac{5}{512} H_{\alpha \beta }{}^{\delta } H^{\alpha \beta \gamma } H_{\gamma }{}^{\epsilon \varepsilon } H_{\epsilon }{}^{\zeta \eta } R_{\zeta \theta \eta \iota } \nabla_{\varepsilon }H_{\delta }{}^{\theta \iota }- \frac{19}{512} H_{\alpha }{}^{\delta \epsilon } H^{\alpha \beta \gamma } H_{\varepsilon }{}^{\theta \iota } H^{\varepsilon \zeta \eta } R_{\epsilon \theta \eta \iota } \nabla_{\zeta }H_{\beta \gamma \delta }\nn\\
&&- \frac{19}{1024} H_{\alpha }{}^{\delta \epsilon } H^{\alpha \beta \gamma } H_{\beta \delta }{}^{\varepsilon } H^{\zeta \eta \theta } R_{\varepsilon \iota \eta \theta } \nabla_{\zeta }H_{\gamma \epsilon }{}^{\iota }- \frac{13}{1024} H_{\alpha \beta }{}^{\delta } H^{\alpha \beta \gamma } H_{\epsilon }{}^{\eta \theta } H^{\epsilon \varepsilon \zeta } R_{\delta \iota \eta \theta } \nabla_{\zeta }H_{\gamma \varepsilon }{}^{\iota }\nn\\
&&+\frac{51}{1024} H_{\alpha }{}^{\delta \epsilon } H^{\alpha \beta \gamma } H_{\beta }{}^{\varepsilon \zeta } H_{\delta }{}^{\eta \theta } R_{\epsilon \iota \eta \theta } \nabla_{\zeta }H_{\gamma \varepsilon }{}^{\iota }+\frac{5}{1024} H_{\alpha \beta }{}^{\delta } H^{\alpha \beta \gamma } H_{\gamma }{}^{\epsilon \varepsilon } H^{\zeta \eta \theta } R_{\varepsilon \iota \eta \theta } \nabla_{\zeta }H_{\delta \epsilon }{}^{\iota }\nn\\
&&+\frac{19}{512} H_{\alpha \beta }{}^{\delta } H^{\alpha \beta \gamma } H_{\gamma }{}^{\epsilon \varepsilon } H_{\epsilon }{}^{\zeta \eta } R_{\varepsilon \theta \eta \iota } \nabla_{\zeta }H_{\delta }{}^{\theta \iota }- \frac{3}{4096} H_{\alpha \beta }{}^{\delta } H^{\alpha \beta \gamma } H_{\gamma }{}^{\epsilon \varepsilon } H^{\zeta \eta \theta } R_{\delta \iota \eta \theta } \nabla_{\zeta }H_{\epsilon \varepsilon }{}^{\iota }\nn\\
&&- \frac{15}{256} H_{\alpha }{}^{\delta \epsilon } H^{\alpha \beta \gamma } H_{\beta \delta }{}^{\varepsilon } H_{\gamma }{}^{\zeta \eta } R_{\varepsilon \theta \eta \iota } \nabla_{\zeta }H_{\epsilon }{}^{\theta \iota }+\frac{9}{1024} H_{\alpha \beta }{}^{\delta } H^{\alpha \beta \gamma } H_{\gamma }{}^{\epsilon \varepsilon } H_{\epsilon }{}^{\zeta \eta } R_{\delta \theta \eta \iota } \nabla_{\zeta }H_{\varepsilon }{}^{\theta \iota }\nn\\
&&+\frac{9}{1024} H_{\alpha \beta }{}^{\delta } H^{\alpha \beta \gamma } H^{\epsilon \varepsilon \zeta } H^{\eta \theta \iota } R_{\delta \theta \zeta \iota } \nabla_{\eta }H_{\gamma \epsilon \varepsilon }+\frac{9}{1024} H_{\alpha \beta }{}^{\delta } H^{\alpha \beta \gamma } H_{\epsilon }{}^{\eta \theta } H^{\epsilon \varepsilon \zeta } R_{\delta \theta \zeta \iota } \nabla_{\eta }H_{\gamma \varepsilon }{}^{\iota }\nn\\
&&- \frac{1}{1024} H_{\alpha \beta }{}^{\delta } H^{\alpha \beta \gamma } H_{\epsilon }{}^{\eta \theta } H^{\epsilon \varepsilon \zeta } R_{\delta \iota \zeta \theta } \nabla_{\eta }H_{\gamma \varepsilon }{}^{\iota }- \frac{13}{256} H_{\alpha }{}^{\delta \epsilon } H^{\alpha \beta \gamma } H_{\beta }{}^{\varepsilon \zeta } H_{\delta }{}^{\eta \theta } R_{\epsilon \theta \zeta \iota } \nabla_{\eta }H_{\gamma \varepsilon }{}^{\iota }\nn\\
&&+\frac{7}{128} H_{\alpha }{}^{\delta \epsilon } H^{\alpha \beta \gamma } H_{\beta \delta }{}^{\varepsilon } H^{\zeta \eta \theta } R_{\epsilon \theta \varepsilon \iota } \nabla_{\eta }H_{\gamma \zeta }{}^{\iota }- \frac{5}{256} H_{\alpha \beta }{}^{\delta } H^{\alpha \beta \gamma } H_{\gamma }{}^{\epsilon \varepsilon } H^{\zeta \eta \theta } R_{\epsilon \theta \varepsilon \iota } \nabla_{\eta }H_{\delta \zeta }{}^{\iota }\nn\\
&&+\frac{13}{256} H_{\alpha \beta }{}^{\delta } H^{\alpha \beta \gamma } H_{\gamma }{}^{\epsilon \varepsilon } H^{\zeta \eta \theta } R_{\delta \theta \varepsilon \iota } \nabla_{\eta }H_{\epsilon \zeta }{}^{\iota }+\frac{35}{1024} H_{\alpha \beta }{}^{\delta } H^{\alpha \beta \gamma } H_{\gamma }{}^{\epsilon \varepsilon } H^{\zeta \eta \theta } R_{\delta \iota \varepsilon \theta } \nabla_{\eta }H_{\epsilon \zeta }{}^{\iota }\nn\\
&&- \frac{9}{1024} H_{\alpha \beta }{}^{\delta } H^{\alpha \beta \gamma } H_{\epsilon }{}^{\eta \theta } H^{\epsilon \varepsilon \zeta } R_{\gamma \theta \delta \iota } \nabla_{\eta }H_{\varepsilon \zeta }{}^{\iota }+\frac{5}{1024} H_{\alpha }{}^{\delta \epsilon } H^{\alpha \beta \gamma } H_{\beta }{}^{\varepsilon \zeta } H_{\delta }{}^{\eta \theta } R_{\gamma \theta \epsilon \iota } \nabla_{\eta }H_{\varepsilon \zeta }{}^{\iota }\nn\\
&&- \frac{5}{512} H_{\alpha }{}^{\delta \epsilon } H^{\alpha \beta \gamma } H_{\beta }{}^{\varepsilon \zeta } H_{\delta }{}^{\eta \theta } R_{\gamma \iota \epsilon \theta } \nabla_{\eta }H_{\varepsilon \zeta }{}^{\iota }- \frac{9}{256} H_{\alpha }{}^{\delta \epsilon } H^{\alpha \beta \gamma } H_{\varepsilon }{}^{\theta \iota } H^{\varepsilon \zeta \eta } R_{\delta \eta \epsilon \iota } \nabla_{\theta }H_{\beta \gamma \zeta }\nn\\
&&+\frac{1}{256} H_{\alpha }{}^{\delta \epsilon } H^{\alpha \beta \gamma } H_{\beta }{}^{\varepsilon \zeta } H^{\eta \theta \iota } R_{\epsilon \iota \varepsilon \zeta } \nabla_{\theta }H_{\gamma \delta \eta }+\frac{1}{256} H_{\alpha }{}^{\delta \epsilon } H^{\alpha \beta \gamma } H_{\beta }{}^{\varepsilon \zeta } H_{\delta }{}^{\eta \theta } R_{\epsilon \iota \varepsilon \zeta } \nabla_{\theta }H_{\gamma \eta }{}^{\iota }\nn\\
&&- \frac{23}{2048} H_{\alpha }{}^{\delta \epsilon } H^{\alpha \beta \gamma } H_{\beta }{}^{\varepsilon \zeta } H^{\eta \theta \iota } R_{\gamma \iota \varepsilon \zeta } \nabla_{\theta }H_{\delta \epsilon \eta }+\frac{1}{512} H_{\alpha }{}^{\delta \epsilon } H^{\alpha \beta \gamma } H_{\beta }{}^{\varepsilon \zeta } H^{\eta \theta \iota } R_{\gamma \epsilon \zeta \iota } \nabla_{\theta }H_{\delta \varepsilon \eta }\nn\\
&&+\frac{19}{1024} H_{\alpha }{}^{\delta \epsilon } H^{\alpha \beta \gamma } H_{\beta \delta }{}^{\varepsilon } H^{\zeta \eta \theta } R_{\varepsilon \iota \eta \theta } \nabla^{\iota }H_{\gamma \epsilon \zeta }- \frac{13}{2048} H_{\alpha }{}^{\delta \epsilon } H^{\alpha \beta \gamma } H_{\beta }{}^{\varepsilon \zeta } H_{\delta }{}^{\eta \theta } R_{\epsilon \iota \eta \theta } \nabla^{\iota }H_{\gamma \varepsilon \zeta }\nn\\
&&+\frac{1}{256} H_{\alpha }{}^{\delta \epsilon } H^{\alpha \beta \gamma } H_{\beta }{}^{\varepsilon \zeta } H_{\delta }{}^{\eta \theta } R_{\epsilon \theta \zeta \iota } \nabla^{\iota }H_{\gamma \varepsilon \eta }- \frac{5}{1024} H_{\alpha \beta }{}^{\delta } H^{\alpha \beta \gamma } H_{\gamma }{}^{\epsilon \varepsilon } H^{\zeta \eta \theta } R_{\varepsilon \iota \eta \theta } \nabla^{\iota }H_{\delta \epsilon \zeta }\nn\\
&&- \frac{21}{1024} H_{\alpha \beta }{}^{\delta } H^{\alpha \beta \gamma } H_{\gamma }{}^{\epsilon \varepsilon } H^{\zeta \eta \theta } R_{\epsilon \theta \varepsilon \iota } \nabla^{\iota }H_{\delta \zeta \eta }- \frac{3}{512} H_{\alpha \beta }{}^{\delta } H^{\alpha \beta \gamma } H_{\gamma }{}^{\epsilon \varepsilon } H_{\epsilon }{}^{\zeta \eta } R_{\varepsilon \iota \eta \theta } \nabla^{\iota }H_{\delta \zeta }{}^{\theta }\nn\\
&&- \frac{7}{2048} H_{\alpha \beta }{}^{\delta } H^{\alpha \beta \gamma } H_{\gamma }{}^{\epsilon \varepsilon } H_{\epsilon \varepsilon }{}^{\zeta } R_{\zeta \iota \eta \theta } \nabla^{\iota }H_{\delta }{}^{\eta \theta }- \frac{3}{4096} H_{\alpha \beta }{}^{\delta } H^{\alpha \beta \gamma } H_{\gamma }{}^{\epsilon \varepsilon } H^{\zeta \eta \theta } R_{\delta \iota \eta \theta } \nabla^{\iota }H_{\epsilon \varepsilon \zeta }\nn\\
&&- \frac{5}{1024} H_{\alpha \beta }{}^{\delta } H^{\alpha \beta \gamma } H_{\gamma }{}^{\epsilon \varepsilon } H_{\delta }{}^{\zeta \eta } R_{\zeta \theta \eta \iota } \nabla^{\iota }H_{\epsilon \varepsilon }{}^{\theta }- \frac{15}{512} H_{\alpha \beta }{}^{\delta } H^{\alpha \beta \gamma } H_{\gamma }{}^{\epsilon \varepsilon } H^{\zeta \eta \theta } R_{\delta \theta \varepsilon \iota } \nabla^{\iota }H_{\epsilon \zeta \eta }\nn\\
&&- \frac{9}{1024} H_{\alpha \beta }{}^{\delta } H^{\alpha \beta \gamma } H_{\gamma }{}^{\epsilon \varepsilon } H^{\zeta \eta \theta } R_{\delta \iota \varepsilon \theta } \nabla^{\iota }H_{\epsilon \zeta \eta }- \frac{19}{256} H_{\alpha }{}^{\delta \epsilon } H^{\alpha \beta \gamma } H_{\beta \delta }{}^{\varepsilon } H_{\gamma }{}^{\zeta \eta } R_{\varepsilon \theta \eta \iota } \nabla^{\iota }H_{\epsilon \zeta }{}^{\theta }\nn\\
&&+\frac{1}{256} H_{\alpha \beta }{}^{\delta } H^{\alpha \beta \gamma } H_{\gamma }{}^{\epsilon \varepsilon } H_{\delta }{}^{\zeta \eta } R_{\varepsilon \theta \eta \iota } \nabla^{\iota }H_{\epsilon \zeta }{}^{\theta }+\frac{11}{128} H_{\alpha }{}^{\delta \epsilon } H^{\alpha \beta \gamma } H_{\beta \delta }{}^{\varepsilon } H_{\gamma }{}^{\zeta \eta } R_{\varepsilon \iota \eta \theta } \nabla^{\iota }H_{\epsilon \zeta }{}^{\theta }\nn\\
&&- \frac{3}{512} H_{\alpha }{}^{\delta \epsilon } H^{\alpha \beta \gamma } H_{\beta }{}^{\varepsilon \zeta } H_{\delta }{}^{\eta \theta } R_{\gamma \theta \epsilon \iota } \nabla^{\iota }H_{\varepsilon \zeta \eta }+\frac{1}{256} H_{\alpha }{}^{\delta \epsilon } H^{\alpha \beta \gamma } H_{\beta }{}^{\varepsilon \zeta } H_{\delta }{}^{\eta \theta } R_{\gamma \iota \epsilon \theta } \nabla^{\iota }H_{\varepsilon \zeta \eta }\nn\\
&&+\frac{1}{1024} H_{\alpha \beta }{}^{\delta } H^{\alpha \beta \gamma } H_{\gamma }{}^{\epsilon \varepsilon } H_{\epsilon }{}^{\zeta \eta } R_{\delta \theta \eta \iota } \nabla^{\iota }H_{\varepsilon \zeta }{}^{\theta }- \frac{1}{1024} H_{\alpha \beta }{}^{\delta } H^{\alpha \beta \gamma } H_{\gamma }{}^{\epsilon \varepsilon } H_{\epsilon }{}^{\zeta \eta } R_{\delta \iota \eta \theta } \nabla^{\iota }H_{\varepsilon \zeta }{}^{\theta }\nn\\
&&+\frac{1}{256} H_{\alpha }{}^{\delta \epsilon } H^{\alpha \beta \gamma } H_{\beta \delta }{}^{\varepsilon } H_{\gamma \epsilon }{}^{\zeta } R_{\zeta \iota \eta \theta } \nabla^{\iota }H_{\varepsilon }{}^{\eta \theta }+\frac{1}{256} H_{\alpha \beta }{}^{\delta } H^{\alpha \beta \gamma } H_{\gamma }{}^{\epsilon \varepsilon } H_{\delta \epsilon }{}^{\zeta } R_{\zeta \iota \eta \theta } \nabla^{\iota }H_{\varepsilon }{}^{\eta \theta }\nn\\
&&- \frac{3}{2048} H_{\alpha \beta }{}^{\delta } H^{\alpha \beta \gamma } H_{\gamma }{}^{\epsilon \varepsilon } H^{\zeta \eta \theta } R_{\delta \iota \epsilon \varepsilon } \nabla^{\iota }H_{\zeta \eta \theta }- \frac{13}{1024} H_{\alpha \beta }{}^{\delta } H^{\alpha \beta \gamma } H_{\gamma }{}^{\epsilon \varepsilon } H_{\epsilon }{}^{\zeta \eta } R_{\delta \theta \varepsilon \iota } \nabla^{\iota }H_{\zeta \eta }{}^{\theta }\nn\\
&&+\frac{3}{256} H_{\alpha \beta }{}^{\delta } H^{\alpha \beta \gamma } H_{\gamma }{}^{\epsilon \varepsilon } H_{\epsilon }{}^{\zeta \eta } R_{\delta \iota \varepsilon \theta } \nabla^{\iota }H_{\zeta \eta }{}^{\theta }+\frac{5}{512} H_{\alpha }{}^{\delta \epsilon } H^{\alpha \beta \gamma } H_{\beta \delta }{}^{\varepsilon } H_{\gamma }{}^{\zeta \eta } R_{\epsilon \theta \varepsilon \iota } \nabla^{\iota }H_{\zeta \eta }{}^{\theta }\,.
\eeqa

\end{document}